\documentclass[12pt]{article}%
\usepackage[sort]{cite}
\usepackage{graphicx}
\usepackage{multicol}
\usepackage{amsfonts}
\usepackage{amssymb}
\usepackage{amsmath}
\usepackage{heck}
\usepackage{afterpage}
\usepackage{setspace}
\usepackage{verbatim}
\usepackage{color}
\usepackage{longtable}
\usepackage{standalone}
\usepackage{float}
\usepackage{subcaption}
\usepackage{epsfig}
\usepackage{adjustbox}
\usepackage{tikz}
\usepackage{soul}
\usepackage{mathrsfs}
\usepackage{xypic}
\usepackage[margin=1in]{geometry}
\usepackage{titletoc}
\usepackage{hyperref}%
\providecommand{\U}[1]{\protect\rule{.1in}{.1in}}
\pdfoutput=1
\newsavebox{\mysavebox}

\hypersetup{colorlinks,citecolor=black,filecolor=black,linkcolor=black,urlcolor=black,pdftex}
\numberwithin{equation}{section}

\newcommand{\ba}{\begin{eqnarray}}
\newcommand{\ea}{\end{eqnarray}}

\newcommand{\be}{\begin{equation}}
\newcommand{\ee}{\end{equation}}
\newcommand\MAT[1]{ \left(\begin{matrix} #1 \end{matrix}\right) }

\usepackage{tikz}
\usetikzlibrary{arrows}
\usetikzlibrary{arrows.meta}
\usetikzlibrary{shapes.geometric,calc,arrows, positioning,shapes.misc,decorations.markings}
\tikzset{
  big arrow/.style={
    decoration={markings,mark=at position 1 with {\arrow[scale=2,#1]{>}}},
    postaction={decorate},
    shorten >=0.4pt},
  big arrow/.default=black}

\pgfdeclarelayer{edgelayer}
\pgfdeclarelayer{nodelayer}
\pgfsetlayers{edgelayer,nodelayer,main}
\tikzstyle{none}=[inner sep=0pt]

\tikzstyle{NodeCross}=[draw, shape=circle, cross out, inner sep=0pt, minimum size=6pt,line width=0.25mm]
\tikzstyle{NodeCrossRed}=[draw, shape=circle, red, cross out, inner sep=0pt, minimum size=6pt,line width=0.25mm]
\tikzstyle{Circle}=[draw, shape=circle, black, fill=black, inner sep=0pt, minimum size=6pt]
\tikzstyle{Star}=[draw, shape=star, fill=black, star points=8, inner sep=0pt, minimum size=8pt]
\tikzstyle{CircleRed}=[draw, shape=circle, black, fill=red, inner sep=0pt, minimum size=4pt]
\tikzstyle{StarP}=[draw={rgb,255: red,128; green,0; blue,128}, shape=star, fill={rgb,256: red,128; green,0; blue,128}, star points=8, inner sep=0pt, minimum size=12pt]

\tikzstyle{DashedLine}=[-, densely dashed, line width=0.25mm]
\tikzstyle{DottedLine}=[-, dotted, line width=0.25mm]
\tikzstyle{ThickLine}=[-, line width=0.25mm]
\tikzstyle{ArrowLineRight}=[-, -{Stealth[scale=1.75]}, line width=0.1mm, scale=5]
\tikzstyle{ArrowLineRed}=[-, draw={rgb,255: red,191; green,0; blue,0}, -{Stealth[scale=1.75]}, line width=0.1mm, scale=5]
\tikzstyle{RedLine}=[-, draw={rgb,255: red,191; green,0; blue,0}, fill=none, line width=0.25mm]
\tikzstyle{DashedLineThin}=[-, densely dashed, line width=0.125mm, fill=none, draw=black]
\tikzstyle{DottedRed}=[-, dotted, draw={rgb,255: red,191; green,0; blue,0}, fill=none, line width=0.25mm]
\tikzstyle{DashedRed}=[-, densely dashed, draw={rgb,255: red,191; green,0; blue,0}, fill=none, line width=0.25mm]
\tikzstyle{BlueLine}=[-, draw={rgb,255: red,0; green,0; blue,191}, fill=none, line width=0.25mm]

\tikzstyle{LightBlue}=[-, draw={rgb,255: red,0; green,0; blue,191}, fill=none, line width=0.1mm]
\tikzstyle{BlueDottedLight}=[-, dotted, draw={rgb,255: red,0; green,0; blue,191}, fill=none, line width=0.1mm]

\begin{document}

\date{April 2023}

\begin{flushright}
KCL-PH-TH/2023-22
\end{flushright}
\title{Junctions, Edge Modes,\\[4mm]and $G_2$-Holonomy Orbifolds}

\institution{ICTP}{\centerline{$^{1}$Abdus Salam International Centre for Theoretical Physics, 34151, Trieste, Italy}}
\institution{KINGS}{\centerline{$^{2}$Department of Physics, Kings College London, London, WC2R 2LS, UK}}
\institution{UPPSALAMATH}{\centerline{$^{3}$Department of Mathematics, Uppsala University, Uppsala, Sweden}}
\institution{UPPSALAPHYS}{\centerline{$^{4}$Department of Physics and Astronomy, Uppsala University, Uppsala, Sweden}}
\institution{PENN}{\centerline{$^{5}$Department of Physics and Astronomy, University of Pennsylvania, Philadelphia, PA 19104, USA}}
\institution{PENNmath}{\centerline{$^{6}$Department of Mathematics, University of Pennsylvania, Philadelphia, PA 19104, USA}}

\authors{
Bobby S. Acharya\worksat{\ICTP,\KINGS}\footnote{e-mail: \texttt{bacharya@ictp.it}},
Michele Del Zotto\worksat{\UPPSALAMATH,\UPPSALAPHYS}\footnote{e-mail: \texttt{michele.delzotto@math.uu.se}},\\[4mm]
Jonathan J. Heckman\worksat{\PENN,\PENNmath}\footnote{e-mail: \texttt{jheckman@sas.upenn.edu}},
Max H\"ubner\worksat{\PENN}\footnote{e-mail: \texttt{hmax@sas.upenn.edu}}, and
Ethan Torres\worksat{\PENN}\footnote{e-mail: \texttt{emtorres@sas.upenn.edu}}
}

\abstract{One of the general strategies for realizing a wide class of interacting QFTs is via
junctions and intersections of higher-dimensional bulk theories. In the context of string/M-theory, this includes
many $D > 4$ superconformal field theories (SCFTs) coupled to an IR free bulk. Gauging the flavor symmetries of these theories and allowing position dependent gauge couplings provides a general strategy for realizing novel higher-dimensional junctions of theories coupled to localized edge modes. Here, we show that M-theory on singular, asymptotically conical $G_2$-holonomy orbifolds provides a general template for realizing strongly coupled 5D bulk theories with 4D $\mathcal{N} = 1$ edge modes. This geometric approach also shows how bulk generalized symmetries are inherited in the boundary system.}

\maketitle

\enlargethispage{\baselineskip}

\setcounter{tocdepth}{2}

\tableofcontents

\newpage

\section{Introduction} \label{sec:INTRO}

A general theme in much recent work in high energy theory are bulk / boundary correspondences. For example,
the anomalies and the global structures of $D$-dimensional quantum field theory can be understood in terms of 
$D+1$-dimensional topological field theories, and in many cases this can be used to obtained important information on both the symmetries as well as degrees of freedom localized as ``edge modes'' of a system
(see e.g., \cite{Moore:1988qv,Freed:2012bs,Freed:2022qnc,Gaiotto:2014kfa,Wan:2018bns,Thorngren:2020yht,
Gaiotto:2020iye,Gukov:2020btk,Apruzzi:2021nmk,DelZotto:2022ras,Apruzzi:2022dlm,Kaidi:2022cpf,Bashmakov:2022uek}
for a partial list of references). More broadly, this has of course been an important theme in the study of a wide variety of systems, ranging from condensed matter phenomena, to quantum gravity / holography. It is also the template for engineering a wide variety of quantum field theories in string / M-theory, where the QFT degrees of freedom are localized excitations in a higher-dimensional gravitational theory.

In this latter context one usually considers special kinds of singularities, which could be in the metric and/or field profiles of the bulk supergravity theory, or via the presence of brane probes / solitonic objects. In many cases of interest, the global symmetries of
various QFT excitations are then realized in the higher-dimensional bulk system either via pure geometry or via additional higher-dimensional ``flavor branes''. Giving a uniform characterization of such symmetries is an important question, especially in systems that evade conventional Lagrangian descriptions, and there has recently been much progress in understanding various generalized symmetries in this, and related settings.\footnote{For a partial list
of recent work in this direction see e.g.,
\cite{Gaiotto:2010be,Kapustin:2013qsa,Kapustin:2013uxa,Aharony:2013hda,Gaiotto:2014kfa,
DelZotto:2015isa,Sharpe:2015mja, Heckman:2017uxe, Tachikawa:2017gyf,
Cordova:2018cvg,Benini:2018reh,Hsin:2018vcg,Wan:2018bns,
Thorngren:2019iar,GarciaEtxebarria:2019caf,Eckhard:2019jgg,Wan:2019soo,Bergman:2020ifi,Morrison:2020ool,
Albertini:2020mdx,Hsin:2020nts,Bah:2020uev,DelZotto:2020esg,Hason:2020yqf,Bhardwaj:2020phs,
Apruzzi:2020zot,Cordova:2020tij,Thorngren:2020aph,DelZotto:2020sop,BenettiGenolini:2020doj,
Yu:2020twi,Bhardwaj:2020ymp,DeWolfe:2020uzb,Gukov:2020btk,Iqbal:2020lrt,Hidaka:2020izy,
Brennan:2020ehu,Komargodski:2020mxz,Closset:2020afy,Thorngren:2020yht,Closset:2020scj,
Bhardwaj:2021pfz,Nguyen:2021naa,Heidenreich:2021xpr,Apruzzi:2021phx,Apruzzi:2021vcu,
Hosseini:2021ged,Cvetic:2021sxm,Buican:2021xhs,Bhardwaj:2021zrt,Iqbal:2021rkn,Braun:2021sex,
Cvetic:2021maf,Closset:2021lhd,Thorngren:2021yso,Sharpe:2021srf,Bhardwaj:2021wif,Hidaka:2021mml,
Lee:2021obi,Lee:2021crt,Hidaka:2021kkf,Koide:2021zxj,Apruzzi:2021mlh,Kaidi:2021xfk,Choi:2021kmx,
Bah:2021brs,Gukov:2021swm,Closset:2021lwy,Yu:2021zmu,Apruzzi:2021nmk,Beratto:2021xmn,Bhardwaj:2021mzl,
Debray:2021vob, Wang:2021vki,
Cvetic:2022uuu,DelZotto:2022fnw,Cvetic:2022imb,DelZotto:2022joo,DelZotto:2022ras,Bhardwaj:2022yxj,Hayashi:2022fkw,
Kaidi:2022uux,Roumpedakis:2022aik,Choi:2022jqy,
Choi:2022zal,Arias-Tamargo:2022nlf,Cordova:2022ieu, Bhardwaj:2022dyt,
Benedetti:2022zbb, Bhardwaj:2022scy,Antinucci:2022eat,Carta:2022spy,
Apruzzi:2022dlm, Heckman:2022suy, Baume:2022cot, Choi:2022rfe,
Bhardwaj:2022lsg, Lin:2022xod, Bartsch:2022mpm, Apruzzi:2022rei,
GarciaEtxebarria:2022vzq, Heckman:2022muc, Heckman:2022xgu, Cherman:2022eml, Lu:2022ver, Niro:2022ctq, Kaidi:2022cpf,
Mekareeya:2022spm, vanBeest:2022fss, Antinucci:2022vyk, Giaccari:2022xgs, Bashmakov:2022uek,Cordova:2022fhg,
GarciaEtxebarria:2022jky, Choi:2022fgx, Robbins:2022wlr, Bhardwaj:2022kot, Bhardwaj:2022maz, Bartsch:2022ytj, Gaiotto:2020iye,Agrawal:2015dbf, Robbins:2021ibx, Robbins:2021xce,Huang:2021zvu,
Inamura:2021szw, Cherman:2021nox,Sharpe:2022ene,Bashmakov:2022jtl, Inamura:2022lun, Damia:2022bcd, Lin:2022dhv,Burbano:2021loy, Damia:2022rxw, Apte:2022xtu, Nawata:2023rdx, Bhardwaj:2023zix, Kaidi:2023maf, Debray:2023yrs, DelZotto:2023ahf, Etheredge:2023ler, Lin:2023uvm, Amariti:2023hev, Carta:2023bqn, Koide:2023rqd} and \cite{Cordova:2022ruw} for a recent review.}
 Especially in the context of QFTs which resist a conventional weakly coupled Lagrangian description, the characterization of bulk / boundary correspondences, and the associated symmetries often relies on structures which can be extracted from the extra-dimensional geometry of a string compactification.

At this point it is worth noting that a large number of stringy realizations of QFTs can be reinterpreted as degrees of freedom coupled to a higher-dimensional bulk QFT. The lower-dimensional system can arise from intersections, i.e., the local geometry of each bulk QFT is supported on a copy of $\mathbb{R}^{n}$). It can also arise from junctions, i.e., the QFT is supported on a manifold with a corner / edge which is locally of the form $\mathbb{R}^{m} \times \mathbb{R}^{n}_{\geq 0}$, which is then glued to other QFTs supported on manifolds with corners / edges. To give a few examples, 6D conformal matter arises as an M5-brane probe of 7D Super Yang-Mills (SYM) theory, but can also be realized as the intersection of 7-branes in F-theory \cite{Gaiotto:2014lca,DelZotto:2014hpa}, and the 5D $\mathcal{T}_N$ theory (see e.g., \cite{Benini:2009gi}) realized via M-theory on $\mathbb{C}^3 / \mathbb{Z}_N \times \mathbb{Z}_N$ can be viewed as a trivalent junction of three 7D Super Yang-Mills theories. Many other examples are known, but the general property they all share is a QFT realized as a defect in a bulk system. This should not surprise our readers: it is in a sense a generalization of the standard ``intersecting branes'' picture heavily used in the string / M-theory realization of many quiver gauge theories. Of course, branes and geometrical singularities are often dual to one another, with the difference that geometrical singularities have access to a slightly broader spectrum of degrees of freedom, including some which do not admit perturbative descriptions.

In this regard, a natural starting point is to consider a bulk system which is itself a $D > 4$ interacting conformal field theory (CFT).
Higher-dimensional supersymmetric conformal field theories (SCFTs) do not have exactly marginal deformations \cite{Louis:2015mka, Cordova:2016xhm} and therefore we can think of them as intrinsically strongly coupled (meaning that they cannot be reached by perturbations of a Gaussian fixed point): these systems are known to provide a rich class of new phenomena in the study of supersymmetric QFTs (SQFTs). From this starting point, we can ask whether, when viewed as a bulk theory, intersections as well as junctions will support localized degrees of freedom along lower-dimensional spaces. Our main focus will be on 4D $\mathcal{N} = 1$ edge modes coupled to 5D interacting bulk theories, but clearly there are many natural generalizations one might contemplate.

The study of this question from a bottom up perspective is quite challenging. First of all, even finding vacua which retain 4D $\mathcal{N} = 1$ supersymmetry is by no means guaranteed. Additionally, compared with the cases of an IR free bulk, we can expect that there will be 4D/5D couplings which do not permit a decoupling limit. In this sense, the 4D theories thus constructed are better viewed as localized degrees of freedom inside a strongly coupled bulk theory. Indeed, one expects on general grounds that the resulting 4D theory generically supports interacting degrees of freedom deep in the IR, and as such, is scale invariant. On the other hand, the presence of couplings to the bulk means that it does not have a well-defined stress energy tensor, and so is not quite a conformal field theory in its own right; it is better viewed as providing localized defect modes. To emphasize these points, we refer to these 4D systems as ``quasi-SCFTs''.\footnote{To the best of our knowledge, this usage is compatible with related phenomena encountered in purely field theoretic settings.}

Our aim in this paper will be to provide a broad template for engineering such theories using the geometry of M-theory on singular $G_2$-holonomy spaces. We remind the reader that 7-dimensional $G_2$-holonomy spaces are models for the seven extra dimensions in M-theory and give rise to physical theories in four spacetime dimensions, see \cite{Acharya:2004qe} for a review.
In such a framework, much of the interesting physics is localized near very special kinds of singularities in the seven extra dimensions, e.g., codimension four orbifold singularities give rise to non-Abelian gauge symmetries \cite{Acharya:2000gb} and particular codimension seven conical singularities support chiral fermions \cite{Acharya:2001gy}. We will utilize the fact that many known, complete, non-compact $G_2$-holonomy spaces have orbifold singularities and thus, given such a space $X$ it comes readily equipped with a configuration of intersecting codimension four, six and, sometimes, codimension seven singularities. $X$ typically contains a closed compact submanifold, $M$, whose volume can be varied and in 
the limit where $M$ shrinks to zero size, $X$ develops a conical singularity and these QFT sectors are instead supported on manifolds with corners, so are best viewed as specifying junctions of theories rather than just intersections. As we will see from examples, the configurations that can arise are quite rich and elaborate and would have been difficult to envision without the foresight provided by M-theory. By instead considering Type IIA or IIB theory on such spaces one obtains three-dimensional theories.

Our main illustrative examples involve taking singular limits of geometries obtained from discrete quotients of smooth $G_2$-holonomy manifolds.
We mainly focus on the original examples of Bryant and Salamon \cite{bryant1989} where $X$ is the bundle of anti-self-dual 2-forms over $S^4$ or $\mathbb{CP}^2$, namely $X = \Lambda^{2}_{\mathrm{ASD}}(S^4)$ or $X = \Lambda^{2}_{\mathrm{ASD}}(\mathbb{CP}^2)$. Taking a quotient $X / \Gamma$ with $\Gamma$ a finite subgroup of the isometries of $X$ will, in general, result in an orbifold with various singularities localized on codimension four, six and seven subspaces. In the limit where the $S^4$ or the $\mathbb{CP}^2$ 4-cycle in the orbifold is of infinite size, each of the codimension six singularities can be viewed as engineering a 5D SCFT with flavor symmetries controlled by 7D SYM sectors localized on codimension four orbifold loci. In the limit where the $S^4$ or the $\mathbb{CP}^2$ are of finite size, we find 5D SCFTs localized at special points (e.g., the North and South poles of the $S^4$), with certain subgroups of their mutual flavor symmetries diagonally gauged due to the finite volume of the 4-cycle. This gauging procedure only preserves 4D $\mathcal{N} = 1$ supersymmetry because the coupling itself depends non-trivially on the position with the radial direction of the $G_2$ cone. This procedure also lifts some of the directions of the 5D moduli space: blowup moduli of the local Calabi-Yau (used to define each 5D SCFT) do not survive in the local $G_2$ geometry. In the limit where the 4-cycle collapses to zero size, these 5D SCFT sectors form a configuration of junctions supported on real half-lines $\mathbb{R}_{\geq 0}$.

In field theory terms, these geometries provide a general method for engineering rather complicated bulk systems with localized defect modes. Implicit in our considerations is the choice of specific boundary conditions for the bulk modes near the defect. This implicit choice amounts to trivial boundary conditions (i.e., no bulk field profiles switched on) as we approach the boundary. One could in principle consider introducing further non-geometric decorations such as ``T-brane data'' (see e.g., \cite{Cecotti:2010bp,Barbosa:2019bgh}), but we defer this possibility to future investigations.

\begin{figure}
\begin{center}
\includegraphics[scale=0.5]{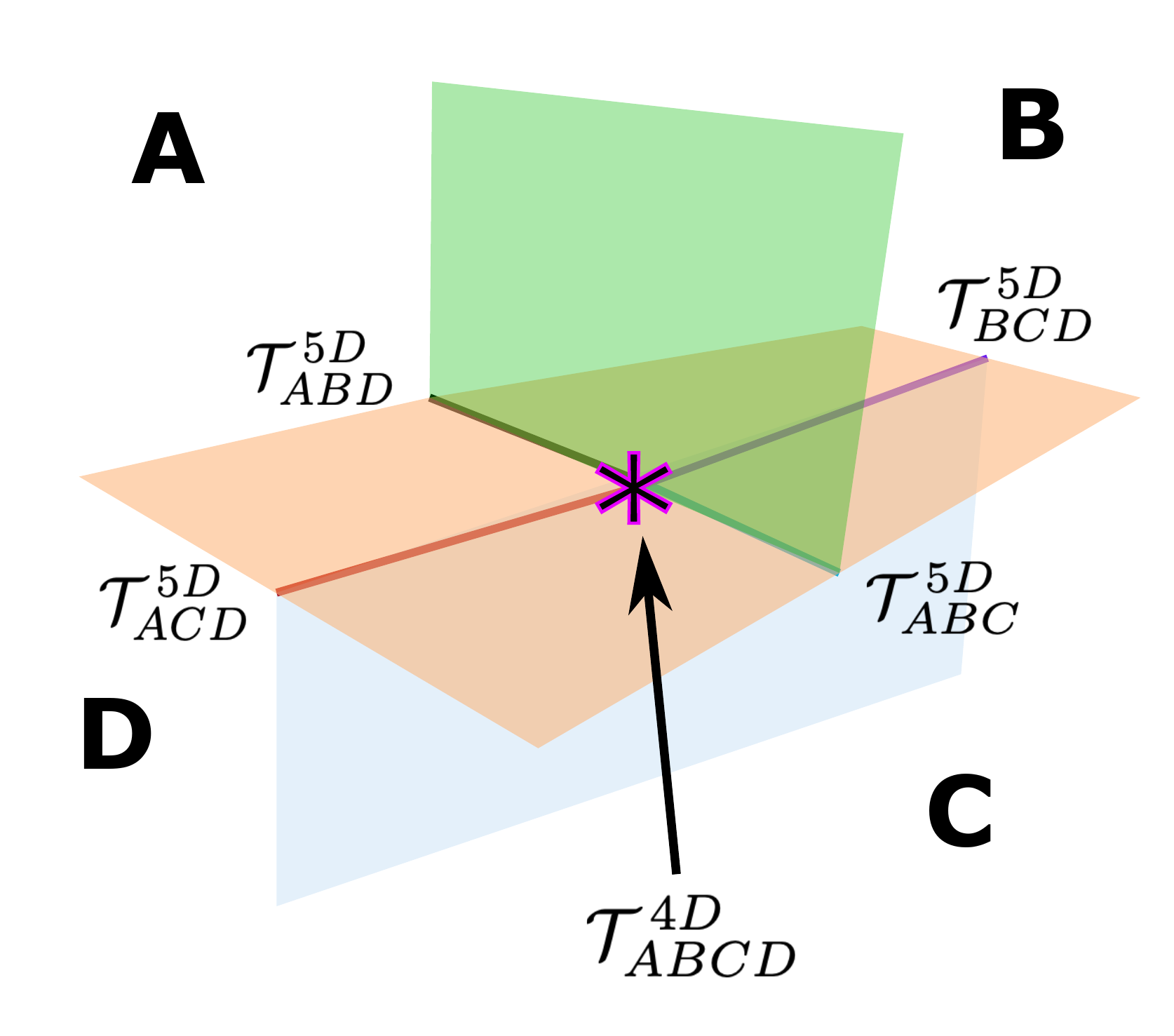}
\end{center}
\caption{Example of a 4D $\mathcal N=1$ quasi-SCFT obtained at a junction of four 7D theories, labeled {\bf A, B, C}, and {\bf D} above.}\label{fig:sticazzi}
\end{figure}

In this setting, then, we can use geometry to study some basic features of the 4D localized defect modes. For example, since we have an explicit group action, we can use this to identify the non-Abelian (zero-form) flavor symmetry algebra from the singularities which extend out to the boundary of our $G_2$-holonomy space. In the $S^4$ case, which we use as our main running example in the paper, $X$ is a metric cone over $\partial X = \mathbb{CP}^3$ when the $S^4$ is contracted to a point, and the group action naturally extends to a quotient space $\mathbb{CP}^3 / \Gamma$. Viewing the bulk 5D SCFTs as the junction of 7D theories, the resulting 4D theories can \textit{also} be viewed as a junction of 7D systems (see Figure \ref{fig:sticazzi}). This helps to illustrate that there is a non-trivial bulk / boundary coupling since a dynamical modulus field (associated with the size of the $S^4$) triggers a spontaneous breaking of the flavor symmetry which is present in the collapsed limit.

The topology of the boundary geometry allows one to understand more refined structures, such as the global form of the non-Abelian flavor symmetry group (0-form symmetries), discrete higher-form symmetries, as well as possible higher-group structures which entwine these structures. Compared with other examples in the literature which have analyzed related phenomena (see e.g., \cite{DelZotto:2015isa, GarciaEtxebarria:2019caf, Albertini:2020mdx, Morrison:2020ool, Tian:2021cif, DelZotto:2022fnw,Cvetic:2022imb, DelZotto:2022joo}), the way in which defects get screened is a bit more subtle because the 4-cycle at the tip of the cone is not a genuine cycle in the boundary.

 That being said, we find that the boundary topology still accurately accounts for the higher-symmetries both of the bulk 5D SCFT sectors as well as the 4D quasi-SCFT sector.

Asides from quotients of the form $\Lambda^{2}_{\mathrm{ASD}}(S^4) / \Gamma$, we also consider the 4D theories that arise from complete Bryant-Salamon metrics on $G_2$-holonomy spaces of the form $X = \Lambda^{2}_{\mathrm{ASD}}(M)$ where $M$ is a compact self-dual Einstein 4-orbifold as in the case $M=\mathbb{WCP}^{2}$ or the more general examples discussed in \cite{Acharya:2001gy}. We show the latter to give rise to a rather intricate generalized polygonal quiver structure.

The main new features of the resulting systems we obtain are the following:
\begin{itemize}
   
    \item \textit{Position dependent couplings.} Often the codimension 4 loci associated to 7D gauge theories have the topology of $S^2 \times \mathbb R_{\geq 0}$, where $S^2$ varies along the $\mathbb R_{\geq 0}$ direction, corresponding to the $G_2$ cone radial direction. We interpret these as a position dependent gauge coupling for the various 5D SCFTs involved, effectively breaking 5D Lorentz invariance. This gives rise to novel types of generalized quivers, where the conformal matter is 5D, but the gaugings involve position dependent couplings.
    \item \textit{Quasi-SCFTs.} For most of the 4D / 5D systems, we see that they arise from junctions of 5D SCFTs along a common 4D subspace. The bulk theories are strongly coupled, and so we expect the edge modes to also support interacting degrees of freedom. As the edge modes do not fully decouple from the bulk, we call them quasi-SCFTs;
    \item \textit{Symmetry inheritance.} When extended charged bulk operators can end on the edge, the interacting edge mode theory inherits the corresponding higher-symmetry (with a corresponding shift in degree);
    \item \textit{Symmetry breaking.} For the 4D / 5D systems corresponding to $G_2$ cones, we can track spontaneous symmetry breaking by comparing the symmetries of the singularity link, with the symmetries of the resolved singularity;
   
\end{itemize}

The rest of this paper is organized as follows. In section \ref{sec:geometric_interfaces} we explain how a number of known geometrically engineered SCFTs can be viewed as localized modes of a higher-dimensional bulk system. After this, in section \ref{sec:CONEHEAD} we turn to a detailed analysis of orbifolds of the form $X / \Gamma$ where $X = \Lambda^{2}_{\mathrm{ASD}}(S^4)$. We then illustrate some of the theories obtained in this way in section \ref{sec:EXAMP}. In section \ref{sec:GEN} we study some of the generalized symmetries of these quasi-SCFTs. In section \ref{sec:Rest} we give examples of interfaces arising from additional explicit examples of more general $G_2$-holonomy cones. This includes examples with a closed polygonal generalised quiver structure with an arbitrary number of nodes. We present our conclusions in section \ref{sec:CONC}. Some additional technical details on the higher-symmetry computations are deferred to the Appendices. This includes a brief discussion of continuous $k$-form symmetries, more general quotient spaces, as well as a number of additional details on various homology group calculations.

\section{Geometric Engineering of Bulk / Boundary Systems}\label{sec:geometric_interfaces}

To frame the analysis of $G_2$-holonomy orbifolds to follow, in this section we reinterpret a number of well-known
$D \geq 4$ supersymmetric theories as defects in a higher-dimensional system of bulk QFTs. There are two generic cases we can expect to encounter. In the first case the localized degrees of freedom decouple from the bulk, in this situation we reach an SCFT. In the second case the bulk couples non-trivially to the interface, so that we really have a quasi-SCFT, i.e., it cannot be defined without also referencing the bulk.

\subsection{Engineering SCFTs and Quasi-SCFTs}

Here we want to consider some general features of geometric engineering of a given quantum field theory via a suitable background in string / M-theory. We therefore introduce the notation $\mathscr{S}$ and $D_{\mathscr{S}}$ where $\mathscr S$ is a shorthand for IIA, IIB, M and $D_{\mathscr S}$ is the corresponding spacetime dimension.

To geometrically engineer a quantum field theory, we specify a non-compact $d$-dimensional background $ X^d$ with a localized singularity of codimension $d$; it can also support non-isolated singularities, but the interacting degrees of freedom will thus be located on a lower-dimensional subspace of dimension $D_{\mathscr{S}} - d$. Working on $X^d$ non-compact means that the lower-dimensional system is decoupled from lower-dimensional gravity.\footnote{Here for ease of exposition we are assuming the background has a trivial global structure: if this is not the case, $\mathscr S$-theory would assign to $ X^d$ a Hilbert space of theories having the same local dynamics, but distinct spectra of extended operators \cite{DelZotto:2015isa, Garcia-Etxebarria:2019cnb,Albertini:2020mdx}.} This gives a dictionary:
\be
\mathscr S \text{ on } \mathbb R^{1,D-1} \times  X^{d} \quad\longleftrightarrow\quad \mathcal{T}^{\textnormal{\tiny\!  $(\mathscr{S})$}}_{{   X^d }}\:\!\! \in \textnormal{QFT}_{D} \qquad D = D_{\mathscr S} - d.
\ee
In particular, whenever the background $ X^{d}$ is such that it preserves some supercharges, the theory $\mathcal{T}^{\textnormal{\tiny\!  $(\mathscr{S})$}}_{{   X^d }}$ is supersymmetric. This is the case for instance when $ X^d$ is a local manifold with special holonomy. In this paper we restrict our attention to the latter case, since our main focus are interfaces that can be described exploiting $G_2$-holonomy spaces.

We shall often be interested in backgrounds which have a singularity. At this singularity, we can expect localized interacting degrees of freedom.\footnote{It could happen that in the deep IR we just have free fields, but this is typically not the case.} A broad class of examples admit a metric cone of the form:
\be
 X^d = \mathcal C ( Y^{d-1}) \qquad\qquad ds^2_{ X^d} = dr^2 + r^2 ds^2_{ Y^{d-1}}.
\ee
Proceeding in this fashion, we encounter two possible situations:
\begin{itemize}
\item \textbf{${ X}^d$ is an isolated singularity}: the singularity at the tip of the cone is isolated of codimension $d$. In this case by geometric engineering we obtain an actual SCFT;
\item \textbf{${ X}^d$ is a non-isolated singularity}: the singularity at the tip of the cone arise at the intersection of singularities  supported on loci of higher codimension. In field theory terms, we can often view these singularities as supported on a manifold with corners / edges, and as such we get a junction of theories. The localized degrees of freedom at the tip of the cone may or may not decouple. We consider two possibilities here:
    \begin{itemize}
    \item[$\circ$] \textbf{Localized SCFTs}: If the bulk singularity is IR free, then the localized modes decouple in the IR and we reach an SCFT;
    \item[$\circ$] \textbf{Localized Quasi-SCFTs}: If the bulk singularity supports an SCFT in the IR, then it can happen that the localized modes do not completely decouple from the bulk, we refer to this as a quasi-SCFT.
    \end{itemize}
\end{itemize}

The localized degrees of freedom might be free or interacting: proper criteria to distinguish a non-trivial dynamics depend on $\mathscr S$ and on $ X^d$ (i.e., on the kind of background considered for $\mathscr S$-theory), and need to be studied on a case by case basis. An example of a criterion that can be used to identify a non-trivial fixed point is that mutually non-local BPS excitations become simultaneously massless. For example, in all known 6D SCFTs (see e.g., \cite{Heckman:2018jxk, Argyres:2022mnu} for recent reviews), the tensor branch moduli space has dyonic BPS strings which approach vanishing tension at the fixed point \cite{Strominger:1995ac, Witten:1995zh, Seiberg:1996vs}. Similarly, Argyres-Douglas fixed points \cite{Argyres:1995jj} in four dimensions are characterized by the fact that mutually non-local BPS dyons simultaneously become massless.

\subsection{Some Well-Known Examples}\label{sec:firstex}

To illustrate the above considerations, we now turn to some explicit examples.

\subsubsection{6D Examples}\label{sec:6dexamples}

\begin{figure}
    \centering
    \scalebox{0.9}{
    \begin{tikzpicture}
	\begin{pgfonlayer}{nodelayer}
		\node [style=none] (0) at (-7, 2) {};
		\node [style=none] (1) at (-2, 2) {};
		\node [style=none] (2) at (-2, 5) {};
		\node [style=none] (3) at (-7, 5) {};
		\node [style=none] (4) at (2, 2) {};
		\node [style=none] (5) at (7, 2) {};
		\node [style=none] (6) at (7, 5) {};
		\node [style=none] (7) at (2, 5) {};
		\node [style=none] (8) at (-1, 3.5) {};
		\node [style=none] (9) at (1, 3.5) {};
		\node [style=none] (14) at (-5.72, 3.75) {};
		\node [style=none] (15) at (-5.48, 3.75) {};
		\node [style=none] (16) at (-5.48, 3.5) {};
		\node [style=none] (17) at (-5.72, 3.5) {};
		\node [style=none] (18) at (-5.37, 3.75) {};
		\node [style=none] (19) at (-5.12, 3.75) {};
		\node [style=none] (20) at (-5.12, 3.5) {};
		\node [style=none] (21) at (-5.37, 3.5) {};
		\node [style=none] (22) at (-3.875, 3.75) {};
		\node [style=none] (23) at (-3.625, 3.75) {};
		\node [style=none] (24) at (-3.625, 3.5) {};
		\node [style=none] (25) at (-3.875, 3.5) {};
		\node [style=none] (26) at (-3.52, 3.75) {};
		\node [style=none] (27) at (-3.27, 3.75) {};
		\node [style=none] (28) at (-3.27, 3.5) {};
		\node [style=none] (29) at (-3.52, 3.5) {};
		\node [style=none] (30) at (-4.75, 3.625) {};
		\node [style=none] (31) at (-4.25, 3.625) {};
		\node [style=none] (32) at (-6.5, 5) {};
		\node [style=none] (33) at (-5, 2) {};
		\node [style=none] (34) at (-4, 2) {};
		\node [style=none] (35) at (-2.5, 5) {};
		\node [style=none] (36) at (3.75, 5) {};
		\node [style=none] (37) at (5.25, 2) {};
		\node [style=none] (38) at (3.75, 2) {};
		\node [style=none] (39) at (5.25, 5) {};
		\node [style=none] (40) at (-4.5, 1.5) {};
		\node [style=none] (41) at (-4.5, 0.5) {};
		\node [style=none] (42) at (4.5, 1.5) {};
		\node [style=none] (43) at (4.5, 0.5) {};
		\node [style=none] (44) at (-7, -0.5) {};
		\node [style=none] (45) at (-2, -0.5) {};
		\node [style=none] (46) at (2, -0.5) {};
		\node [style=none] (47) at (7, -0.5) {};
		\node [style=NodeCrossRed] (48) at (-5.375, -0.5) {};
		\node [style=NodeCrossRed] (49) at (-5., -0.5) {};
		\node [style=NodeCrossRed] (51) at (-4, -0.5) {};
		\node [style=NodeCrossRed] (52) at (-3.625, -0.5) {};
		\node [style=none] (53) at (-5.5, -1) {};
		\node [style=none] (54) at (-3.5, -1) {};
		\node [style=none] (55) at (-3.5, 0) {};
		\node [style=none] (56) at (-5.5, 0) {};
		\node [style=CircleRed] (57) at (4.5, -0.5) {};
		\node [style=CircleRed] (58) at (4.5, 3.5) {};
		\node [style=none] (59) at (0, 4) {shrink};
		\node [style=none] (60) at (-3.25, 1) {M-theory};
		\node [style=none] (61) at (3.25, 1) {M-theory};
		\node [style=none] (62) at (-1.25, -0.5) {$\mathbb{C}^2/\Gamma$};
		\node [style=none] (63) at (1.25, -0.5) {$\mathbb{C}^2/\Gamma$};
		\node [style=none] (64) at (-4.5, -1.5) {Interface};
		\node [style=none] (65) at (-3.25, -1.25) {};
		\node [style=none] (66) at (-2.5, -2.375) {};
		\node [style=none] (67) at (-2.5, -2.625) {\color{red} Fractionated M5's};
		\node [style=none] (68) at (4.5, -1) {Interface};
		\node [style=none] (69) at (-3.25, 2.5) {$\mathfrak{g}_\Gamma$};
		\node [style=none] (70) at (-5.75, 2.5) {$\mathfrak{g}_\Gamma$};
		\node [style=none] (71) at (5.5, 2.5) {$\mathfrak{g}_\Gamma$};
		\node [style=none] (72) at (3.5, 2.5) {$\mathfrak{g}_\Gamma$};
        \node [style=none] (73) at (0,-3) {};
        \node [style=none] (78) at (-4.5, -2) {Thickness};
        \node [style=none] (79) at (-4.75, -0.6) {};
		\node [style=none] (80) at (-4.25, -0.6) {};
         \node [style=none] (81) at (-4.75, -0.4) {};
		\node [style=none] (82) at (-4.25, -0.4) {};
	\end{pgfonlayer}
	\begin{pgfonlayer}{edgelayer}
		\draw [style=DottedLine] (3.center) to (0.center);
		\draw [style=DottedLine] (0.center) to (1.center);
		\draw [style=DottedLine] (1.center) to (2.center);
		\draw [style=DottedLine] (2.center) to (3.center);
		\draw [style=DottedLine] (7.center) to (6.center);
		\draw [style=DottedLine] (6.center) to (5.center);
		\draw [style=DottedLine] (5.center) to (4.center);
		\draw [style=DottedLine] (4.center) to (7.center);
		\draw [style=ArrowLineRight] (8.center) to (9.center);
		\draw [style=ThickLine, bend right=45] (14.center) to (17.center);
		\draw [style=ThickLine, bend right=45] (17.center) to (16.center);
		\draw [style=ThickLine, bend right=45] (16.center) to (15.center);
		\draw [style=ThickLine, bend left=315] (15.center) to (14.center);
		\draw [style=ThickLine, bend right=45] (18.center) to (21.center);
		\draw [style=ThickLine, bend right=45] (21.center) to (20.center);
		\draw [style=ThickLine, bend right=45] (20.center) to (19.center);
		\draw [style=ThickLine, bend left=315] (19.center) to (18.center);
		\draw [style=ThickLine, bend right=45] (22.center) to (25.center);
		\draw [style=ThickLine, bend right=45] (25.center) to (24.center);
		\draw [style=ThickLine, bend right=45] (24.center) to (23.center);
		\draw [style=ThickLine, bend left=315] (23.center) to (22.center);
		\draw [style=ThickLine, bend right=45] (26.center) to (29.center);
		\draw [style=ThickLine, bend right=45] (29.center) to (28.center);
		\draw [style=ThickLine, bend right=45] (28.center) to (27.center);
		\draw [style=ThickLine, bend left=315] (27.center) to (26.center);
		\draw [style=DottedLine] (30.center) to (31.center);
		\draw [style=ThickLine] (32.center) to (33.center);
		\draw [style=ThickLine] (35.center) to (34.center);
		\draw [style=ThickLine] (36.center) to (37.center);
		\draw [style=ThickLine] (39.center) to (38.center);
		\draw [style=ArrowLineRight] (40.center) to (41.center);
		\draw [style=ArrowLineRight] (42.center) to (43.center);
		\draw [style=ThickLine] (44.center) to (45.center);
		\draw [style=ThickLine] (46.center) to (47.center);
		\draw [style=DashedLineThin] (56.center) to (55.center);
		\draw [style=DashedLineThin] (55.center) to (54.center);
		\draw [style=DashedLineThin] (54.center) to (53.center);
		\draw [style=DashedLineThin] (53.center) to (56.center);
		\draw [style=ArrowLineRight] (66.center) to (65.center);
        \draw [style=DottedRed] (79.center) to (80.center);
	\end{pgfonlayer}
\end{tikzpicture}
    }
    \caption{6D conformal matter as an example of an interface.}
    \label{fig:conformatta}
\end{figure}
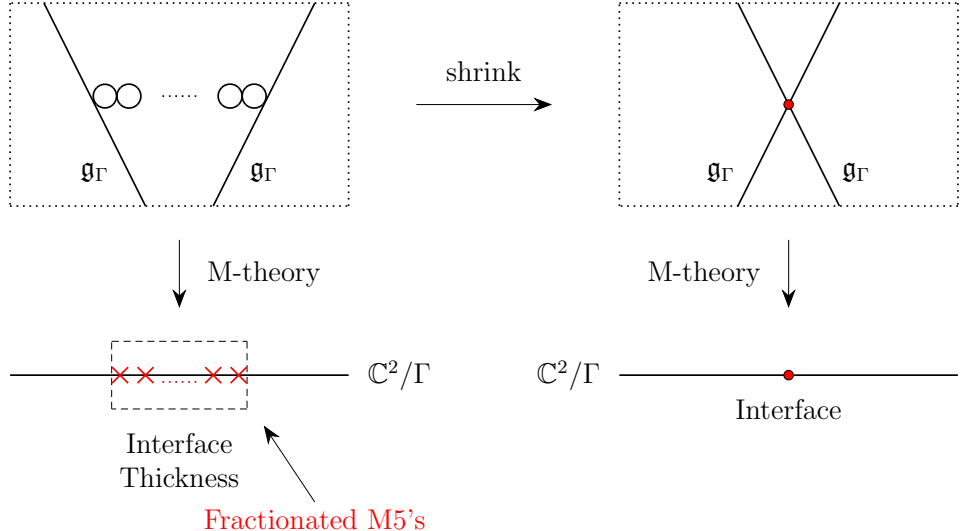

The geometric engineering of 6D SCFTs has been the subject of a plethora of recent developments --- see
\cite{Heckman:2018jxk, Argyres:2022mnu} for recent reviews.

Perhaps the simplest examples of isolated singularities giving rise to non-trivial interacting 6D systems are the ADE singularities, $\mathbb{C}^2 / \Gamma_{ADE}$ for $\Gamma_{ADE} \subset SU(2)$ a finite subgroup of ADE type in Type IIB string theory.

F-theory allows for richer backgrounds and lower supersymmetry, giving rise to extensive lists of 6D $\mathcal{N} = (1,0)$ theories.
A classification of all elliptically fibered Calabi-Yau threefolds which can support a 6D SCFT was carried out in references  \cite{Heckman:2013pva,Heckman:2015bfa}.\footnote{All known 6D SCFTs can be realized in this fashion, though the interpretation of a Calabi-Yau geometry can sometimes have ambiguities, as captured by possible ``frozen'' singularities. See e.g., \cite{Tachikawa:2015wka,Bhardwaj:2018jgp, Bhardwaj:2015oru} for further discussion.} The basic building blocks of such 6D SCFTs are the conformal matter theories \cite{DelZotto:2014hpa}, which arise at the intersection of a pair of non-compact singularities in F-theory. More precisely, we have a geometry in which the elliptic fibration degenerates over two non-compact curves (giving rise to a pair of Kodaira singularities) intersecting transversally at a point, where the conformal matter is located. By F-theory / M-theory duality \cite{Vafa:1996xn} the conformal matter theories can be understood as interfaces between two copies of the 7D SYM theory with Lie algebra $\mathfrak g_\Gamma$. The seven-dimensional gauge theory is realized in M-theory by $\mathcal{T}^{\textnormal{\tiny\!  (M)}}_{{ \mathbb{C}^2/\Gamma}}$, whose locus is a seven-dimensional plane. Six out of the seven directions of this plane are occupied by M5-branes (which can fractionate because of the singularity). The location of the M5-branes along the seventh direction parameterizes the thickness of the interface, which in the 6D field theory is in turn interpreted as a tensor branch (see Figure \ref{fig:conformatta}). In geometry, we can also find a similar interpretation. It is sufficient to consider the phase in which the interface has a thickness, obtained by blowing up the singularity in the base of the F-theory model (see Figure \ref{fig:conformatta} top left). There we can zoom to a neighborhood of the locus where one of the non-compact curves is intersecting one of the compact ones to recover the geometry which is engineering the bulk gauge theory.

The 6D flavor symmetry of conformal matter theories \cite{DelZotto:2014hpa} can also be understood in terms of the fact that these models arise as interfaces between two copies of the theory $\mathcal{T}^{\textnormal{\tiny\!  (M)}}_{{ \mathbb{C}^2/\Gamma}}$. This is also the case for conformal matter of $(\mathfrak g, {\mathfrak g}^\prime)$ type \cite{DelZotto:2014hpa}, which corresponds to interfaces of the schematic form
\be\label{eq:interfaccia6d}
\left(\mathcal{T}^{\textnormal{\tiny\!  (M)}}_{{ {X}^{4}_\Gamma}}\right)_{7D} \quad \Big|\Big|\,\,\big(\mathfrak g, {\mathfrak g}^\prime\big)_{6d}\,\,\Big|\Big| \quad\left(\mathcal{T}^{\textnormal{\tiny\!  (M)}}_{{{X}^{4}_{\Gamma'}}}\right)_{7D}\,.
\ee
From this perspective, the fission and fusion of \cite{Heckman:2018pqx} (see also \cite{DelZotto:2018tcj}) can be interpreted as operations at the level of interfaces among higher-dimensional gauge theories.

\subsubsection{5D Examples}\label{sec:examples5D}
Another well-studied example is M-theory on $X^{6}$ a Calabi-Yau threefold with a canonical singularity\footnote{One necessary criterion to produce a conformal fixed point is that we reach the singularity at finite distance in Calabi-Yau moduli space. However, it is important to discuss the $M$-theory space-time metric as
some non-compact manifolds possess an asymptotically conical Calabi-Yau metric, while others only admit an incomplete metric. Canonical singularities for which the existence of complete metrics are obstructed are known \cite{Gauntlett:2006vf, Collins2015SasakiEinsteinMA}. The orbifolds $\mathbb{C}^3/\Gamma$ with finite $\Gamma\subset SU(3)$ constitute large class of examples with AC complete Calabi-Yau metrics \cite{Ito1994CrepantRO, Roan1996MinimalRO, Coevering2008RicciflatKM}.}, in which case the theory $\mathcal{T}^{\textnormal{\tiny\!  (M)}}_{X^6}$ is a 5D $\mathcal N=1$ SCFT \cite{Morrison:1996xf,Douglas:1996xp, Intriligator:1997pq} (see also \cite{Xie:2017pfl,Closset:2018bjz}). Many well-known examples of both kinds occur in this context. Here we will highlight some that are relevant for the following sections.

Examples of isolated singularities include the series of 5D $\mathcal N=1$ $E_N$ SCFTs where $N=0,...,8$ (see \cite{Seiberg:1996bd, Morrison:1996xf, Douglas:1996xp}). The latter is obtained in M-theory via an isolated singularity that we denote $ X^6_{(N)}$. The singularity $ X^6_{(N)}$ is obtained from the smooth local Calabi-Yau threefold $K \rightarrow \mathrm{dP}_N$, where $\mathrm{dP}_N$ denotes $\mathbb{CP}^2$ with $N$ blowups at general position, and $K$ refers to the canonical bundle. In the limit where we shrink the base $\text{dP}_N$ surface to zero volume, we get a 5D SCFT. The case $N=0$ is an example of a model  without a gauge theory phase. The cases with $1 \leq N \leq 8$ correspond to an $SU(2)$ gauge theory with $N_f = N-1$ fundamental flavors, in which the gauge coupling is formally tuned to infinite strength. In the case $N=9$ one obtains the 5D KK theory for the 6D E-string --- see \cite{Jefferson:2018irk} for a recent detailed review of the geometric engineering of 5D SCFTs.

\begin{figure}
$$\begin{gathered}
\xymatrix{
&&&&&\bullet_{p,1}\\
&&&&\circ&\bullet_{p-1,1}\ar@{-}[u]\\
&&&&\vdots&\vdots\ar@{-}[u]\\
&&\circ&\dots&\circ_{2,2}&\bullet_{2,1}\ar@{-}[u]\\
\bullet_{1,q}\ar@{-}[uuuurrrrr]\ar@{-}[r]&\bullet_{1,q-1}\ar@{-}[r]&\bullet_{1,q-2}\ar@{-}[r]&\cdots\ar@{-}[r]&\bullet_{1,2}\ar@{-}[r]&\bullet_{1,1}\ar@{-}[u]\\ }
\end{gathered}
$$
\caption{Toric diagram of the $ X^6_{p,q}$ singularity.}\label{fig:pqtoric}
\end{figure}
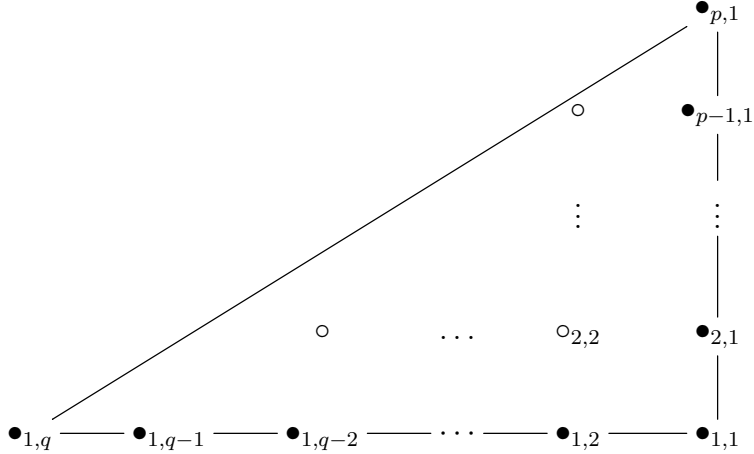

A class of examples of non-isolated singularity is provided by the $CY_3$ singularities
\be
 X^6_{p,q} \equiv \mathbb C^3 / (\mathbb Z_p \times \mathbb Z_q) \,,
\ee
where the two actions of the cyclic groups are given by
\be
\mathbb Z_p \, \colon \, \MAT{\alpha \\ &\alpha^{-1}&\\ &&1}\qquad \qquad \mathbb Z_q \, \colon \, \MAT{1 \\ &\beta&\\ &&\beta^{-1}}
\ee
and $\alpha = \exp(2 \pi i / p)$ and $\beta = \exp(2 \pi i / q)$ are primite $p$th and $q$th roots of unity. In this geometry, along the line $\ell_3 = \{ z_1 = z_2 = 0\}$ we have a $\mathbb C^2 / \mathbb Z_p$ singularity, along the line $\ell_1 = \{ z_2 = z_3 = 0\}$ we have a $\mathbb C^2 / \mathbb Z_q$ singularity, and along the line $\ell_2 = \{ z_1 = z_3 = 0\}$ we have a $\mathbb C^2 / \mathbb Z_{g}$ singularity with $g=\gcd(p,q)$. See Figure \ref{fig:pqtoric} for a depiction of the toric diagram.

The theory $\mathcal{T}^{\textnormal{\tiny\!  (M)}}_{\mathbb C^3 / (\mathbb Z_p \times \mathbb Z_q)}$ arises at the junction of three distinct
seven-dimensional gauge theories. In the covering space $\mathbb{C}^3$, each of the 7D SYM sectors is supported on a cone $S^1 \times \mathbb{R}_{\geq 0}$, and these meet to form a trivalent junction:
\be
\scalebox{0.8}{\begin{tikzpicture}
	\begin{pgfonlayer}{nodelayer}
		\node [style=none] (1) at (-2.5, 2.5) {};
		\node [style=none] (2) at (3.5, 0) {};
		\node [style=none] (3) at (0, -3.5) {};
		\node [style=none] (4) at (0.75, -2) {$\ell_3,z_3$};
		\node [style=none] (5) at (2, -0.5) {$\ell_1,z_1$};
		\node [style=none] (6) at (-2, 1) {$\ell_2,z_2$};
		\node [style=none] (7) at (-0.625, -0.625) {};
		\node [style=none] (8) at (-0.625, 0.625) {};
		\node [style=none] (9) at (0.625, 0.625) {};
		\node [style=none] (10) at (0.625, -0.625) {};
		\node [style=none] (11) at (0.625, 0) {};
		\node [style=none] (12) at (0, -0.625) {};
		\node [style=none] (13) at (-3, 3) {$\mathcal{T}^{\textnormal{\tiny\!  (M)}}_{X^4_{{\mathbb Z}_g}}$};
		\node [style=none] (14) at (4.25, 0) {$\mathcal{T}^{\textnormal{\tiny\!  (M)}}_{X^4_{{\mathbb Z}_p}}$};
		\node [style=none] (15) at (0, -4.125) {$\mathcal{T}^{\textnormal{\tiny\!  (M)}}_{X^4_{{\mathbb Z}_q}}$};
		\node [style=none] (16) at (0, 0) {$\mathcal{T}^{\textnormal{\tiny\!  (M)}}_{X^6_{p,q}}$};
	\end{pgfonlayer}
	\begin{pgfonlayer}{edgelayer}
		\draw [style=ThickLine] (8.center) to (1.center);
		\draw [style=ThickLine] (11.center) to (2.center);
		\draw [style=ThickLine] (12.center) to (3.center);
	\end{pgfonlayer}
\end{tikzpicture}}
\ee
In our explicit examples involving orbifolds of $G_2$ cones, we will often encounter the special case where $p = q = N$, and we refer to this as the $\mathcal{T}_N$ theory:
\be
\mathcal T_N \equiv \mathcal{T}^{\textnormal{\tiny\!  (M)}}_{X^6_{N,N}}\,.
\ee

It is interesting to remark that for all the models $\mathcal{T}^{\textnormal{\tiny\!  (M)}}_{X^6_{p,q}}$ the higher-form symmetry groups are trivial \cite{Morrison:2020ool,Albertini:2020mdx}. Exploiting more general singularities, systems with higher-symmetries can be obtained \cite{Apruzzi:2021vcu,Apruzzi:2021mlh,DelZotto:2022fnw,Tian:2021cif, Cvetic:2022imb, DelZotto:2022joo}, and we will encounter examples of these cases as well. As a final comment, also in 5D, the gauging of 5D flavor symmetries discussed e.g., in \cite{Hayashi:2019fsa} (see also \cite{Apruzzi:2019kgb}) can be interpreted in terms of the gauging of 5D flavor symmetries, leading to a ``fusion'' of different theories (in the sense of \cite{Heckman:2018pqx}).

\subsubsection{4D Examples}\label{sec:4Dfirst}
We can obtain similar structures in a four-dimensional setting if we consider backgrounds for type II superstrings on the same $CY_3$ singularities. Here we just illustrate this in the context of the $CY_3$ singularities of type $ X^6_{p,q}$.

Consider first type IIA backgrounds on $ X^6_{p,q}$. Since IIA is obtained as the circle reduction of M-theory, we see that quite literally, we can just start from the previously discussed 5D theories, and descend to 4D. In an appropriate decoupling limit, the bulk 6D theories are associated with three 6D $\mathcal{N} = (1,1)$ super Yang-Mills theories. 
In particular, if we consider the case $p=q=N$, the theory on the junction
is the 4D $\mathcal N=2$ $\mathcal{T}_N$ theory \cite{Gaiotto:2009we}, now realized
as the theory supported at a junction of three 6D $\mathcal{N} = (1,1)$ $\mathfrak{su}(N)$ SYMs.\footnote{As pointed out recently in \cite{Closset:2020scj}, upon further reduction on a circle one ends up with 3D $\mathcal N=4$ theories that are mirror dual to each other. This hints at a possible application of T-duality of junctions of LSTs to the topic of dualities of 3D $\mathcal N=2$ theories, a subject we leave for future work.}

Consider next type IIB on the same geometry $X^{6}_{p,q}$. In this context we have:
\be
6d \, (2,0) \,\mathfrak{a}_{k-1} \text{ theory} \quad \longleftrightarrow \quad \mathcal{T}^{\textnormal{\tiny\!  (IIB)}}_{{  X^4_{\mathbb Z_k} }}
\ee
so we get families of 4D $\mathcal N=2$ ``edge mode'' theories at a junction of 6D $\mathcal{N} = (2,0)$ theories:
\be
\scalebox{0.8}{\begin{tikzpicture}
	\begin{pgfonlayer}{nodelayer}
		\node [style=none] (1) at (-2.5, 2.5) {};
		\node [style=none] (2) at (3.5, 0) {};
		\node [style=none] (3) at (0, -3.5) {};
		\node [style=none] (4) at (0.75, -2) {$\ell_3,z_3$};
		\node [style=none] (5) at (2, -0.5) {$\ell_1,z_1$};
		\node [style=none] (6) at (-2, 1) {$\ell_2,z_2$};
		\node [style=none] (7) at (-0.625, -0.625) {};
		\node [style=none] (8) at (-0.625, 0.625) {};
		\node [style=none] (9) at (0.625, 0.625) {};
		\node [style=none] (10) at (0.625, -0.625) {};
		\node [style=none] (11) at (0.625, 0) {};
		\node [style=none] (12) at (0, -0.625) {};
		\node [style=none] (13) at (-3.125, 3) {$6d \, (2,0) \, \mathfrak{a}_{g-1}$};
		\node [style=none] (14) at (5, 0) {$6d \, (2,0) \, \mathfrak{a}_{p-1}$};
		\node [style=none] (15) at (0, -4) {$6d \, (2,0) \, \mathfrak{a}_{q-1}$};
		\node [style=none] (16) at (0, 0) {$\mathcal{T}^{\textnormal{\tiny\!  (IIB)}}_{X^6_{p,q}}$};
	\end{pgfonlayer}
	\begin{pgfonlayer}{edgelayer}
		\draw [style=ThickLine] (8.center) to (1.center);
		\draw [style=ThickLine] (11.center) to (2.center);
		\draw [style=ThickLine] (12.center) to (3.center);
	\end{pgfonlayer}
\end{tikzpicture}}
\ee
Little is known about the theories $\mathcal{T}^{\textnormal{\tiny\!  (IIB)}}_{{  X^6_{p,q} }}$ so far, for some preliminary analysis we refer the interested reader to \cite{Closset:2020scj,Closset:2021lwy,Closset:2021lhd}.

\section{4D$\ \mathcal{N}=1$ Edge Modes via $G_{2}$ Cones \label{sec:CONEHEAD}}

In the previous section we provided various examples illustrating how
higher-dimensional theories can realize interacting lower-dimensional degrees of
freedom. One of the general features of
these examples is that the \textquotedblleft bulk modes\textquotedblright\
are often relatively simple in the sense that they are characterized at long
distances by a higher-dimensional Lagrangian field theory, and often do not
realize interacting quantum field theories in the infrared. It is natural to
ask whether one can produce more general bulk / boundary correspondences by
considering bulk theories which also admit strong coupling
dynamics. One way to accomplish this is by taking intersections and junctions of such strongly coupled theories.

As mentioned earlier, from a bottom up perspective, the construction of such
interfaces is rather challenging, if only because one is now considering
bulk theories which are themselves strongly coupled.
From a geometric standpoint, however, there is not much difference,
and can in fact provide a general guide for how to build such theories.
In this section we will demonstrate that the framework of M-theory with the seven extra dimensions modelled on
a space $X$ with a complete metric with holonomy group $G_2$, automatically generates the appropriate intersections and junctions
of 5D SCFTs with other 5D SCFTs and 7D flavor QFTs.

To construct the simplest such examples, our starting point will be to consider $X_a$, a smooth, non-compact
$G_2$-holonomy manifold whose metric $ds^2(X_a)$ is asymptotic to a metric cone. $X_a$ contains a closed, compact four-dimensional submanifold, $M$ whose volume is set by $a$ which is a parameter of the background metric. When discussing topological features we often simply write $X$ for $X_a$ with $a\neq 0$. In the limit $a\rightarrow 0$, $X_0$ has an isolated conical singularity. In the simplest cases, $ds^2(X_a)$ will have a continuous group of isometries, Isom$(X_a)$ and we can consider M-theory on ${X_a}/\Gamma$ with $\Gamma $ a finite subgroup of Isom$(X_a)$.
The quotient space ${X_a}/\Gamma $ retains the same set of
supercharges as the parent, but it now has various orbifold singularities.
Some of these singularities will be localized on the collapsing cycle, but
others will stretch along non-compact subspaces of $X_a$. We
interpret these non-compact loci as higher-dimensional bulk theories with
possibly non-trivial dynamics of their own. The tip of the cone when $a\rightarrow 0$ will be
interpreted as the 4D\ \textquotedblleft edge mode\textquotedblright\
realized as the quotient of the $G_{2}$ cone $X_0$. Let us mention at the outset
that singularities in non-compact $G_{2}$-holonomy spaces have been encountered
before in many cases, but usually to construct examples of Lagrangian field
theories (e.g., gauge theories localized on codimension-4 singularities,
vector-like matter on codimension-6 singularities, and chiral matter on
codimension-7 singularities \cite{Acharya:2001gy}). Here, our aim will
be to consider a broader class of singularities,
including their further intersections.

To begin, then, we briefly review the geometry of the relevant $%
G_{2}$ cones prior to applying a quotient. Letting $r$ denote the radial
direction of the cone, these geometries have an asymptotic profile for the
metric of the form:%
\begin{equation}
ds_{X}^{2}=dr^{2}+r^{2}ds_{\partial X}^{2}\text{ \ \ as \ \ }r\rightarrow
\infty \text{.}
\end{equation}%
The first examples of such AC $G_2$-holonomy manifolds were discovered by Bryant and Salamon \cite{bryant1989},
where the total space is of one of three types:

\begin{itemize}
\item $X=\mathbb{S}(S^{3})$, the spinor bundle of $S^{3}$

\item $X=\Lambda _{\text{ASD}}^{2}(S^{4})$, the bundle of anti-self-dual $2$%
-forms on $S^{4}$

\item $X=\Lambda _{\text{ASD}}^{2}(\mathbb{CP}^{2})$, the bundle of
anti-self-dual $2$-forms on $\mathbb{CP}^{2}$

\end{itemize}

See, e.g., \cite{Cvetic:2001zx, Cvetic:2001sr, Chong:2002yh, Cvetic:2002kj, Corti:2012kd, Foscolo:2015vqa, Foscolo:2017vzf, Foscolo:2018mfs} for further related discussions and examples. Let us also
mention that if one permits orbifold singularities in the collapsing 4-cycle itself,
one can also consider the bundle of anti-self-dual 2-forms on any self-dual Einstein 4-orbifold \cite{bryant1989}
and we will consider such examples in what follows (see e.g., \cite{Acharya:2001gy} and references therein).

For technical reasons, we mainly focus on cases where we can use the
classical geometry to reliably read off some features of the 4D\ interface
theories. In particular, this simplifying assumption means that we shall
mostly omit cases where we have compact 3-cycles, since Euclidean M2-branes are
expected to produce quantum corrections to the classical geometry as well
as the field theory dynamics.\footnote{%
See e.g., \cite{Apruzzi:2018oge} for a related discussion in the context of
\textquotedblleft conformal Yukawas\textquotedblright\ in F-theory, where
Euclidean D3-brane corrections modify the geometry of colliding bulk
theories realized from triple intersections of 7-branes.}

In what follows, we primarily focus on
$X_a=\Lambda _{\text{ASD}}^{2}(S^{4})$ since most of
the relevant physical phenomena already appear in singular quotients of this space.
For completeness, however, in Section \ref{sec:Rest} we treat some examples where we
construct quotients of $\Lambda _{\text{ASD}}^{2}(\mathbb{CP}^{2})$ and treat the case of
$\Lambda _{\text{ASD}}^{2}(\mathbb{WCP}^{2})$ together with some further generalizations.

The rest of this section is organized as follows. We begin by briefly
reviewing the geometry and physics of M-theory on  $X_a=\Lambda _{\text{ASD%
}}^{2}(S^{4})$ studied by Atiyah and Witten in \cite{Atiyah:2001qf}.\ Then, we
turn to the construction of quotients of the form ${X_a}/\Gamma $. Depending on
the choice of $\Gamma $, we find
a plethora of possibilities for the configurations of orbifold singularities in ${X_a}/\Gamma $,
leading to physically quite distinct theories with edge modes.
Our aim in this section will be to give a general characterization,
primarily illustrating this by way of a particular example. In section \ref{sec:EXAMP} we
present additional examples which exhibit a range of related phenomena.

\subsection{Geometry and Physics of $\Lambda _{\text{ASD}}^{2}(S^{4})$}

To frame the discussion to follow, let us briefly review the geometry and
physics of M-theory on $X_a=\Lambda _{\text{ASD}}^{2}(S^{4})$, the bundle of anti-self-dual (ASD)
2-forms over $S^{4}$. Our discussion closely follows the discussion in
\cite{Atiyah:2001qf}, see there for additional details. As found in
\cite{bryant1989}, this space admits a complete $G_{2}$-holonomy metric which is
asymptotically conical (AC):%
\begin{equation}
ds(X_a)^{2}=\frac{dr^{2}}{1-\left( \frac{a}{r}\right) ^{4}}+\frac{r^{2}}{4}%
\left( 1-\left( \frac{a}{r}\right) ^{4}\right) \left\vert D_{A}u\right\vert
^{2}+\frac{r^{2}}{2}ds_{S^{4}}^{2},  \label{Xmetric}
\end{equation}%
with $u_{i}$ the three coordinates on the fibres of $\Lambda_{\text{ASD}}$ subject to $\Sigma_i u_i u_i=1$. $D_{A}u_{i}=du_{i}+\varepsilon
_{ijk}A_{j}u_{k}$ is the connection induced by the metric on $S^4$ i.e., $A_{j}$ is the $SO(3)$ gauge connection
associated with the positive chirality spin connection on the $S^{4}$. The
round metric on $S^{4}$ given by $ds_{S^{4}}^{2}$ is fixed to have
scalar curvature $R=12$ and the radial coordinate runs from $a\leq r<\infty $. Note that replacing $ds_{S^{4}}^{2}$ in the Bryant-Salamon metric by any complete self-dual Einstein metric with $R=12$, gives a complete $G_2$-holonomy metric. However, since $\mathbb{CP}^2$ (with its standard K\"ahler metric) is the only other smooth, oriented 4-manifold with such a metric,
all other examples of Bryant-Salamon AC cones of this type are based on Einstein 4-orbifolds.
At the boundary of $X_a$, $r \rightarrow \infty $ and the Bryant-Salamon metric asymptotes to a metric cone. The cross-section of
the cone is the space of unit anti-self-dual 2-forms which is also known as
the twistor
space of $S^{4}$, i.e., we have:$\ \partial X_a=$ Tw$(S^{4})=\mathbb{CP}^{3}$.
The parameter $a$ in the metric clearly fixes the size of the $S^4$ at the center of $X_a$, which is naturally the
zero section of the bundle of ASD 2-forms. In the limit $a \rightarrow 0$, the $S^4$ shrinks to a point and $X_0$ itself becomes conically singular in this limit. Hence, the conical singularity with cross section $\mathbb{CP}^3$ is resolved or desingularized by gluing in a finite size 4-sphere.

Consider next the 4D physics realized by this model. As found by
\cite{Atiyah:2001qf}, an important feature of this metric is that the volume
modulus parameterized by $a$ is an $L^{2}$ normalizable zero-mode. As such, this
mode descends to the 4D\ theory as a massless, dynamical real scalar field. There is a related
normalizable 3-form \cite{Cvetic:2001ma}, and its harmonic
representative describes the corresponding zero mode of the M-theory
3-form $C$. Combined, one obtains a complex scalar, as required by 4D\ $%
\mathcal{N}=1$ supersymmetry. More precisely, the complex scalar $\Phi$ takes the
form:%
\begin{equation}
\Phi =V_{S^{4}}\exp \left( i\underset{F}{\int }C\right) ,
\end{equation}%
where $V_{S^{4}}$ denotes the volume of the $S^{4}$ at $r=a$
(the zero-section), and the 3-form potential is integrated over the
fibers $F$ of the bundle.

As explained in \cite{Atiyah:2001qf}, $\Phi $ is charged under a global $U(1)$
symmetry which originates in $C$-field gauge transformations. A non-zero $%
\Phi $ vacuum expectation value (vev)\ , which occurs when $a \neq 0$ and $X$ is smooth,
breaks this symmetry and is
consistent with the fact that the second Betti number $b^{2}(\partial
X)=b^{2}(X)+1$. Reference \cite{Atiyah:2001qf} thus proposed that the 4D\ dynamics of
this $G_{2}$ cone is simply that of a free chiral superfield.\ It is also interesting
to note that one can wrap an M5-brane over the $S^{4}$ at the tip of the
cone. This results in a string-like excitation in the 4D\ theory which
couples to a 2-form potential associated with the reduction of the
magnetic dual 6-form potential of M-theory reduced over the $S^{4}$. This,
however, does not result in any new dynamics in the 4D\ theory; it is
already accounted for by spontaneous symmetry breaking of the $U(1)$.
Indeed, to get additional degrees of freedom we would need to have electric
and / or magnetic particles becoming light, something which does not
occur in this example.

\subsection{Isometries}
\label{sec:Isometries}

Having introduced the geometry $X_a=\Lambda _{\text{ASD}}^{2}(S^{4})$, we now
turn to its isometries. We will use this to determine the singularities of
the quotient space $X_a/\Gamma $ for $\Gamma $ a finite subgroup of Isom$(X_a)$.
To begin, we note that in equation (\ref{Xmetric}), the metric $ds_{S^4}^{2}$
enjoys an $SO(5) \simeq Sp(2)/\mathbb{Z}_2$ isometry group,
corresponding to rotations of the $S^{4}$. For any finite $\Gamma \subset SO(5)$,
we observe that, since $\Gamma$ is an orientation preserving isometry,
the associated 3-form $G_2$-structure is also invariant and hence, $X_a/\Gamma $ is also a $%
G_{2} $-holonomy space, so M-theory on this background leads to a 4D\ $\mathcal{N}=1$
theory. To study the
resulting geometries and then extract information about the corresponding effective field
theories, we therefore now turn to a more detailed account of the geometries
generated by such finite quotients. We will then be in a position to
interpret the different physical theories constructed from this starting
point.

To proceed further, we now describe the group action of $%
SO(5)$ on both the zero section $S^{4}$ as well as the boundary $\mathbb{CP%
}^{3}=\partial X$ and establish some notation. We
describe $S^{4}$ as $\mathbb{HP}^{1}$, the quaternionic
projective line, where in our conventions, the quaternionic generators will
be labeled as $i,j,k$ so a general quaternion $q \in \mathbb{H}$ will be presented as:

\begin{equation}
q = q_0 + q_1 i + q_2 j + q_3 k, \,\,\, \text{with} \,\,\, q_i \in \mathbb{R}\,.
\end{equation}
Then both $\mathbb{CP}^{3}$ and $S^{4}$ arise
as suitable quotients of $\mathbb{C}^{4} \backslash \{0\} \simeq \mathbb{H}^2 \backslash \{0\}$
in terms of identifications specified by the homogeneous coordinates:%
\begin{eqnarray}
\mathbb{CP}^{3} &:& [Z_{1},Z_{2},Z_{3},Z_{4}]\sim \lbrack \lambda
Z_{1},\lambda Z_{2},\lambda Z_{3},\lambda Z_{4}]\text{ \ \ }(\text{for }%
\lambda \in \mathbb{C}^{\ast }) \\
\mathbb{HP}^{1} &:& [Q_{1},Q_{2}]\sim \lbrack \lambda
Q_{1},\lambda Q_{2}]\text{ \ \ }(\text{for }\lambda \in \mathbb{H}^{\ast })\,.
\end{eqnarray}%
The advantage of this
presentation is that the twistor space fibration is manifest. Indeed, in the
fibration $S^{2}\hookrightarrow \mathbb{CP}^{3}\rightarrow S^{4}$, the
projection to the base is accomplished by treating each complex
$Z_{r}=\mathrm{Re}Z_{r}+i\mathrm{Im}Z_{r}$ as a quaternion and performing the projection on
homogeneous coordinates:%
\begin{eqnarray}
\mathbb{CP}^{3} &\rightarrow &\mathbb{HP}^{1} \\
\lbrack Z_{1},Z_{2},Z_{3},Z_{4}] &\rightarrow &[Z_{1}+Z_{2}j,Z_{3}+Z_{4}j]\,.
\label{homocoords}
\end{eqnarray}%

By inspection, there is a manifest $SU(4)$ group action on the homogeneous
coordinates of $\mathbb{CP}^{3}$ (treated as a row vector, in our
conventions). By a similar token, we also observe that there is a $U(2,%
\mathbb{H})=Sp(2)$ group action on the quaternionic homogeneous coordinates
of $\mathbb{HP}^{1}$. Our convention for this is right multiplication,
in order to remain compatible with the complex
structure chosen for our $\mathbb{CP}^{3}$. Given an element of $U(2,\mathbb{%
H})$ (i.e., a $2\times 2$ matrix whose elements are quaternions subject to
the condition that its hermitian conjugate is also its inverse) the explicit
group action is:%
\begin{equation}
\lbrack Q_{1},Q_{2}]\rightarrow \lbrack Q_{1}\lambda _{11}+Q_{2}\lambda
_{21},Q_{1}\lambda _{12}+Q_{2}\lambda _{22}],  \label{Sp2Action}
\end{equation}%
in the obvious notation. This naturally lifts to a group action
on the homogeneous coordinates of $\mathbb{CP}^{3}$.\footnote{%
To see this, observe that right multiplication on a quaternion $\left(
Z_{1}+Z_{2}j\right) $ by a general quaternion $(\Lambda _{1}+\Lambda _{2}j)$
with $Z_{r}$ and $\Lambda _{r}$ complex numbers can be written as $\left(
Z_{1}+Z_{2}j\right) (\Lambda _{1}+\Lambda _{2}j)=(Z_{1}\Lambda
_{1}-Z_{2}\Lambda _{2}^{\ast })+(Z_{2}\Lambda _{1}^{\ast }+Z_{1}\Lambda
_{2})j$, namely it is a linear map on the doublet $Z_{r}$'s.}
Notice that only an $SO(5) \simeq Sp(2)/\mathbb{Z}_2$ acts faithfully on the geometry.

There is also a natural $Sp(1)_{(1)} \times Sp(1)_{(2)} \subset Sp(2)$
subgroup generated by right multiplication on the individual quaternionic coordinates:
\begin{equation}
\lbrack Q_{1},Q_{2}]\rightarrow \lbrack Q_{1}\lambda _{1},Q_{2}\lambda _{2}],  \label{Sp1Sp1Action}
\end{equation}%
in the obvious notation. To fully specify the action on $X_a$, we also need to
indicate how $SO(5)$ acts on the bundle of anti-self-dual 2-forms. For our purposes, it will suffice to specify
this in the two affine patches parameterized by the local quaternionic coordinates
$q_{\mathrm{North}} = Q_2^{-1} Q_1$ and $q_{\mathrm{South}} = Q_{1}^{-1} Q_2$ (for the patches where $Q_2 \neq 0$ and $Q_1 \neq 0$).
Here, it is helpful to note that on $\mathbb{R}^4$, the self-dual and anti-self-dual 2-forms respectively transform in the $(\mathbf{3},\mathbf{1})$ and $(\mathbf{1}, \mathbf{3})$ of $Sp(1)_{(1)} \times Sp(1)_{(2)} \simeq Spin(4)$. With this in mind, observe that in our local patches, the two quaternionic coordinates transform under the group action of line (\ref{Sp1Sp1Action}) as:
\begin{align}\label{eq:action}
q_{\mathrm{North}} & \rightarrow \lambda_{2}^{-1} q_{\mathrm{North}} \lambda_{1} \\
q_{\mathrm{South}} & \rightarrow \lambda_{1}^{-1} q_{\mathrm{South}} \lambda_{2}.
\end{align}
As such, first observe that the 2-form $dq_{\mathrm{North}} \wedge dq^{\ast}_{\mathrm{North}}$ is invariant under $Sp(1)_{(1)}$,
and $dq^{\ast}_{\mathrm{North}} \wedge dq_{\mathrm{North}}$ is invariant under $Sp(1)_{(2)}$ (since the group action is by unit norm
quaternions). We can build up the full set of anti-self-dual 2-forms by sweeping out the full orbit under the corresponding $Sp(1)$ group action. Note that the roles are reversed on the South pole, i.e., $dq_{\mathrm{South}} \wedge dq^{\ast}_{\mathrm{South}}$ is instead invariant under $Sp(1)_{(2)}$. Locally, it will often prove convenient to write the local $\mathbb{R}^7$ patch as $\mathbb{C}^3 \times \mathbb{R}$, but the complex structure will be different in the North and South poles. We write:
\be \label{eq:coordsquat}
q_{\textnormal{North}}=v_1+v_2j\,, \quad q_{\textnormal{South}}=v_1'+v_2'j\,,
\ee with complex coordinates $v_1,v_2,v_1',v_2'\in \mathbb{C}$. In each patch the anti-self-dual 2-forms can then be parameterized by the pairs $v_3,t$ and $v_3',t'$ where $v_3,v_3'\in\mathbb{C}$ and $t,t'\in\mathbb{R}$. The identification $\mathbb{R}^7=\mathbb{C}^3 \times \mathbb{R}$ is then manifest, and the patches are parameterized by
\be \begin{aligned}
\textnormal{North}:&\quad  (v_1,v_2,v_3,t) \\
\textnormal{South}:&\quad (v_1',v_2',v_3',t')
\end{aligned} \ee
and $\Gamma$ locally acts transverse to the lines parameterized by $t,t'$.

As already mentioned, we note that in each local patch, we have a 5D orbifold SCFT, as specified by a quotient of the form $\mathbb{C}^3 / \Gamma$, though the choice of complex structure and group action is different at the North and South poles. In the purely 5D limit, we have a theory with eight real supercharges, and a corresponding Coulomb branch of vacua. Geometrically, these are associated with blowup moduli. It is important to note that in the $G_2$ setting, some of these blowups are now obstructed. At some level, this is to be expected simply because we now only retain four real supercharges. It also means that possible ``gauge theory phases'' of these 5D systems may end up being inaccessible in the full system.

\subsection{The View from the Bulk}

A common feature of many of the examples we will be considering is the appearance of an interacting bulk which couples to an edge mode at the tip of the $G_2$ cone. To study this, we first consider the geometry with a finite size tip (finite volume zero section), and then explain what happens in the collapsed limit.

In the geometry with a finite size $S^4$, the resulting quotient ${X_a} / \Gamma$ will result in singularities at both the North and the South pole of the $S^4$, with additional singularities possibly stretched between the two poles. For illustrative purposes, we now focus on the fixed point loci of Abelian $\Gamma \subset Sp(1)_{(1)}\times Sp(1)_{(2)}$, given by multiplication by complex phases.\footnote{We briefly discuss some additional phenomena, such as compact codimension-6 loci, associated with non-Abelian group actions in Appendix \ref{sec:ExtraQuotients}.}

For such group actions, we find the following types of fixed point loci in $X_a$:
\begin{enumerate}
\item Codimension-6 loci of topology $\mathbb{R}$: There are up to two disconnected fixed point loci of this type. Runing radially in $X_a=\Lambda _{\text{ASD}}^{2}(S^{4})$ they are contained in a single fiber projecting to either the North or South pole of $S^4$ and give bulk 5D SCFTs in $X_a/\Gamma$.

\item Codimension-4 loci of topology $\mathbb{C}\times \mathbb{R}$: There are up to two disconnected fixed point loci of this type. They are contained in the full fiber of $X_a=\Lambda _{\text{ASD}}^{2}(S^{4})$ above either the North or South pole of $S^4$. In $X_a/\Gamma$ they give singularities engineering 7D SYM bulk modes with A-type gauge groups.

\item Codimension-4 loci $\mathbb{C}\times\mathbb{R}$ with codimension-6 enhancement at $\{0\}\times \mathbb{R}$: This case is a combination of the previous two cases. In $\Lambda _{\text{ASD}}^{2}(S^{4})/\Gamma$ this geometry descends to a 5D SCFT with A-type flavor symmetries.

\item Codimension-4 loci $S^2\times \mathbb{R}$ with codimension-6 enhancement at $(\{\text{North}\}\cup \{\text{South}\}) \times \mathbb{R}$: Here $S^2\subset S^4$ and $\{\text{North}\}, \{\text{South}\}$ denote their common North and South poles. The codimension-6 loci are as in case 1 along $\mathbb{R}$. The codimension-4 loci at fixed radius are compact, disjoint 2-spheres running between the codimension-6 loci. In $X_a/\Gamma$ these give A-type flavor symmetries branes which are gauged in 5D.

\end{enumerate}

We emphasize that the 5D bulk interpretation of codimension-4 singular loci in $X_a/\Gamma$ as 5D gauge or flavor depends on whether the fixed point components normal to the 5D SCFT are compact or non-compact respectively. They are (non)-compact when (vertical) horizontal in $X_a\rightarrow S^{4}/\Gamma$ as SCFT loci necessarily run radially/vertically in Abelian examples.

Since we have a preferred radial slicing for the geometry, we can visualize all of these 5D theories as extending out in that direction. From the perspective of the zero-section $S^4 / \Gamma$, these 5D theories are locally specified by geometries of the form $\mathbb{C}^3 / \Gamma \times \mathbb{R}$. In the limit where the zero section collapses, however, it can happen that these 5D theories are ``cut in two'' along the real line factor. As such, it is more appropriate to view the geometry as building up a junction, the structure of which is smoothed out to various intersections when the zero-section has non-zero size. In this picture, there are 4D degrees of freedom---edge modes---localized at the tip of the conical geometry.

Summarizing, then, we often expect to encounter individual 5D SCFTs with their own flavor symmetries. In the case of Abelian group actions these are concentrated at the North and South pole, so we shall denote them by $\mathcal{T}^{\mathrm{North}}$ and $\mathcal{T}^{\mathrm{South}}$. The compact $S^4$ couples the two theories together. This results in some of their common flavor symmetries being gauged since the flavor locus of the local model now extends along a compact subspace in the $S^4$.

The strength of the 5D\ gauge coupling is specified by $V(r)$,
the volume of the cycle wrapped by the 7D SYM theory:
\begin{equation}
\frac{1}{g_{5D}^{2}}\sim V(r),
\end{equation}
in natural units ignoring factors of $2$ and $\pi$.
This volume depends on the radial direction precisely because
the volume of the cycle wrapped by the SYM theory expands as we go to larger values of $r$.
Observe that as $r \rightarrow \infty$, the 5D gauge symmetry
turns back into a flavor symmetry since $V \rightarrow \infty$.
The function has a minimum at $r=a$, hence the $a \rightarrow 0$ limit is formally at
strong 5D gauge coupling, a key hint that the physics at the conical singularity supports
non-trivial interactions.

We can thus regard the 5D gauge theory as being
defined on a warped space, with warping confined to just the radial direction of the metric.
This is reflected in the standard action:%
\begin{equation}\label{eqn:5Daction}
S_{5D}\supset - \frac{1}{4}\int d^{4}x\int dr\text{ }\sqrt{\det g}\text{ }V(r)\text{Tr}%
(F_{\mu\nu}F^{\mu\nu}),
\end{equation}
with the understanding that the metric is non-trivial in the $r$ direction.

So far, we have focused on the structure of the bulk 5D theory. Note,
however, that there is an end to our space at $r=a$, and (as we reviewed above) that there is a complexified
volume modulus which can be interpreted as the vev of a 4D field $\Phi$ \cite{Atiyah:2001qf}. Returning
to equation (\ref{eqn:5Daction}), this in turn implies the existence of a coupling between the 5D bulk and the 4D edge mode.
We see a remnant of this by promoting $V(r)$ to a non-trivial function of the field $\Phi$.

Let us now analyze in more detail the resulting 4D system coupling to this
5D\ bulk. Following the general procedure outlined in \cite{Marcus:1983wb, ArkaniHamed:2001tb},
we can view 5D fields as a collection of 4D\ fields
labelled by points of an extra dimension with local coordinate $r$ in the radial direction of the cone.
From this perspective, we view the 5D $\mathcal{N}=1$ vector multiplets as
producing an infinite tower of 4D $\mathcal{N}=2$ vector multiplets with a 4D
marginal coupling $\tau$. What we are proposing to do is promote this marginal
coupling to a 4D $\mathcal{N}=1$ chiral superfield. Doing so, we
get a minimal coupling between the 5D\ and 4D\ system:\footnote{There is a small
subtlety here because the chiral superfield is localized at the tip of the singular cone.
Nevertheless, Goldstone's theorem ensures that such a
massless mode persists and will couple to the bulk near $r = a$.}
\begin{equation}
S_{\text{4D / 5D}}\supset\int d^{4}x\int dr\sqrt{\det g}\int d^{2}\theta\text{
}\delta(r-a)h\left(\frac{\Phi(x,r)}{f} \right)\text{Tr}(\mathcal{W}^{2})+h.c.,
\end{equation}
or, written slightly differently:%
\begin{equation}
S_{\text{4D / 5D}}\supset\int d^{4}x\int d^{2}\theta\text{ }h\left(\frac{\Phi}{f} \right)%
\text{Tr}(\mathcal{W}^{2})|_{r=a}+h.c., \label{L4D}%
\end{equation}
in the obvious notation. In the above, we have introduce a scale $f$, as
required by dimensional analysis. This background value is set by the background vev we expand around, i.e.,
it is dictated by the minimal volume of the $S^4 / \Gamma$.

As a point of clarification, we note that really, we should treat the
fluctuations localized at the end of the 5D\ system as distinct from
the one which sources the couplings of the 5D\ system. Note, however, that due
to the minimal coupling between the two, we are free to conflate the sources /
boundary terms, and in what follows we shall freely do so. In any event, it is
clear that we have a 5D\ system with a gauge coupling,
and this system couples to a localized fluctuation trapped on the
boundary. As we have already noted, in the singular limit where $\left\langle
\Phi\right\rangle \rightarrow0$, the 5D\ gauge theory is being extrapolated to
infinite coupling. So, we can think of this as a 5D bulk which is becoming
progressively more strongly coupled near the localized codimension-one defect.

\subsection{4D Quasi-SCFTs}

In the limit where the zero-section collapses to zero size, we have a 4D edge mode localized at the tip of the $G_2$ cone. Since this tip is singular, we expect to encounter localized degrees of freedom. There is no guarantee that these can be decoupled from the bulk modes, and so in general, we expect 5D / 4D couplings. Indeed, we already explained that the 4D modulus $\Phi$ couples to the 5D gauge theory factors via a function of the form $h(\Phi / f) \mathrm{Tr} F^{2}$.

Even though the interface supports gapless excitations, the coupling to the 5D bulk means that the 4D theory may not have a well-defined stress energy tensor defined independently of the bulk. One way to argue that this occurs generically is to observe that insertions of $\mathcal{O}_{4D}$ states will necessarily create 5D stress-energy, which in turn prevents one from defining a conserved 4D stress tensor.

Additionally, because the bulk 5D SCFTs are always interacting (i.e. their 3-point OPE coefficients are non-trivial) the 3-point coefficients of the 4D theory must also be non-trivial if the 4D boundary is not decoupled from the bulk. This is because we can always bring a bulk operator on the boundary giving us a bulk-boundary expansion as \cite{Billo:2016cpy}
\begin{equation}
\lim_{r \rightarrow 0}  \mathcal{O}_{5D}(x,r)\sim  \sum_{i}\langle \mathcal{O}_{5D}\mathcal{O}^i_{4D}\rangle  \mathcal{O}^i_{4D}(x)
\end{equation}
where $i$ labels all 4D local operators along the boundary, $r$ is the distance from the boundary in 5D and $x$ is a point on the 4D boundary. Now if the 5D OPE relations are non-trivial and the bulk-boundary 2-point functions $\langle \mathcal{O}_{5D}\mathcal{O}^i_{4D}\rangle$ are non-zero then this entails the 4D OPE relations must necessarily be non-trivial.

A general comment here is that in many string constructions of SCFTs, one often considers defects / edge modes of a higher-dimensional bulk theory. The difference from the cases considered in those situations and the present case is that the higher-dimensional bulk is often trivial in the IR, namely it has trivial three-point functions.\footnote{For example, the E-string theory, as realized via probe M5-branes of an $E_8$ 9-brane, or the 5D $\mathcal{T}_N$ theory as the junction of three 7D SYM theories.} In the present case, the bulk is typically an interacting gapless system. As such, it is unclear that there is a decoupling limit which retains only the 4D modes of the system.

\subsection{Symmetry Enhancement}

Let us provide some further evidence for the existence of non-trivial bulk / boundary couplings by studying the symmetries of these systems, and how they act on the edge modes. Consider the structure of the flavor symmetry factors of our 5D / 4D system. A rather common feature of many of these orbifold geometries is that the singularities in local patches of ${X_a}/\Gamma = \Lambda_{\mathrm{ASD}}^{2}(S^4) / \Gamma$ often comprise a subset of the singularities which appear in ${X_0}/\Gamma$, which is a cone over $\mathbb{CP}^{3} / \Gamma$. Geometrically, this is an indication of spontaneous symmetry breaking, with the breaking scale set by $\langle \Phi \rangle$, the volume modulus of the $S^4 / \Gamma$. As we tune the volume modulus to zero size, then, we should expect to see additional non-compact flavor loci emerge. We will indeed repeatedly encounter this phenomenon, and we interpret it as the distinction between the UV and IR flavor symmetries carried by the singularities.

In the case at hand, there is a distinctive enhancement pattern in which the 4D UV flavor symmetry consists of ``two copies'' of the 5D IR gauge symmetry plus one copy of the 5D IR flavor symmetry. This happens because our flavor symmetries are realized via 7D SYM theories which are supported on non-compact 3-cycles of the form $\mathbb{R}_{\geq 0} \times S^{2}$ and $\mathbb{R} \times S^{2}$, where the $S^{2}$ is either concentrated in the ASD fiber direction of $X_a / \Gamma$, or instead resides on a 2-cycle inside the $S^4 / \Gamma$ base respectively. In the latter cases, observe that the real line factor of $\mathbb{R} \times S^{2}$ filled by the 7D SYM theory splits as $\mathbb{R} = \mathbb{R}^{+} \cup \mathbb{R}^{-}$ and the $S^2$ collapses when $a\rightarrow 0$, splitting these loci in two. When the zero-section has finite volume, this results in a single factor of the flavor symmetry $\mathfrak{g}^{\mathrm{(diag)}}$, but in the collapsed limit it breaks up into two distinct copies, $\mathfrak{g}^{(+)} \times \mathfrak{g}^{(-)}$. No such doubling occurs for the 7D SYM factors. We thus conclude that the IR flavor symmetry is of the general form:
\begin{equation}
\mathfrak{g}_{IR} = \mathfrak{g}_{7D} \times \mathfrak{g}^{\mathrm{(diag)}}_{5D},
\end{equation}
while the UV flavor symmetry is of the form:
\begin{equation}
\mathfrak{g}_{UV} = \mathfrak{g}_{7D} \times \mathfrak{g}^{(+)}_{5D} \times \mathfrak{g}^{(-)}_{5D},
\end{equation}
in the obvious notation.

\begin{figure}
\begin{center}
\includegraphics[scale=0.35]{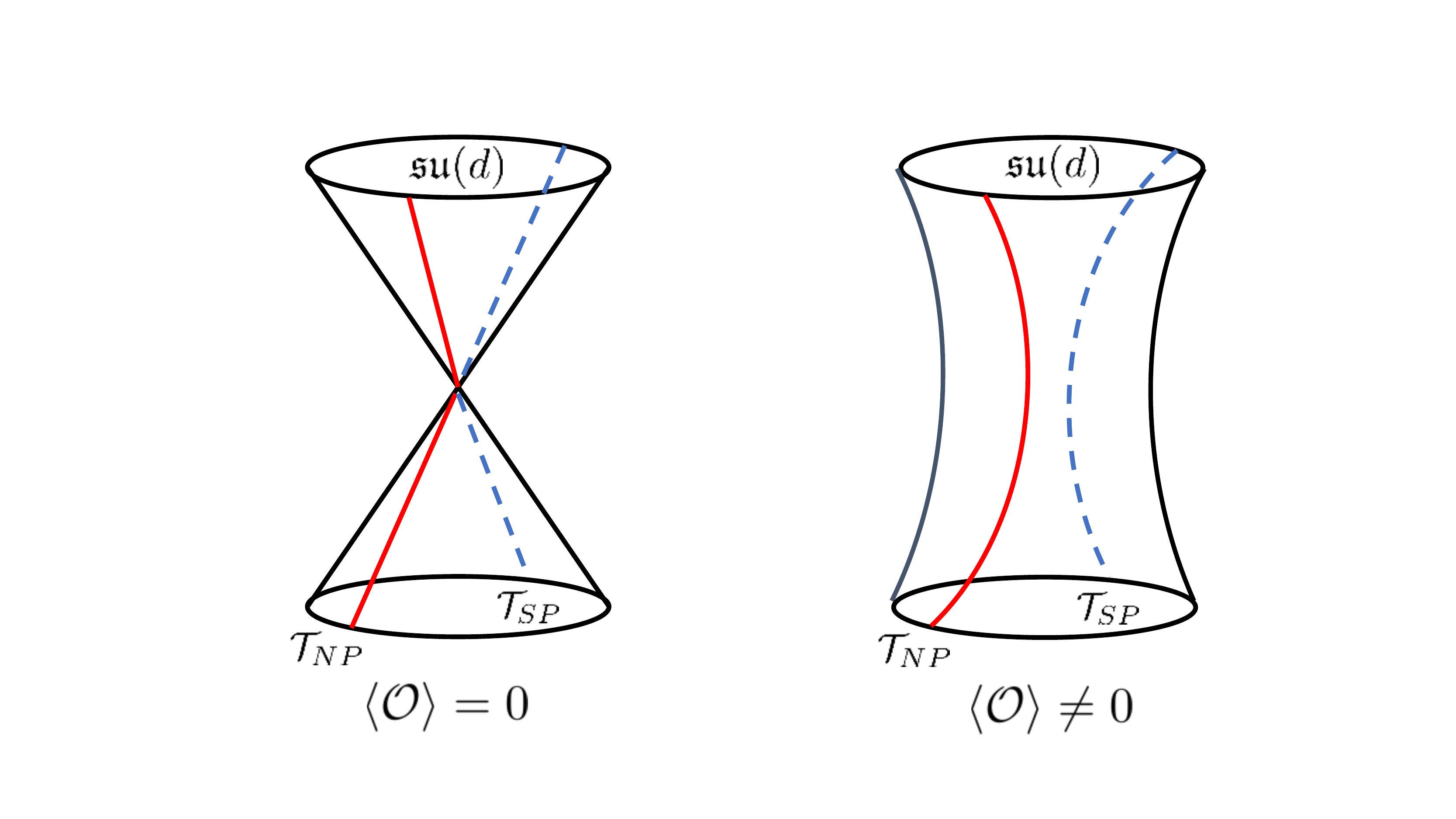}
\end{center}
\caption{Geometry of Quadrion coupled to bifundamental matter. The black locus denotes an $A_{d-1}$ singularity and the red/blue lines indicate the 5D SCFTs $\mathcal{T}^{\mathrm{North}}$ and $\mathcal{T}^{\mathrm{South}}$ respectively. Turning on a vev for a bifundamental operator $\mathcal{O}$ breaks the flavor symmetry algebra from $\mathfrak{su}(d)\times \mathfrak{su}(d)$ to $\mathfrak{su}(d)$.}\label{fig:quartionbifund}
\end{figure}

At this point, it is worth remembering that in spontaneous symmetry breaking of $\mathfrak{g}_{UV} \rightarrow \mathfrak{g}_{IR}$ there is a sense in which the UV symmetry is always present, and is simply packaged in terms of appropriate WZ terms. One can see this by moving out in the radial slicing of the $G_2$ cone, where both $\mathfrak{g}^{(+)}_{5D}$ and $\mathfrak{g}^{(-)}_{5D}$ are present, and extend to the boundary $\mathbb{CP}^3$. This is also an indication that the degree of freedom responsible for breaking the bulk symmetries is a 4D mode localized near the tip of the cone. See Figure \ref{fig:quartionbifund} for a depiction of this symmetry breaking in a $G_2$ cone geometry.

With these considerations in place, we can in principle turn to explicit finite subgroups of $SO(5)$ and construct the corresponding quotients $X_a /\Gamma$. We carry this out for a number of different examples in section \ref{sec:EXAMP}, primarily focusing on a rich class of Abelian group actions, though we do consider more general possibilities in section \ref{sec:Rest} and Appendix \ref{sec:ExtraQuotients}. In the remainder of this section we consider an illustrative example.

\subsection{Illustrative Example: Coupled $\mathcal{T}_N$ Theories}

We now present an illustrative example to exhibit some of the general considerations presented above. Our starting point is the 5D $\mathcal{T}_N$ theory as obtained from M-theory on the orbifold ${\mathbb C^3/({\mathbb Z_N \times \mathbb Z_N})}$, where the action on ${\mathbb C^3}$ is generated by diagonal matrices $\textnormal{diag} (1, \eta^{-1},\eta)$ and $\textnormal{diag} (\omega^{-1},1,\omega)$ with $\omega,\eta$ being primitive $N$th roots of unity. The action on the two patches of $\Lambda_{\text{ASD}}^2(S^4)$ are:
\begin{eqnarray}\label{eq:TN}
(v_{1},v_{2},v_{3},t) &\rightarrow &(\omega v_{1},\eta
^{-1}v_{2},\eta ^{}\omega ^{-1}v_{3},t) \\
(v_{1}^{\prime },v_{2}^{\prime },v_{3}^{\prime },t^{\prime }) &\rightarrow
&(\omega ^{-1}v_{1}^{\prime },\eta ^{-1}v_{2}^{\prime },\eta ^{}\omega v_{3}^{\prime },t^{\prime })\,.
\end{eqnarray}
From this we see that the fixed points in $X_a$ are as follows
\begin{itemize}
\item[(i.)] a copy of ${S^2 \times \mathbb R}$ from elements of the form $\eta^k$,
\item[(ii.)] a copy of ${S^2 \times \mathbb R}$ from elements of the form $\omega^k$,
\item[(iii.)] two copies of ${\mathbb R^3}$ from elements of the form $(\eta\omega)^k$ and  $(\eta^{-1}\omega)^k$,
\item[(iv.)] two copies of ${\mathbb R}$ from the other elements of the form $(\eta^k\omega^l)$ when $k\neq \pm l$. All of the components above intersect along these particular two lines.
\end{itemize}
When $X_a$ degenerates to the cone $X_0$, the components (i.) and (ii.) each become two copies of ${\mathbb R^3}$ which meet at the origin, and each pair of lines in (iv.) become two copies of ${\mathbb R^+}$ which also meet at the origin. We can thus think of the theory on $X_0$ as
arising from four copies of the 5D $\mathcal{T}_N$ theory which meet at the origin. The structure of singular loci is illustrated in figure \ref{fig:sticazzi}. Four copies of $\mathcal{T}_N$ would normally have $\mathfrak{su}(N)^{12}$ global symmetry, but here we see that the 4D theory we obtain has $\mathfrak{su}(N)^6$ flavor symmetry, so the four theories must be coupled in such a way that they share $\mathfrak{su}(N)$ factors in common. When $X_0$ is deformed to $X_a$ this becomes two copies of the $\mathcal{T}_N$ theory on ${\mathbb R}$ but whose symmetry group is not $\mathfrak{su}(N)^6$ but $\mathfrak{su}(N)^4$ from components (i.),(ii.) and (iii.).

\begin{figure}
    \centering
    \scalebox{0.8}{
    \begin{tikzpicture}
	\begin{pgfonlayer}{nodelayer}
		\node [style=none] (0) at (-2, 10) {};
		\node [style=none] (1) at (-2, 9) {};
		\node [style=none] (2) at (-1, 9) {};
		\node [style=none] (3) at (-1, 10) {};
		\node [style=none] (4) at (1, 9) {};
		\node [style=none] (5) at (1, 10) {};
		\node [style=none] (6) at (2, 10) {};
		\node [style=none] (7) at (2, 9) {};
		\node [style=none] (8) at (3.35, 7.5) {$\mathcal{T}_N^{\;\!\textnormal{south}}$};
		\node [style=none] (9) at (-3.35, 7.5) {$\mathcal{T}_N^{\;\!\textnormal{north}}$};
		\node [style=none] (10) at (-3, 8) {};
		\node [style=none] (11) at (-3, 7) {};
		\node [style=none] (12) at (3, 8) {};
		\node [style=none] (13) at (3, 7) {};
		\node [style=none] (14) at (-2, 6) {};
		\node [style=none] (15) at (-2, 5) {};
		\node [style=none] (16) at (-1, 5) {};
		\node [style=none] (17) at (-1, 6) {};
		\node [style=none] (18) at (1, 6) {};
		\node [style=none] (19) at (1, 5) {};
		\node [style=none] (20) at (2, 5) {};
		\node [style=none] (21) at (2, 6) {};
		\node [style=none] (22) at (5.5, 8) {};
		\node [style=none] (23) at (5.5, 7) {};
		\node [style=none] (24) at (6.5, 7) {};
		\node [style=none] (25) at (6.5, 8) {};
		\node [style=none] (26) at (-6.5, 8) {};
		\node [style=none] (27) at (-6.5, 7) {};
		\node [style=none] (28) at (-5.5, 7) {};
		\node [style=none] (29) at (-5.5, 8) {};
		\node [style=none] (30) at (-4.25, 7.5) {};
		\node [style=none] (31) at (4.25, 7.5) {};
		\node [style=none] (32) at (5.5, 7.5) {};
		\node [style=none] (33) at (-5.5, 7.5) {};
		\node [style=none] (34) at (-1.5, 9.5) {$N$};
		\node [style=none] (35) at (1.5, 9.5) {$N$};
		\node [style=none] (36) at (-1.5, 5.5) {$N$};
		\node [style=none] (37) at (-6, 7.5) {$N$};
		\node [style=none] (38) at (1.5, 5.5) {$N$};
		\node [style=none] (39) at (6, 7.5) {$N$};
		\node [style=none] (40) at (-9, 7.5) {$a=\infty$};
		\node [style=none] (41) at (-0.5, 3.25) {};
		\node [style=none] (42) at (-0.5, 2.25) {};
		\node [style=none] (43) at (0.5, 2.25) {};
		\node [style=none] (44) at (0.5, 3.25) {};
		\node [style=none] (49) at (2.1, 0.75) {$\mathcal{T}_N^{\;\!\textnormal{south}}$};
		\node [style=none] (50) at (-2.1, 0.75) {$\mathcal{T}_N^{\;\!\textnormal{north}}$};
		\node [style=none] (51) at (-1.75, 1.25) {};
		\node [style=none] (52) at (-1.75, 0.25) {};
		\node [style=none] (53) at (1.75, 1.25) {};
		\node [style=none] (54) at (1.75, 0.25) {};
		\node [style=none] (55) at (-0.5, -0.75) {};
		\node [style=none] (56) at (-0.5, -1.75) {};
		\node [style=none] (57) at (0.5, -1.75) {};
		\node [style=none] (58) at (0.5, -0.75) {};
		\node [style=none] (63) at (4.25, 1.25) {};
		\node [style=none] (64) at (4.25, 0.25) {};
		\node [style=none] (65) at (5.25, 0.25) {};
		\node [style=none] (66) at (5.25, 1.25) {};
		\node [style=none] (67) at (-5.25, 1.25) {};
		\node [style=none] (68) at (-5.25, 0.25) {};
		\node [style=none] (69) at (-4.25, 0.25) {};
		\node [style=none] (70) at (-4.25, 1.25) {};
		\node [style=none] (71) at (-3, 0.75) {};
		\node [style=none] (72) at (3, 0.75) {};
		\node [style=none] (73) at (4.25, 0.75) {};
		\node [style=none] (74) at (-4.25, 0.75) {};
		\node [style=none] (75) at (0, 2.75) {$N$};
		\node [style=none] (77) at (0, -1.25) {$N$};
		\node [style=none] (78) at (-4.75, 0.75) {$N$};
		\node [style=none] (80) at (4.75, 0.75) {$N$};
		\node [style=none] (81) at (-9, 0.75) {$0<a<\infty$};
		\node [style=none] (82) at (0, 4) {};
		\node [style=none] (83) at (0, -2.5) {};
		\node [style=none] (84) at (-0.25, -5.5) {};
		\node [style=none] (85) at (-0.25, -6) {};
		\node [style=none] (86) at (0.25, -6) {};
		\node [style=none] (87) at (0.25, -5.5) {};
		\node [style=none] (98) at (4.25, -6) {};
		\node [style=none] (99) at (4.25, -7) {};
		\node [style=none] (100) at (5.25, -7) {};
		\node [style=none] (101) at (5.25, -6) {};
		\node [style=none] (102) at (-5.25, -6) {};
		\node [style=none] (103) at (-5.25, -7) {};
		\node [style=none] (104) at (-4.25, -7) {};
		\node [style=none] (105) at (-4.25, -6) {};
		\node [style=none] (108) at (4.25, -6.5) {};
		\node [style=none] (109) at (-4.25, -6.5) {};
		\node [style=none] (110) at (0, -5.75) {$N$};
		\node [style=none] (112) at (-4.75, -6.5) {$N$};
		\node [style=none] (113) at (4.75, -6.5) {$N$};
		\node [style=none] (114) at (-9, -6.5) {$a=0$};
		\node [style=none] (115) at (0, -5) {};
		\node [style=none] (117) at (2.5, -4.95) {$\mathcal{T}_{N}^{\,\textnormal{south},+}$};
		\node [style=none] (118) at (2, -4.5) {};
		\node [style=none] (119) at (2, -5.5) {};
		\node [style=none] (120) at (3.25, -5) {};
		\node [style=none] (121) at (2.5, -7.95) {$\mathcal{T}_{N}^{\,\textnormal{south},-}$};
		\node [style=none] (122) at (2, -7.5) {};
		\node [style=none] (123) at (2, -8.5) {};
		\node [style=none] (124) at (3.25, -8) {};
		\node [style=none] (125) at (-2.3, -4.95) {$\mathcal{T}_{N}^{\,\textnormal{north},+}$};
		\node [style=none] (126) at (-2, -4.5) {};
		\node [style=none] (127) at (-2, -5.5) {};
		\node [style=none] (128) at (-3.25, -5) {};
		\node [style=none] (129) at (-2.3, -7.95) {$\mathcal{T}_{N}^{\,\textnormal{north},-}$};
		\node [style=none] (130) at (-2, -7.5) {};
		\node [style=none] (131) at (-2, -8.5) {};
		\node [style=none] (132) at (-3.25, -8) {};
		\node [style=none] (139) at (-0.25, -3.75) {};
		\node [style=none] (140) at (-0.25, -4.25) {};
		\node [style=none] (141) at (0.25, -4.25) {};
		\node [style=none] (142) at (0.25, -3.75) {};
		\node [style=none] (143) at (0, -4) {$N$};
		\node [style=none] (144) at (0, -4.75) {};
		\node [style=none] (145) at (-0.25, -8.75) {};
		\node [style=none] (146) at (-0.25, -9.25) {};
		\node [style=none] (147) at (0.25, -9.25) {};
		\node [style=none] (148) at (0.25, -8.75) {};
		\node [style=none] (149) at (0, -9) {$N$};
		\node [style=none] (150) at (0, -8.25) {};
		\node [style=none] (151) at (-0.25, -7) {};
		\node [style=none] (152) at (-0.25, -7.5) {};
		\node [style=none] (153) at (0.25, -7.5) {};
		\node [style=none] (154) at (0.25, -7) {};
		\node [style=none] (155) at (0, -7.25) {$N$};
		\node [style=none] (156) at (0, -8) {};
		\node [style=none] (157) at (-0.5, -9) {};
		\node [style=none] (158) at (0.5, -9) {};
		\node [style=none] (159) at (0.5, -4) {};
		\node [style=none] (160) at (-0.5, -4) {};
		\node [style=none] (161) at (-0.75, -6.43) {};
		\node [style=none] (162) at (-0.93, -6.58) {};
		\node [style=none] (163) at (0.75, -6.43) {};
		\node [style=none] (164) at (0.93, -6.58) {};
        \node [style=none] (165) at (0, -9.75) {};
	\end{pgfonlayer}
	\begin{pgfonlayer}{edgelayer}
		\draw [style=ThickLine] (0.center) to (1.center);
		\draw [style=ThickLine] (2.center) to (1.center);
		\draw [style=ThickLine] (2.center) to (3.center);
		\draw [style=ThickLine] (3.center) to (0.center);
		\draw [style=ThickLine] (5.center) to (6.center);
		\draw [style=ThickLine] (6.center) to (7.center);
		\draw [style=ThickLine] (7.center) to (4.center);
		\draw [style=ThickLine] (4.center) to (5.center);
		\draw [style=ThickLine] (7.center) to (12.center);
		\draw [style=ThickLine] (1.center) to (10.center);
		\draw [style=ThickLine] (14.center) to (15.center);
		\draw [style=ThickLine] (16.center) to (15.center);
		\draw [style=ThickLine] (16.center) to (17.center);
		\draw [style=ThickLine] (17.center) to (14.center);
		\draw [style=ThickLine] (18.center) to (19.center);
		\draw [style=ThickLine] (20.center) to (19.center);
		\draw [style=ThickLine] (20.center) to (21.center);
		\draw [style=ThickLine] (21.center) to (18.center);
		\draw [style=ThickLine] (22.center) to (23.center);
		\draw [style=ThickLine] (24.center) to (23.center);
		\draw [style=ThickLine] (24.center) to (25.center);
		\draw [style=ThickLine] (25.center) to (22.center);
		\draw [style=ThickLine] (26.center) to (27.center);
		\draw [style=ThickLine] (28.center) to (27.center);
		\draw [style=ThickLine] (28.center) to (29.center);
		\draw [style=ThickLine] (29.center) to (26.center);
		\draw [style=ThickLine] (30.center) to (33.center);
		\draw [style=ThickLine] (11.center) to (14.center);
		\draw [style=ThickLine] (21.center) to (13.center);
		\draw [style=ThickLine] (31.center) to (32.center);
		\draw [style=ThickLine, bend right=45] (41.center) to (42.center);
		\draw [style=ThickLine, bend left=45] (43.center) to (42.center);
		\draw [style=ThickLine, bend right=45] (43.center) to (44.center);
		\draw [style=ThickLine, bend left=315] (44.center) to (41.center);
		\draw [style=ThickLine] (42.center) to (51.center);
		\draw [style=ThickLine, bend right=45] (55.center) to (56.center);
		\draw [style=ThickLine, bend left=45] (57.center) to (56.center);
		\draw [style=ThickLine, bend right=45] (57.center) to (58.center);
		\draw [style=ThickLine, bend left=315] (58.center) to (55.center);
		\draw [style=ThickLine] (63.center) to (64.center);
		\draw [style=ThickLine] (65.center) to (64.center);
		\draw [style=ThickLine] (65.center) to (66.center);
		\draw [style=ThickLine] (66.center) to (63.center);
		\draw [style=ThickLine] (67.center) to (68.center);
		\draw [style=ThickLine] (69.center) to (68.center);
		\draw [style=ThickLine] (69.center) to (70.center);
		\draw [style=ThickLine] (70.center) to (67.center);
		\draw [style=ThickLine] (71.center) to (74.center);
		\draw [style=ThickLine] (52.center) to (55.center);
		\draw [style=ThickLine] (72.center) to (73.center);
		\draw [style=ThickLine] (43.center) to (53.center);
		\draw [style=ThickLine] (54.center) to (58.center);
		\draw [style=ThickLine] (56.center) to (83.center);
		\draw [style=ThickLine] (83.center) to (57.center);
		\draw [style=ThickLine] (41.center) to (82.center);
		\draw [style=ThickLine] (82.center) to (44.center);
		\draw [style=ThickLine, bend right=45] (84.center) to (85.center);
		\draw [style=ThickLine, bend left=45] (86.center) to (85.center);
		\draw [style=ThickLine, bend right=45] (86.center) to (87.center);
		\draw [style=ThickLine, bend left=315] (87.center) to (84.center);
		\draw [style=ThickLine] (98.center) to (99.center);
		\draw [style=ThickLine] (100.center) to (99.center);
		\draw [style=ThickLine] (100.center) to (101.center);
		\draw [style=ThickLine] (101.center) to (98.center);
		\draw [style=ThickLine] (102.center) to (103.center);
		\draw [style=ThickLine] (104.center) to (103.center);
		\draw [style=ThickLine] (104.center) to (105.center);
		\draw [style=ThickLine] (105.center) to (102.center);
		\draw [style=ThickLine] (84.center) to (115.center);
		\draw [style=ThickLine] (115.center) to (87.center);
		\draw [style=ThickLine] (124.center) to (99.center);
		\draw [style=ThickLine] (98.center) to (120.center);
		\draw [style=ThickLine] (105.center) to (128.center);
		\draw [style=ThickLine] (104.center) to (132.center);
		\draw [style=ThickLine, bend right=45] (139.center) to (140.center);
		\draw [style=ThickLine, bend left=45] (141.center) to (140.center);
		\draw [style=ThickLine, bend right=45] (141.center) to (142.center);
		\draw [style=ThickLine, bend left=315] (142.center) to (139.center);
		\draw [style=ThickLine] (140.center) to (144.center);
		\draw [style=ThickLine] (144.center) to (141.center);
		\draw [style=ThickLine, bend right=45] (145.center) to (146.center);
		\draw [style=ThickLine, bend left=45] (147.center) to (146.center);
		\draw [style=ThickLine, bend right=45] (147.center) to (148.center);
		\draw [style=ThickLine, bend left=315] (148.center) to (145.center);
		\draw [style=ThickLine] (145.center) to (150.center);
		\draw [style=ThickLine] (150.center) to (148.center);
		\draw [style=ThickLine, bend right=45] (151.center) to (152.center);
		\draw [style=ThickLine, bend left=45] (153.center) to (152.center);
		\draw [style=ThickLine, bend right=45] (153.center) to (154.center);
		\draw [style=ThickLine, bend left=315] (154.center) to (151.center);
		\draw [style=ThickLine] (152.center) to (156.center);
		\draw [style=ThickLine] (156.center) to (153.center);
		\draw [style=ThickLine] (127.center) to (151.center);
		\draw [style=ThickLine] (126.center) to (160.center);
		\draw [style=ThickLine] (159.center) to (118.center);
		\draw [style=ThickLine] (119.center) to (154.center);
		\draw [style=ThickLine] (123.center) to (158.center);
		\draw [style=ThickLine] (157.center) to (131.center);
		\draw [style=ThickLine] (130.center) to (162.center);
		\draw [style=ThickLine] (161.center) to (85.center);
		\draw [style=ThickLine] (86.center) to (163.center);
		\draw [style=ThickLine] (164.center) to (122.center);
	\end{pgfonlayer}
\end{tikzpicture}
    }
    \caption{Three sketches of the 5D bulk theory for different sizes of the zero section Vol$(S^4/\Gamma)\sim a$. Square (circle) nodes denote flavor (gauge) symmetries in 5D. The cone over gauge nodes denotes a gauging which breaks 5D Poincar\'e symmetry via a gauge coupling which depends on one 5D direction (the radial direction in $X_a$). In the 4D transverse directions along which the coupling does not vary 4D $\mathcal{N}=1$ supersymmetry is preserved. $a=\infty:$ two decoupled 5D $\mathcal{T}_N$ theories, formally at the north/south pole of an infinite volume $S^4/\Gamma$. $0<a<\infty:$ flavor loci are compactified transverse to the 5D SCFT locus and the volume of such transverse slices depend on the radial shell of $X_a$ they are contained in. $a=0:$ the $S^4/\Gamma$ collapses and the geometry is conical. The $\mathcal{T}_N$ theories supported on $\mathbb{R}$ over the north and south pole decompose into $\mathcal{T}_N^{\pm}$ supported on two half lines $\mathbb{R}^\pm$. 5D gauge symmetry loci also contain $\mathbb{R}$ and decompose similarly. No such splitting occurs for 7D SYM theory factors which appear as 5D flavor symmetries.}
    \label{fig:my_label}
\end{figure}
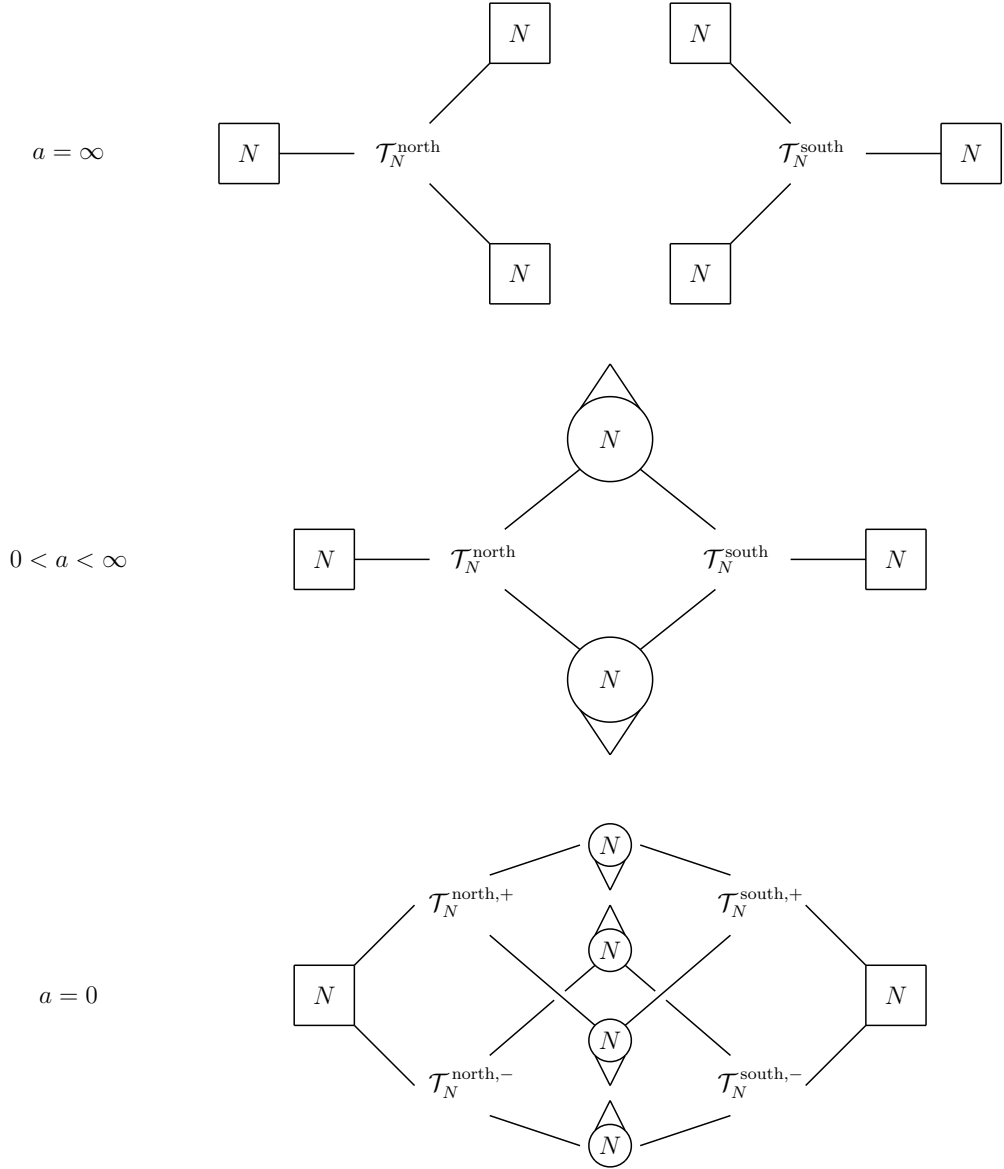

To characterize the physics of this specific example, it will prove helpful to consider various limits for the parameter $a$. In particular, as $a \rightarrow \infty$, the 5D theories have decoupled, when $0 \leq a < \infty$, the theories are now coupled together, and when $a \rightarrow 0$, we have a 5D bulk coupled to a 4D quasi-SCFT. See Figure \ref{fig:my_label} for a depiction of the various theories in this case.

Let us consider the
special case where the volume of the orbifold $S^{4}/\Gamma$ has expanded to
infinite size, i.e., the asymptotic limit $a\rightarrow\infty$. In this limit,
the two fixed points have decoupled, and so it is fruitful to consider each
separately. In this asymptotic limit, the geometry in the vicinity of the
North pole fixed point takes the form:%
\begin{equation}
\text{Decompactified Limit: } \mathbb{C}^{3}/\mathbb Z_N \times \mathbb Z_N \times \mathbb{R},
\label{Northpolelimit}%
\end{equation}
which is the geometry of the 5D $\mathcal{T}_{N}$ theory on the 5D spacetime $\mathbb{R}^{1,3} \times \mathbb{R}$.
Since we have a North and a South pole, we get two such theories. Observe that in the vicinity of the
South pole, the orientation of the Calabi-Yau is reversed relative to the
North pole geometry, but that is compensated by being on the opposite pole of
the $S^{4}$, so each system in this limit preserves the same set of eight real
supercharges. Summarizing, in this limit we have arrived at two 5D theories:%
\begin{equation}
\underset{\text{North Patch}}{\underbrace{\left[  \mathfrak{su}(N)\right]  -\left(
\mathcal{T}_{N}\right)  -\left[  \mathfrak{su}(N)^{2}\right]  }}\times \underset{\text{South
Patch}}{\underbrace{\left[  \mathfrak{su}(N)^{2}\right]  -\left(  \mathcal{T}_{N}\right)  -\left[
\mathfrak{su}(N)\right]  }}. \label{NorthSouthdecoup}%
\end{equation}
where here, the round brackets indicate the conformal fixed point and the
square brackets indicate global symmetries of the 5D\ system. We can view
these 5D flavor symmetries as 7D gauge theories on non-compact spaces.\footnote{In this discussion we neglect subtleties corresponding to the global form of the flavor symmetry group.}

When we have a finite size zero-section, the contributions to the global symmetries from the 7D SYM factors are of the form:
\begin{equation}
\mathfrak{g}_{7D} = \mathfrak{su}(N)_{\mathrm{North}} \times \mathfrak{su}(N)_{\mathrm{South}},
\end{equation}
while those from the gauged 5D SYM sectors are of the form:
\begin{equation}
\mathfrak{g}_{5D} = \mathfrak{su}(N)^2_{\mathrm{diag}},
\end{equation}
so the IR non-Abelian flavor symmetry is:
\begin{equation}
\mathfrak{g}_{IR} = \mathfrak{g}_{7D} \times \mathfrak{g}_{5D} = \mathfrak{su}(N)_{\mathrm{North}} \times \mathfrak{su}(N)^2_{\mathrm{diag}} \times \mathfrak{su}(N)_{\mathrm{South}}.
\end{equation}
In the limit where the zero-section collapses to zero size, we see an effective ``doubling'' in the 5D SYM sectors, but no such doubling for the
7D SYM sectors. Thus, the UV non-Abelian flavor symmetry is:
\begin{equation}
\mathfrak{g}_{UV} = \mathfrak{g}_{7D} \times \mathfrak{g}_{5D}^2 = \mathfrak{su}(N)_{\mathrm{North}} \times \mathfrak{su}(N)^4_{\mathrm{diag}} \times \mathfrak{su}(N)_{\mathrm{South}} = \mathfrak{su}(N)^6.
\end{equation}
This is in accord with the singularities observed in the boundary $\mathbb{CP}^3 / \Gamma$. We find that the group action $\Gamma$ has six $\mathbb{CP}^{1}$'s, each of which supports a local $A_{N-1}$ singularity locally of the form $\mathbb{C}^{2} / \mathbb{Z}_N$. In the $G_2$ cone, two of these $\mathbb{CP}^{1}$'s can be viewed as the boundary $S^2$ of an $\Lambda_{\mathrm{ASD}}^{2} \simeq \mathbb{R}^3$ fiber and as such it remains a flavor symmetry in the 5D theory. The other $\mathbb{CP}^{1}$'s supporting an A-type singularity are instead supported on a compact 2-cycle which extends from the North pole to the South pole of the $S^4 / \Gamma$. Tracking the profile of these boundary $\mathbb{CP}^{1}$'s into the interior of $X_a$, we see that they join up pairwise at the zero-section. The position dependent volume of the 7D SYM theories supported on a compact 2-cycle of $S^4 / \Gamma$ can be viewed as being gauged to a diagonal subgroup of the flavor symmetries localized at the North and South pole.

\section{Examples of Abelian Quotients} \label{sec:EXAMP}

In the previous section we presented a general discussion of quotients of $X_a = \Lambda^{2}_{\mathrm{ASD}}(S^4)$ of the form $X_a / \Gamma$. In this section, we treat in more detail the case of $\Gamma$ an Abelian group\footnote{We also cover an example with non-Abelian group action in appendix \ref{sec:NonAb}. In contrast to Abelian group actions, non-Abelian group actions can produce compact codimension-6 singularities.} group which embeds in the $Sp(2)/\mathbb{Z}_2\simeq SO(5)$ isometries via:
\begin{equation}\label{eq:embedding}
\Gamma \hookrightarrow (Sp(1)_{(1)} \times Sp(1)_{2})/\mathbb{Z}_2\hookrightarrow Sp(2)/\mathbb{Z}_2.
\end{equation}
We focus on the case where the group action is always right multiplication via complex phases on the homogeneous quaternionic coordinates of $\mathbb{HP}^1 = S^4$. Our treatment is general in the sense that we cover all of the Abelian group actions of the form $\Gamma = \mathbb{Z}_K$ and $\Gamma = \mathbb{Z}_K \times \mathbb{Z}_{L}$, though the structure of the resulting singularities can depend in a sensitive way on the divisibility properties of $K,L$ as well as the specific weights of the group actions. For this reason, we mainly focus on representative phenomena which arise in this setting.\footnote{At some level, the question boils down to a systematic treatment of all possible 5D orbifold SCFTs, a topic which has been studied in \cite{Acharya:2021jsp,Tian:2021cif, DelZotto:2022fnw}.} We begin by analyzing the case where $\Gamma$ is a cyclic group, and then turn to the case where $\Gamma$ is a product of two cyclic factors.

\subsection{Single Cyclic Factor with Generic $\Gamma =%
\mathbb{Z}
_{K}$}
\label{sec:GammaAbelian}

With a single cyclic group factor, it suffices to set $\lambda_1 =\zeta ^{a}$
and $\lambda_2 =\zeta ^{b}$ for $\zeta =\exp (2\pi i/K)$ and some $a,b$ both integer weights in \eqref{eq:action}. On the various patches of $\Lambda _{\text{ASD}}^{2}(S^{4})$, the
group action on coordinates introduced in \eqref{eq:coordsquat} then takes the form:%
\begin{eqnarray}
(v_{1},v_{2},v_{3},t) &\rightarrow &(\zeta ^{a-b}v_{1},\zeta
^{-a-b}v_{2},\zeta ^{2b}v_{3},t) \\
(v_{1}^{\prime },v_{2}^{\prime },v_{3}^{\prime },t^{\prime }) &\rightarrow
&(\zeta ^{b-a}v_{1}^{\prime },\zeta ^{-a-b}v_{2}^{\prime },\zeta
^{2a}v_{3}^{\prime },t^{\prime }).
\end{eqnarray}
In both the North and the South pole patch, we see a local singularity of the form $\mathbb{C}^3 / \Gamma \times \mathbb{R}$, albeit,
with respect to different complex structures and group actions. By itself, each 5D SCFT would preserve eight real supercharges, but the combined system preserves only four real supercharges.

To analyze the resulting singularities, it is convenient to write $c=a-b$
and $d=-a-b$. For example, on the patches of $\Lambda _{\text{ASD}%
}^{2}(S^{4})$, we have:%
\begin{eqnarray}
\label{eq:GroupactionPara1}
(v_{1},v_{2},v_{3},t) &\rightarrow &(\zeta ^{c}v_{1},\zeta ^{d}v_{2},\zeta
^{-c-d}v_{3},t) \\ \label{eq:GroupactionPara2}
(v_{1}^{\prime },v_{2}^{\prime },v_{3}^{\prime },t^{\prime }) &\rightarrow
&(\zeta ^{-c}v_{1}^{\prime },\zeta ^{d}v_{2}^{\prime },\zeta
^{c-d}v_{3}^{\prime },t^{\prime }).
\end{eqnarray}%
By itself, M-theory on a $\mathbb{C}^{3}/%
\mathbb{Z}
_{K}$ singularity would lead to a 5D\ SCFT\ with flavor symmetry. At the
North and South pole the local models exhibit flavor symmetries with non-Abelian Lie
algebra factors
\begin{eqnarray}
\mathfrak{su}((\left\vert c\right\vert ,K))_{\text{North}%
}\times \mathfrak{su}((\left\vert d\right\vert ,K))_{\text{North}}\times
\mathfrak{su}((\left\vert c+d\right\vert ,K))_{\text{North}}\\
\mathfrak{su}((\left\vert c\right\vert ,K))_{\text{South}%
}\times \mathfrak{su}((\left\vert d\right\vert ,K))_{\text{South}}\times
\mathfrak{su}((\left\vert c-d\right\vert ,K))_{\text{South}}
\end{eqnarray}
respectively, where we
write $(n,m)$ to denote the greatest common divisor of integers $n$ and $m$. Observe, however, that
while the $v_{3}$ (and $v_{3}^{\prime }$) directions are always non-compact,
$v_{1}$ and $v_{2}$ are coordinates inside the $S^{4}$. In particular, this
means that the loci supporting the pairs
\begin{eqnarray}
    \mathfrak{su}((\left\vert c\right\vert ,K))_{\text{North}}, \,\mathfrak{su}((\left\vert c\right\vert ,K))_{\text{South}}~\rightarrow ~ \mathfrak{su}((\left\vert c\right\vert ,K))\\  \mathfrak{su}((\left\vert d\right\vert ,K))_{\text{North}},\, \mathfrak{su}((\left\vert d\right\vert ,K))_{\text{South}}~\rightarrow ~ \mathfrak{su}((\left\vert d\right\vert ,K))
\end{eqnarray}
are identified in going between the two quaternionic patches and combine into a single locus. This compactification into a common locus means that the 5D symmetries are gauged into a diagonal subgroup. So, out of the original six flavor
symmetry factors observed in each local patch, we only retain a $\mathfrak{su%
}((\left\vert c+d\right\vert ,K))_{\text{North}}\times \mathfrak{su}%
((\left\vert c-d\right\vert ,K))_{\text{South}}$ flavor symmetry in 5D.

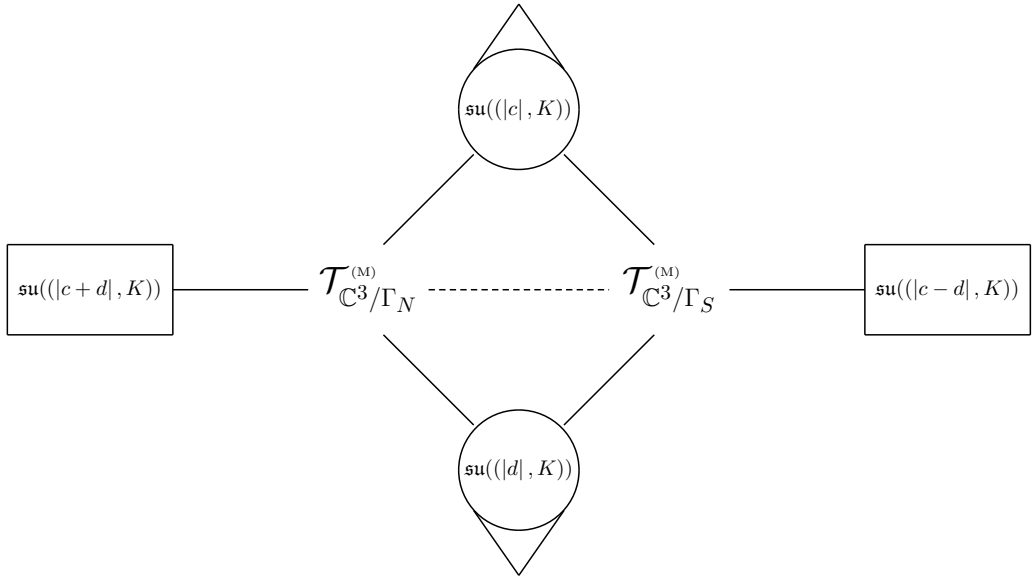
\begin{figure}
    \centering
    \scalebox{0.8}{\begin{tikzpicture}
	\begin{pgfonlayer}{nodelayer}
		\node [style=none] (0) at (0, 4) {};
		\node [style=none] (1) at (0, 2) {};
		\node [style=none] (2) at (1, 3) {};
		\node [style=none] (3) at (-1, 3) {};
		\node [style=none] (4) at (0, -2) {};
		\node [style=none] (5) at (0, -4) {};
		\node [style=none] (6) at (1, -3) {};
		\node [style=none] (7) at (-1, -3) {};
		\node [style=none] (8) at (5.75, 0.75) {};
		\node [style=none] (9) at (8.5, 0.75) {};
		\node [style=none] (10) at (8.5, -0.75) {};
		\node [style=none] (11) at (5.75, -0.75) {};
		\node [style=none] (12) at (-8.5, 0.75) {};
		\node [style=none] (13) at (-5.75, 0.75) {};
		\node [style=none] (14) at (-5.75, -0.75) {};
		\node [style=none] (15) at (-8.5, -0.75) {};
		\node [style=none] (20) at (0.75, 2.25) {};
		\node [style=none] (21) at (0.75, -2.25) {};
		\node [style=none] (22) at (-0.75, -2.25) {};
		\node [style=none] (23) at (-0.75, 2.25) {};
		\node [style=none] (24) at (2.25, 0.75) {};
		\node [style=none] (25) at (2.25, -0.75) {};
		\node [style=none] (26) at (-2.25, 0.75) {};
		\node [style=none] (27) at (-2.25, -0.75) {};
		\node [style=none] (28) at (5.75, 0) {};
		\node [style=none] (29) at (-5.75, 0) {};
		\node [style=none] (30) at (-3.5, 0) {};
		\node [style=none] (31) at (3.5, 0) {};
		\node [style=none] (32) at (0, 3) {\footnotesize $\mathfrak{su}((\left\vert c\right\vert ,K))$};
		\node [style=none] (33) at (0, -3) {\footnotesize $\mathfrak{su}((\left\vert d\right\vert ,K))$};
		\node [style=none] (34) at (-2.5, 0) {\Large $\mathcal{T}^{\textnormal{\tiny\!  (M)}}_{\mathbb{C}^3/\Gamma_N}$};
		\node [style=none] (35) at (2.5, 0) {\Large $\mathcal{T}^{\textnormal{\tiny\!  (M)}}_{\mathbb{C}^3/\Gamma_S}$};
		\node [style=none] (36) at (-7.125, 0) {\footnotesize $\mathfrak{su}((\left\vert c+d\right\vert ,K))$};
		\node [style=none] (37) at (7.125, 0) {\footnotesize $\mathfrak{su}((\left\vert c-d\right\vert ,K))$};
  \node [style=none] (38) at (-1.5, 0) {};
		\node [style=none] (39) at (1.5, 0) {};
        \node [style=none] (40) at (0, 4.75) {};
		\node [style=none] (41) at (-0.75, 3.6725) {};
        \node [style=none] (42) at (0.75, 3.6725) {};
        \node [style=none] (43) at (0, -4.75) {};
		\node [style=none] (44) at (-0.75, -3.6725) {};
        \node [style=none] (45) at (0.75, -3.6725) {};
	\end{pgfonlayer}
	\begin{pgfonlayer}{edgelayer}
		\draw [style=ThickLine, bend right=45] (1.center) to (2.center);
		\draw [style=ThickLine, bend right=45] (2.center) to (0.center);
		\draw [style=ThickLine, in=90, out=-180] (0.center) to (3.center);
		\draw [style=ThickLine, bend right=45] (3.center) to (1.center);
		\draw [style=ThickLine, bend right=45] (5.center) to (6.center);
		\draw [style=ThickLine, bend right=45] (6.center) to (4.center);
		\draw [style=ThickLine, in=90, out=-180] (4.center) to (7.center);
		\draw [style=ThickLine, bend right=45] (7.center) to (5.center);
		\draw [style=ThickLine] (9.center) to (10.center);
		\draw [style=ThickLine] (10.center) to (11.center);
		\draw [style=ThickLine] (11.center) to (8.center);
		\draw [style=ThickLine] (8.center) to (9.center);
		\draw [style=ThickLine] (13.center) to (14.center);
		\draw [style=ThickLine] (14.center) to (15.center);
		\draw [style=ThickLine] (15.center) to (12.center);
		\draw [style=ThickLine] (12.center) to (13.center);
		\draw [style=ThickLine] (20.center) to (24.center);
		\draw [style=ThickLine] (25.center) to (21.center);
		\draw [style=ThickLine] (26.center) to (23.center);
		\draw [style=ThickLine] (27.center) to (22.center);
		\draw [style=ThickLine] (29.center) to (30.center);
		\draw [style=ThickLine] (31.center) to (28.center);
  \draw [style=ThickLine] (40.center) to (41.center);
  \draw [style=ThickLine] (40.center) to (42.center);
  \draw [style=DashedLine] (38.center) to (39.center);
  \draw [style=ThickLine] (43.center) to (44.center);
  \draw [style=ThickLine] (43.center) to (45.center);
	\end{pgfonlayer}
\end{tikzpicture}
}
    \caption{General 5D theory with $\Gamma=\mathbb{Z}_K$. Two pairs of flavor symmetry loci between two 5D SCFT sectors are gauged, circular nodes, with one pair of flavor symmetries remaining, square nodes. Clearly, $\Gamma=\Gamma_N=\Gamma_S$ with the subscripts distinguish the group actions at the North and South pole. The dashed line denotes a massive mode resulting from an M2-brane wrapped on a possible 2-cycle of $S^4/\Gamma$. The cones of the gauge nodes denote a 5D gauging with gauge coupling depending on the radial coordinate $r$ of the $G_2$-holonomy space such that 4D $\mathcal{N}=1$ supersymmetry is preserved.}
    \label{fig:QFT}
\end{figure}

Schematically, we represent the 5D setup as in Figure \ref{fig:QFT}. In addition to this basic structure we should keep in mind that the 5D gauge couplings associated with
$\mathfrak{su}((\left\vert c\right\vert ,K))\times \mathfrak{su}((\left\vert d\right\vert ,K))$ are fibered over the radial direction as determined by the volume of the flavor locus on a given radial shell. Furthermore, since $S^4/\Gamma$ has
non-trivial torsional 2-cycles (see Appendix \ref{app:homologygrps}) we obtain additional massive modes by wrapping an M2-brane on these.
Tuning the vev $\langle \Phi \rangle \rightarrow 0$, these states become light  and lead to additional dynamics compared to the smooth case considered in \cite{Atiyah:2001qf}.

\subsection{Single Cyclic Factor with Specialized $\Gamma =%
\mathbb{Z}
_{K}$}
We now cover some of the quite varied landscape of theories by making various choices of the integers $K$, $a$, and $b$ which specify a $\mathbb{Z}_K$ quotient of the $G_2$ cone $X$. We do not carry out an exhaustive analysis, but simply illustrate that different choices lead to modifications in the number of flavor symmetry factors, as well as various sub-diagrams of the diagram in Figure \ref{fig:QFT}.

\paragraph{Example: Quadrion Theories.} For this class of examples we specialize to $\Gamma=\mathbb{Z}_K$ and group action as parameterized in lines \eqref{eq:GroupactionPara1} and \eqref{eq:GroupactionPara2}, with non-vanishing weights satisfying
\begin{equation}\label{eq:quartcond}
    (\left\vert c\right\vert,K)=(\left\vert d\right\vert, K)=(\left\vert c+d\right\vert, K)=(\left\vert c-d\right\vert, K)=1.
\end{equation}
With these constraints the local models $\mathbb{C}^3/\Gamma$ are isolated singularities and the associated 5D SCFTs have no apparent non-Abelian flavor symmetries. Within the $G_2$-holonomy space we find two lines of codimension-6 singularities parameterized by the local coordinates $t,t'$ in fibers over the North and South pole. The base $S^4/\Gamma$ contains torsional 2-cycles $H_2(S^4/\Gamma)\cong\mathbb{Z}_K$ and we obtain additional massive states from wrapped branes.

\begin{figure}
\begin{center}
\includegraphics[scale=0.35]{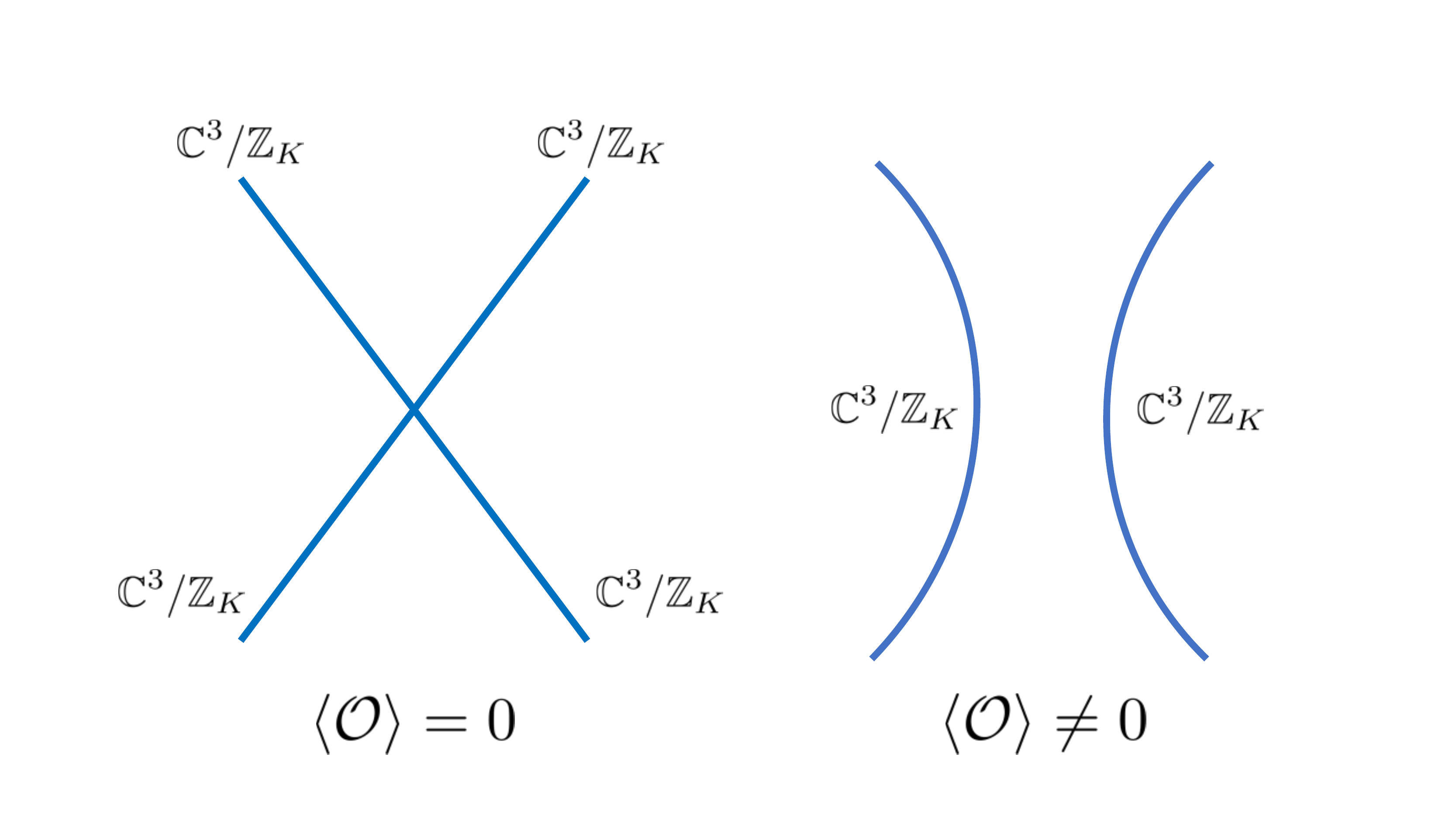}
\end{center}
\caption{Breaking the Quadrion. \textsc{left:} The singularity corresponding to a Quadrion theory, a fourvalent junction of 5d SCFTs. \textsc{right:} Resolving the tip of the cone triggers a breaking pattern to a pair of 5D SCFTs. }\label{fig:ommerdah}
\end{figure}

In the conical limit $\mathrm{Vol}(S^4/\Gamma)\rightarrow 0$, the pair of singular lines is deformed into a collection of four codimension-6 singularities, each with radial worldvolume $\mathbb{R}_+$ and transverse geometry $\mathbb{C}^3/\mathbb{Z}_K$. The four lines intersect in codimension-7 at the tip of the $G_2$ cone, see Figure \ref{fig:ommerdah} (left). In this limit the torsional 2-cycle $H_2(S^4/\Gamma)$ is contracted and we obtain additional light degrees of freedom.

We refer to such 4D $\mathcal{N}=1$ theories as \textit{Quadrion} theories. They are quasi-SCFTs in the sense that they couple to the bulk 5D theories $\mathcal{T}^{\textnormal{\tiny\!  (M)}}_{\mathbb{C}^3/\mathbb{Z}_K}$. The fact that we have a four-valent junction (see Figure \ref{fig:sticazzi}) means that there can be localized edge modes.

When $\mathrm{Vol}(S^4/\Gamma)=\langle \Phi \rangle \neq 0$, the $S^4 / \Gamma$ and the geometry becomes the one in Figure \ref{fig:ommerdah} (right), namely, we are left with just two 5D SCFTs. We interpret this effect as a flow from the Quadrion theory to a different recombined interface in which some of the flavor symmetries have been broken to a diagonal subgroup, and in which the 5D SCFTs decouple in the deep IR.

\paragraph{Example: Pure Flavor / Bifundamental Matter}
\smallskip

While the Quadrion orbifold geometry only contains codimension-6 singularities, here we consider the other extreme where the orbifold only contains codimension-4 singularities in the geometry. These were previously considered in \cite{Acharya:2001gy, Atiyah:2001qf} and are dual to the Type IIA background given by two stacks, each containing $K$ D6-branes. The low energy theory then consists of a chiral multiplet $\mathcal{O}$ in the bifundamental representation $(\overline{\mathbf{K}},\mathbf{K})$. Giving a vev to this operator triggers a breaking pattern which also resolves the tip of the cone. This is packaged in terms of a flavor neutral combination of operators.

Explicitly, this $\mathbb{Z}_K$ action is given by
\begin{equation}
    [Z_1,Z_2,Z_3,Z_4] \rightarrow [\lambda Z_1, \lambda Z_2, \lambda^{-1}Z_3, \lambda^{-1}Z_4]
\end{equation}
where $\lambda$ is a $K$th root of unity if $K$ is odd and $2K$ if $K$ is even.\footnote{In which case $\mathbb{Z}_{2K}/\mathbb{Z}_2 \simeq \mathbb{Z}_K$ is what acts effectively on the geometry.} There are two fixed $S^2$'s, one located at $Z_1=Z_2=0$ and the other at $Z_3=Z_4=0$ which project to a common $S^2$ in the $S^4$ base. This means that the flavor locus topology is $S^2\times \mathbb{R}$ in the resolved phase and $\mathbb{R}^3\cup \mathbb{R}^3$ in the unresolved phase. The 4D flavor algebra therefore changes from $\mathfrak{su}(K)\times \mathfrak{su}(K)$ to $\mathfrak{su}(K)$.

\paragraph{Example: Quadrions Coupled to Bifundamental Matter.}\smallskip

For this class of examples we specialize to $\Gamma=\mathbb{Z}_K$ with group action as in lines \eqref{eq:GroupactionPara1} and \eqref{eq:GroupactionPara2}, which we reproduce here:
\begin{eqnarray}\label{eq:BifundamentalMatter1}
(v_{1},v_{2},v_{3},t) &\rightarrow &(\zeta ^{c}v_{1},\zeta ^{d}v_{2},\zeta
^{-c-d}v_{3},t) \\\label{eq:BifundamentalMatter2}
(v_{1}^{\prime },v_{2}^{\prime },v_{3}^{\prime },t^{\prime }) &\rightarrow
&(\zeta ^{-c}v_{1}^{\prime },\zeta ^{d}v_{2}^{\prime },\zeta
^{c-d}v_{3}^{\prime },t^{\prime}).
\end{eqnarray}%
We relax the Quadrion constraints \eqref{eq:quartcond} with non-vanishing weights to
\begin{equation}
    (\left\vert c\right\vert,K)=(\left\vert c+d\right\vert, K)=(\left\vert c-d\right\vert, K)=1, \quad K=md,
\end{equation}
for some integers $m,d>1$. Within the $G_2$-holonomy orbifold we again have codimension-6 singularities supporting 5D SCFTs as in the case of the Quadrion (see Figure \ref{fig:ommerdah}), but now with local 5D SCFT sectors exhibiting a simple, non-Abelian flavor symmetry algebra $\mathfrak{su}(d)$ along $v_1,v_1',v_3,v_3'=0$ parameterized by local coordinates $v_2,v_2',t,t'$. The base $S^4/\Gamma$ again contains torsional 2-cycles $H_2(S^4/\Gamma)\cong\mathbb{Z}_m$ and one has massive states from wrapped branes.

So, in both the North and South pole patches we have a 5D SCFT with an $\mathfrak{su}(d)$ flavor symmetry. In the compact $S^4 / \Gamma$, this is supported on a compact two-dimensional subspace. To establish this, it is helpful to consider the group action on the boundary geometry $\mathbb{CP}^3$. Let $\rho\in \mathbb{Z}_K$
be a generating element. The action of $\rho^m$ on the homogeneous coordinates is:
\begin{equation}
    \rho^m:[Z_1,Z_2,Z_3,Z_4]\rightarrow [\zeta^{am}Z_1,\zeta^{-am}Z_2,\zeta^{bm}Z_3,\zeta^{-bm}Z_4].
\end{equation}
where $c+d=-2b$ and $c-d=2a$. By assumption, $K=md=-m(a+b)$ so $\zeta^{am}=\zeta^{-bm}$ and the action becomes, with $\xi=\zeta^a$,
\begin{equation}\label{eq:AW}
    \rho^m:[Z_1,Z_2,Z_3,Z_4]\rightarrow [\xi^{m}Z_1,\xi^{-m}Z_2,\xi^{m}Z_3,\xi^{-m}Z_4].
\end{equation}
Therefore, we have two codimension-4 loci that intersect the boundary
\begin{equation}
    \textnormal{Order\,$d$ fixed point loci in $\mathbb{CP}^3$}: \;\;  (Z_1=Z_4=0), \; \; \;  \; (Z_2=Z_3=0)\,,
\end{equation}
and $\mathbb{CP}^3/\mathbb{Z}_d$ with $\mathbb{Z}_d\cong \langle x^m\rangle$ has two spheres worth of $\mathfrak{su}(d)$ singularities which are further quotiented in $\mathbb{CP}^3/\mathbb{Z}_K$.

The fixed point loci project to the same $S^2$ in the $S^4$ base, as there is only one $A_{d-1}$ singularity in each of the North and South pole patches. This means that in the resolved phase, $\mathrm{Vol}(S^4)\neq 0$, the fixed point locus is connected with topology $\mathbb{R}\times S^2$, and intersects the boundary at $\{\pm \infty \}\times S^2$. In the 4D theory, this means we have an $\mathfrak{su}(d)$ flavor symmetry.

Much as in the other examples, there is a further enhancement once we collapse the zero-section. Indeed, in the boundary $\mathbb{CP}^3 / \Gamma$ we observe that there are two $\mathbb{CP}^{1}$'s, each of which locally supports a $\mathbb{C}^{2} / \mathbb{Z}_d$ singularity. So, we conclude that there is a 4D flavor symmetry enhancement to $\mathfrak{su}(d) \times \mathfrak{su}(d)$. The physical interpretation is that there is 4D bifundamental operator $\mathcal{O}_{4D}$ which picks up a vev, triggering the breaking pattern to the diagonal $\mathfrak{su}(d)$. One can view this as a generalization of the previous example with a free bifundamental $\Phi$, again with a flat direction associated to giving a finite volume to the $S^4$.

\paragraph{Example: 5D SCFT Intersecting a Flavor Stack.}It is also of interest to consider the special case where one of the $Sp(1)$
factors acts trivially on the geometry. It suffices to set $\lambda_2 =1$ and
set $\lambda_1 =\zeta =\exp (2\pi i/K)$, a primitive $K$th root of unity. On the various patches of $\Lambda _{\text{ASD}}^{2}(S^{4})$, the
group action takes the form:%
\begin{eqnarray}
(v_{1},v_{2},v_{3},t) &\rightarrow &(\zeta v_{1},\zeta ^{-1}v_{2},v_{3},t) \\
(v_{1}^{\prime },v_{2}^{\prime },v_{3}^{\prime },t^{\prime }) &\rightarrow
&(\zeta ^{-1}v_{1}^{\prime },\zeta ^{-1}v_{2}^{\prime },\zeta
^{2}v_{3}^{\prime },t^{\prime })
\end{eqnarray}
while on the asymptotic $\mathbb{CP}^3$ we have the action
\begin{equation}\label{eq:bdry}
   [Z_1,Z_2,Z_3,Z_4]\rightarrow [\zeta Z_1,\zeta^{-1}Z_2,Z_3,Z_4].
\end{equation}

So in this case, we observe a 7D Super Yang-Mills theory at the North pole
with gauge algebra $\mathfrak{su}(K)$ (i.e., it fills the $v_{3}$ and $t$
directions). At the South pole, we still have a singularity which would be
characterized by a 5D SCFT. In this case, the flavor symmetry at the South
pole depends on whether $K$ is even or odd. When $K$ is even, the flavor
symmetry is $\mathfrak{su}(2)_{\text{South}}$, supported along $%
v_{1}^{\prime }=v_{2}^{\prime }=0$. As such, it remains a flavor symmetry
(not gauged in 5D)\ in the full geometry. When $K$ is odd, there is no
flavor symmetry factor.

\begin{figure}
\begin{center}
\includegraphics[scale=0.35]{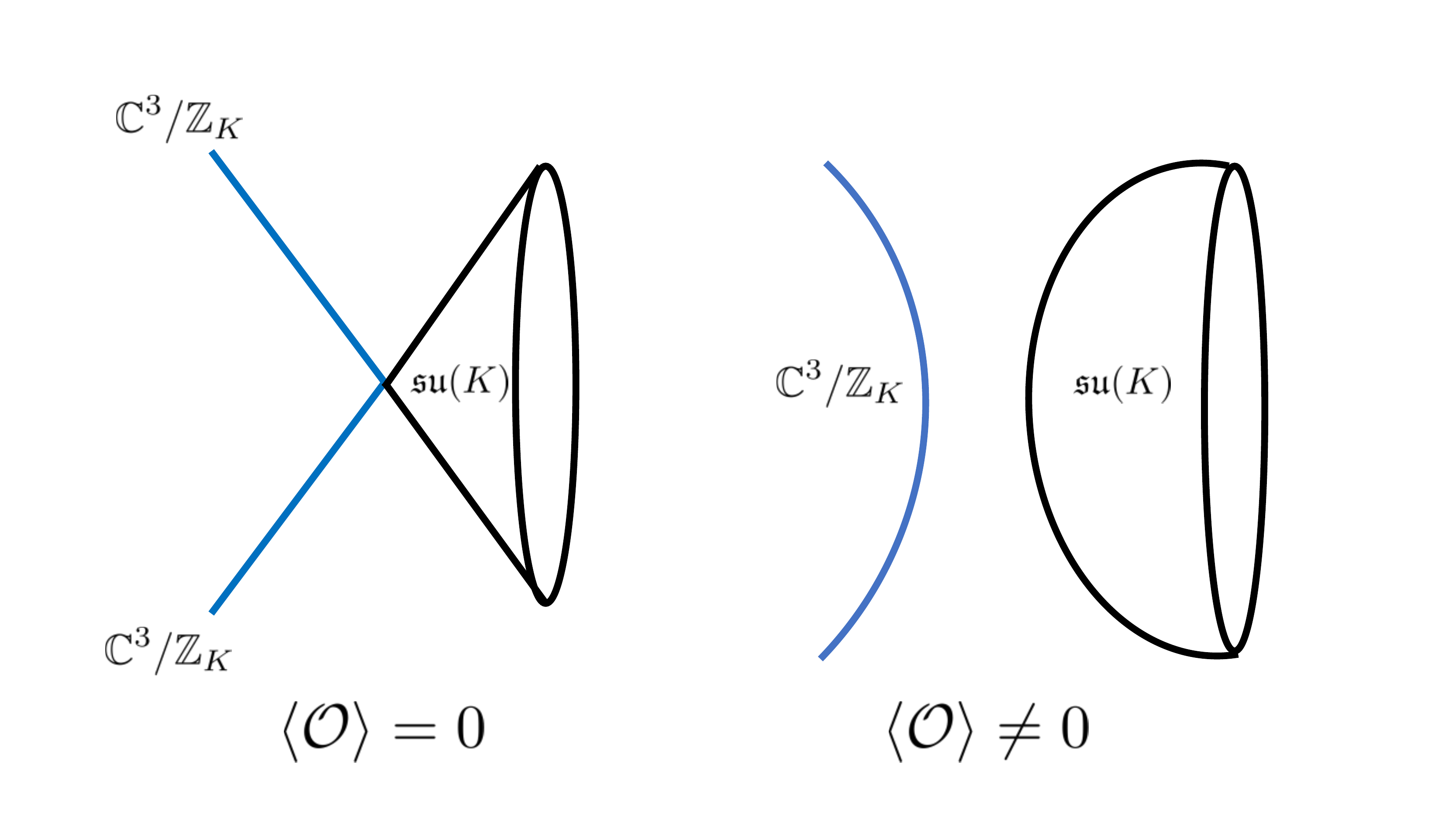}
\end{center}
\caption{On the left, we have the geometry of a 5D SCFT localized at a $\mathbb{C}^3/\mathbb{Z}_K$ singularity which intersects a $\mathfrak{su}(K)$ flavor locus in the unresolved phase, $\langle \mathcal{O}\rangle=0$. On the right, we have the resolved phase $\langle \mathcal{O}\rangle\neq0$ }\label{fig:5dscftplusflavor}
\end{figure}

We can argue similarly from the boundary geometry \eqref{eq:bdry}. Here we have a sphere of $A_{K-1}$ singularities at $Z_1=Z_2=0$. On the other hand when $Z_3=Z_4=0$ we find a trivially acting $\mathbb{Z}_2\subset \mathbb{Z}_K$ for even $K$ which indicates the $A_1$ flavor locus described before. In addition $Z_2=Z_3=Z_4=0$ and $Z_1=Z_3=Z_4=0$ are acted on trivially by the full $\mathbb{Z}_K$ and mark the intersection of the codimension-6 loci with the boundary. See Figure \ref{fig:5dscftplusflavor} for the case of $K$ odd. As a final comment, in this case we have $H_2(S^4/\Gamma)=0$, so there are no additional supersymmetric states available from wrapped M2-branes.

\subsection{Two Cyclic Factors with Generic $\Gamma =%
\mathbb{Z}
_{K}\times
\mathbb{Z}
_{L}$}

Consider in complete generality $\Gamma =%
\mathbb{Z}
_{K}\times
\mathbb{Z}
_{L}$ and
denote by $\eta,\omega $ a
primitive $K$th, $L$th root of unity, respectively.
We take $\lambda_1 =\eta ^{a}\omega ^{b}$ and $\lambda_2 =\eta ^{c}\omega ^{d}$ with $a,b,c,d$ integers.
Then, the group action on the various patches of $\Lambda _{\text{ASD}%
}^{2}(S^{4})$, takes the form:%
\begin{eqnarray}
(v_{1},v_{2},v_{3},t) &\rightarrow &(\eta ^{a-c}\omega ^{b-d}v_{1},\eta
^{-a-c}\omega ^{-b-d}v_{2},\eta ^{2c}\omega ^{2d}v_{3},t) \\
(v_{1}^{\prime },v_{2}^{\prime },v_{3}^{\prime },t^{\prime }) &\rightarrow
&(\eta ^{c-a}\omega ^{d-b}v_{1}^{\prime },\eta ^{-a-c}\omega
^{-b-d}v_{2}^{\prime },\eta ^{2a}\omega ^{2b}v_{3}^{\prime },t^{\prime }).
\end{eqnarray}%
In both the North and South pole patch, we recall that the geometry takes
the general form $%
\mathbb{C}
^{3}/\Gamma \times \mathbb{R}$, although the specific theory so obtained
depends sensitively on the specific divisibility properties for both $K$, $L$
as well as the exponents $a,b,c,d$ which, although not uniquely, determine the embedding \eqref{eq:embedding}. The general point, however, is that much
as in the case of a single cyclic group factor, we get some flavor
symmetries which are gauged, i.e., when they do not sit at $v_{1}=v_{2}=0$.

\paragraph{Example: $\mathcal{N}=1$ Trinion-like Pairs.} As an illustrative case study,
let us restrict to $a=c$ and $b=-d$. Then, the general action specializes to:
\begin{eqnarray}\label{eq:TrinionLikePairs1}
(v_{1},v_{2},v_{3},t) &\rightarrow &(\omega ^{2b}v_{1},\eta
^{-2a}v_{2},\eta ^{2a}\omega ^{-2b}v_{3},t) \\ \label{eq:TrinionLikePairs2}
(v_{1}^{\prime },v_{2}^{\prime },v_{3}^{\prime },t^{\prime }) &\rightarrow
&(\omega ^{-2b}v_{1}^{\prime },\eta ^{-2a}v_{2}^{\prime },\eta ^{2a}\omega ^{2b}v_{3}^{\prime },t^{\prime }).
\end{eqnarray}
In the base four-sphere $S^4$ we find planes of fixed points given by $v_i=0$, with $i=1,2$, which are compactified to two-spheres by the corresponding plane $v_i'=0$ upon transitioning between patches. Extending these loci patchwise along $t$ with $v_3,v_3'=0$ we find overall two 4D flavor loci which are gauged from a 5D perspective. Finally setting $v_1=v_2=0$ and  $v_1'=v_2'=0$ we find two more 4D flavor loci which are also flavor loci from a 5D perspective.

Let us now specialize further to $a=b=1$ for which the action takes the form:
\begin{eqnarray}
(v_{1},v_{2},v_{3},t) &\rightarrow &(\omega ^{2}v_{1},\eta
^{-2}v_{2},\eta ^{2}\omega ^{-2}v_{3},t) \\
(v_{1}^{\prime },v_{2}^{\prime },v_{3}^{\prime },t^{\prime }) &\rightarrow
&(\omega ^{-2}v_{1}^{\prime },\eta ^{-2}v_{2}^{\prime },\eta ^{2}\omega ^{2}v_{3}^{\prime },t^{\prime }).
\end{eqnarray}
If $K,L$ is odd (resp. even) we define $N\equiv K$ (resp. $K/2$) and $M=L$ (resp. $L/2$). The group acting effectively on the geometry is now $\Gamma=\mathbb{Z}_N\times \mathbb{Z}_M$. At the North and South pole we now have two singularities of the type $\mathbb{C}^3/(\mathbb{Z}_N\times \mathbb{Z}_M)$. These locally engineer copies of $\mathcal{T}^{\textnormal{\tiny\!  (M)}}_{X^6_{N,M}}$, discussed in section \ref{sec:examples5D}, and, whenever $N=M$, the well-known 5D trinion theory $\mathcal{T}_N$. The quotient space $X / \Gamma$
now couples the two $\mathcal{T}^{\textnormal{\tiny\!  (M)}}_{X^6_{N,M}}$ theories, preserving four real supercharges.

The case of general $a,b$ is quite similar. First, notice that requiring an effective group
action implies $(2a,K)=(2b,L)=1$. If, however, this is not the case, the effectively acting subgroup is given by
\begin{equation}
    \Gamma'=\mathbb{Z}_{K'}\times \mathbb{Z}_{L'}\,, \quad K'=K/(2a,K)\,, \quad L'=L/(2b,L)\,.
\end{equation} The local model $\mathbb{C}^3/(\mathbb{Z}_{K'}\times \mathbb{Z}_{L'})$ is now realized at the North and South pole.
The 5D SCFT engineered by one such singularity has flavor symmetry algebra:
\begin{eqnarray}\label{eq:ConeSing}
    \mathfrak{su}(K') \times \mathfrak{su}(L') \times \mathfrak{su}((K',L')) \,.
\end{eqnarray}
In passing to the compact zero-section, the flavor symmetries
\be
(\mathfrak{su}(K') \times \mathfrak{su}(L'))_{\mathrm{North}} \times (\mathfrak{su}(K') \times \mathfrak{su}(L'))_{\mathrm{South}} ~\rightarrow~ (\mathfrak{su}(K') \times \mathfrak{su}(L'))_{\mathrm{diag}}
\ee
at the North and South pole are gauged to the diagonal. The $\mathfrak{su}((K',L'))_{\mathrm{North}} \times \mathfrak{su}((K',L'))_{\mathrm{South}}$ factors remain flavor symmetries in 5D.

From the perspective of the 4D theory, we observe a flavor symmetry
\be
\mathfrak{su}((K',L'))_{\mathrm{North}} \times (\mathfrak{su}(K') \times \mathfrak{su}(L'))_{\mathrm{diag}} \times \mathfrak{su}((K',L'))_{\mathrm{South}}\,.
\ee
In the limit where the zero-section collapses, there is a further enhancement to
\be
\mathfrak{su}((K',L'))_{\mathrm{North}} \times \mathfrak{su}(K')^2 \times \mathfrak{su}(L')^2 \times \mathfrak{su}((K',L'))_{\mathrm{South}}\,,
\ee
in the obvious notation. Finally note that in this class of examples we have $H_2(S^4/\Gamma)=0$ and we have no additional massless modes from branes on 2-cycles in the singular limit where the zero-section collapses. One can consider more general situations by relaxing the conditions $a=c$ and $b=-d$, and for the most part the geometry of these models is similar.

\section{Generalized Symmetries} \label{sec:GEN}

In the previous section we introduced a number of 4D $\mathcal{N} = 1$ edge mode theories via orbifolds of $G_2$ cones. In this section we
turn to the higher-symmetries of these theories. Compared with cases previously treated in the literature, there are a number of important subtleties / distinctions. For one, the bulk is already an interacting system. By tracking the spectrum of heavy defects in the higher-dimensional theory, we can ask about their fate in the 4D theory. We refer to this as ``symmetry inheritance''. Some structures such as higher-form symmetries descend in a natural way to the 4D system. One immediate application is that we can use this to read off the global form of the non-Abelian flavor symmetry. On the other hand, some structures such as 2-groups of 5D SCFTs (which involve an entwinement of a 0-form and 1-form symmetry in the 5D theory) do not automatically descend to higher-group structures in the edge mode theory. We leave a more complete treatment of this to future work.

The rest of this section is organized as follows. We begin by giving a brief review of how defects are realized in
geometrically engineered SQFTs, and in particular how this can be used to read off various generalized symmetries.
Since 5D and 7D theories constitute bulk degrees of freedom for our various edge mode theories, we then
briefly review some of the generalized symmetries encountered in these settings. We then turn to the computation of
various higher-form symmetries for the theories realized in the previous section. We find both discrete and continuous higher-symmetries
are generically present in these theories, for the latter see Appendix \ref{sec:appContSym}. Finally we present a number of examples illustrating these features. Here we mainly focus on 0-form symmetries which determine the global structure of flavor symmetries.

\subsection{Review of Symmetries from Wrapped Branes}

Global symmetries\footnote{We restrict to internal symmetries of the SQFTs which derive from the topology of $X$ and do not descend from metric data such as isometries.} of field theories engineered via purely geometric backgrounds $X$ in string theory are specified by topological symmetry operators and their representations, i.e., defect operators. Both are engineered by wrapped branes and in many cases global symmetry data of the system can be extracted from the study of either defect or symmetry operators \cite{DelZotto:2015isa, Morrison:2020ool, Albertini:2020mdx, GarciaEtxebarria:2022vzq, Apruzzi:2022rei, Heckman:2022muc}. We briefly review these two perspectives.

The first perspective focuses on constructing defect operators as branes wrapped on non-compact, relative cycles $\mathbb{E}_*(X,\partial X)$. Here $\mathbb{E}$ is some generalized\footnote{See also \cite{Atiyah:2001qf} where this was pointed out from a background field perspective.} homology theory \cite{DelZotto:2015isa, Albertini:2020mdx, Morrison:2020ool}. For most purposes,
the relevant homology theory in the context of M-theory is singular homology
$\mathbb{E}_*=H_*$ and with this, defect operators are organized into the so-called
defect group \cite{DelZotto:2015isa, GarciaEtxebarria:2019caf, Albertini:2020mdx, Morrison:2020ool}:
\begin{equation}
\begin{aligned}
   \mathbb{D}=\oplus_{k\,} \mathbb{D}^{(k)} \,, \qquad  & \mathbb{D}^{(k)}=\mathbb{D}^{(k)}_{\mathrm M2}\oplus \mathbb{D}^{(k)}_{\mathrm M5}\,.
\end{aligned}
\end{equation}
The defect group contains electric and magnetic defect operators constructed via M2- and M5-branes wrapped on non-compact cycles
\begin{equation}\label{eq:DefectGrp}
\begin{aligned}
\mathbb{D}^{(k)}_{\textnormal{M2}}&\cong \frac{H_{3-k}(X,\partial X)}{H_{3-k}(X)} \cong  H_{3-k-1}(\partial X)\big|_\textnormal{trivial} \\
\mathbb{D}^{(k)}_{\textnormal{M5}}&\cong \frac{H_{6-k}(X,\partial X)}{H_{6-k}(X)} \cong  H_{6-k-1}(\partial X)\big|_\textnormal{trivial}\,.
\end{aligned}
\end{equation}
Here, the subgroups $\mathbb{D}^{(k)}$ collect $k$-dimensional defects acted on by $k$-form symmetries and $|_{\textnormal{trivial} }$ denotes the restriction to the kernel of the inclusion $\partial X \hookrightarrow X$ lifted to homology.

Strictly speaking, working on such a local geometry $X$ results in a relative field theory, in the sense of \cite{Gaiotto:2014kfa, Witten:1998wy, Seiberg:2011dr, Freed:2012bs}. In order to more fully specify the SQFT defined by a given geometric background, one must also provide boundary conditions for the various background fields / fluxes ``at infinity''. Roughly speaking, one views the radial direction of the conical geometry $X$ as a ``timelike'' direction of a topological field theory, and then treats the boundary as a state of this TFT. Specifying a maximally commuting collection of operators in that system amounts to a choice of ``polarization''.\footnote{Due to flux non-commutativity between $G_4$ and $G_7$ \cite{Freed:2006ya, GarciaEtxebarria:2019caf}, it is necessary to pick a maximal isotropic sublattice
of commuting flux operators as geometrized by the linking form between cycles wrapped by electromagnetically dual branes. This specifies a maximal set of mutually local defects and consequently an absolute QFT starting from a relative QFT. Purely electric or magnetic sublattices are automatically mutually local. An important subtlety with this procedure is that the 11D Chern-Simons term for the 3-form potential leads to additional complications in the braiding relations for the magnetic variables specified by $G_7$ fluxes. This is an additional reason why it is typically simpler to work with the electric polarization.}

Absolute theories only follow upon a choice of boundary condition or polarization
\begin{equation}
    \Lambda=\oplus_{k\,} \Lambda_{(k)},, \qquad \Lambda_{(k)}\subset  \mathbb{D}^{(k)}\,, \qquad \mathcal{E}^{(k)}={\Lambda}_{(k)}^\vee\equiv\textnormal{Hom}(\Lambda_{(k)},\mathbb{R}/\mathbb{Z})
\end{equation}
which fixes a maximally mutually local subgroup of defect operators. The $k$-form symmetries $\mathcal{E}^{(k)}$ are then inferred by Pontryagin duality, denoted using $\vee$. The symmetries of all relative theories, deriving from a given absolute theory engineered with geometry $X$, are therefore subgroups $  \oplus_{k\,}\mathcal{E}^{(k)}\subset \mathbb{D}^\vee$ of the Pontryagin dual of the defect group. In the following section we restrict to purely electric polarizations, and all defect operators are constructed from wrapped M2-branes.

The second perspective focuses on constructing the topological symmetry operators, acting on defect operators, by wrapping branes at the asymptotic boundary of spacetime \cite{Heckman:2022muc, GarciaEtxebarria:2022vzq, Apruzzi:2022rei} (see also \cite{Heckman:2022xgu}).
These operators are topological as they are at infinite distance from the dynamical degrees of freedom of the engineered SQFT and arise from branes wrapped on cycles of formally infinite volume. Their action on charged operators is determined from topological linking with the defect operators that make up the defect group $\mathbb{D}$. An important feature of this perspective is that it automatically points to additional structures such as topological field theories attached to each such symmetry operator. For the purposes of the present work, however, we shall mainly content ourselves with identifying the spectrum of defects and the associated higher-symmetries.

Now, whenever the asymptotic boundary $\partial X$ is singular, both perspectives become significantly more intricate.\footnote{One could in principle consider a full resolution of all singularities, including those at the boundary, but this in general obscures the non-Abelian flavor symmetries of the system.} For the $G_2$-holonomy orbifolds collected in section \ref{sec:EXAMP}, the asymptotic boundary contains codimension-4 and/or -6 singular loci. In such situations, as demonstrated in \cite{Cvetic:2022imb,DelZotto:2022joo}, one must also consider M2- and M5-branes that wrap not only $H_*(\partial X)|_\textnormal{trivial}$, but also homology groups that arise in the Mayer-Vietoris sequence associated to cutting out/gluing-in the singular loci into $\partial X$. If we define $\Sigma\subset \partial X$ to be the singular locus, and $T(\Sigma)$ to be a tubular neighborhood thereof, then we interpret\footnote{We will leave implicit the $|_\textnormal{trivial}$ qualification in what follows. Note that this condition is automatically satisfied when the 4D theory is at the SCFT point in its moduli space.} M2-branes wrapped on
\begin{equation}\begin{aligned}\label{eq:Inherit5D}
    H_{3-k}(T(\Sigma),\partial T(\Sigma))&\cong H_{3-k-1}(\partial T(\Sigma))|_\textnormal{trivial}
    \end{aligned}
\end{equation}
as the defect operators charged under the electric $k$-form symmetries of the SCFT
localized along the non-compact singular loci intersecting the boundary in $\Sigma$. The boundary $\partial T(\Sigma)$ is smooth, therefore
M5-branes on
\begin{equation}
    H_{*}(\partial T(\Sigma))
\end{equation}
engineer the symmetry operators of the edge mode theory $\Sigma$ \cite{Heckman:2022muc}.

As a final comment, in the defect group, the torsional contributions to the 0-form symmetry track the centers of non-Abelian flavor symmetry factors, while the free parts track additional $U(1)$ symmetry factors.

\subsection{Symmetry Inheritance}


Since we have localized modes sitting in a bulk theory, we can expect that some of the symmetries of the bulk will act on these edge modes, through a process we refer to as ``inheritance''. We stress that this is a rather general notion, and works whenever we have a bulk / boundary system. 


Starting from a background $X$, the theory $\mathcal{T}_X$ has a defect group $\mathbb{D}(X)$ which encodes the higher-symmetries of this theory. In the context of theories realized via intersections of and junctions of bulk theories with an edge mode, it is helpful to speak of the individual singular loci $V_i \subset X$ which, in isolation, define building blocks $\mathcal{T}_{V_i}$ of the full system $\mathcal{T}_{X}$. Each of these $\mathcal{T}_{V_i}$ has its own defect group $\mathbb{D}(V_i)$, and with it, a set of corresponding higher-form symmetries. In the process of building up $\mathcal{T}_X$, some of these symmetries will be gauged / broken, but there is a clear question as to the fate of the defects / symmetry operators of $\mathcal{T}_{V_i}$ and how they embed in the full theory $\mathcal{T}_X$. Comparing the two, and keeping track of the jumps that arise by pushing symmetry defects on the boundary  (a bulk $p$-form symmetry becomes a $(p-1)$-form symmetry for the boundary), one can determine which collection of the charged operators of the bulk theories $\mathcal T_{V_i}$ can act on the edge modes of $\mathcal T_X$.

\paragraph{Symmetries from the Bulk:} Precisely because we expect some of the bulk generalized symmetries to descend to our edge modes, we now provide a brief review of the sorts of structures we can expect to encounter. Higher-symmetries of 5D and 7D theories have been studied for example in \cite{Albertini:2020mdx, Morrison:2020ool, Apruzzi:2021vcu, Tian:2021cif, DelZotto:2022fnw, Cvetic:2022imb, DelZotto:2022joo}. Our treatment will follow that given in \cite{Cvetic:2022imb, DelZotto:2022joo}.

To begin, the 7D theories under consideration arise from local geometries of the type $\mathbb{C}^2/\Gamma$ where $\Gamma\subset SU(2)$ are finite groups with ADE classification. If we denote by $G_\Gamma$ the simply connected Lie group associated to $\Gamma$ we then find the defect group of lines to be isomorphic to the center $Z_{G_\Gamma}$. These lines are constructed from M2-branes wrapped on cones over 1-cycles $H_1(S^3/\Gamma)\cong Z_{G_\Gamma}$ of the asymptotic boundary. This gives the complete set of electric defect operators acted on by 1-form symmetries.

The 5D theories under consideration arise from local geometries of the type $\mathbb{C}^3/\Gamma$ where $\Gamma\subset SU(3)$ are finite groups (as classified in \cite{gps1916, osti_4056933, Fairbairn1982SomeCO}). Here, electric defects arise from M2-branes wrapped on non-compact 3-cycles and 2-cycles. These give local/line operators in the 5D theory and are acted on by 0-form and 1-form symmetry groups. The relevant cycles are in fact torsional and are cones over elements of $H_2(S^5/\Gamma)$ and $H_1(S^5/\Gamma)$, respectively.

These 0-form and 1-form symmetries can combine to a 2-group. Denote by $(S^5/\Gamma)^\circ$ the quotient $S^5/\Gamma$ with singularities $\Sigma$ removed. Then the Mayer-Vietoris sequence with respect to a covering given by $(S^5/\Gamma)^\circ$ and the tube $T(\Sigma)$ contains the exact subsequence
\be\label{eq:ExactSequence}
0~\rightarrow~H_2(S^5/\Gamma)~\rightarrow~H_1(\partial T(\Sigma))~\rightarrow~ H_1((S^5/\Gamma)^\circ)~\rightarrow~H_1(S^5/\Gamma)~\rightarrow~0
\ee
and detects operators in projective representations of the flavor symmetry group and therefore the extension of the 0-form symmetry group by a 1-form symmetry group to a 2-group \cite{Lee:2021crt, Bhardwaj:2021wif}.
Non-trivial 2-groups are then characterized by this sequence not splitting at the third entry.

We now reinterpret this result about 2-groups in 5D theories, in terms of symmetry inheritance. Let us consider for simplicity a concrete example, the $SU(p)_p$ SYM theories, corresponding to the singularities $\mathbb C^3/\mathbb Z_{2p}$. In the electric frame, we have a 2-group mixing the $\mathbb Z_p^{(1)}$ form symmetry of the 5d theory with the $SO(3)^{(0)}$ 0-form symmetry of the model. The latter arises from a bulk 7D SYM theory, corresponding to a non-compact $A_1$ locus in the $\mathbb C^3/\mathbb Z_{2p}$ orbifold. The 7D theory has a $\mathbb Z_2^{(1)}$ center symmetry, which induces a $\mathbb Z_2^{(0)}$ 0-form symmetry on the 5D boundary. The latter is parameterizing the charges of the line operators in 5D that cannot be screened, and thus transform in projective representations. This leads to the exact sequence
\be\label{eq:exactseq}
0 \to \mathbb Z_2 \to \mathbb Z_{2p} \to \mathbb Z_p \to 0
\ee
where $\mathbb Z_{2p}$ correspond to the charges of line operators screened by operators transforming in definite representations of the global symmetry group \cite{Lee:2021crt}. This is encoded in the Mayer-Vietoris sequence on the boundary $S^5/\mathbb Z_{2p}$ where the $\mathbb C^2/\mathbb Z_2$ fixed locus is excised. This is an example of symmetry inheritance: the compatible boundary conditions for the operators on the edge are parameterized by line \eqref{eq:exactseq}. Indeed, for the Mayer-Vietoris argument of \cite{Cvetic:2022imb,DelZotto:2022joo}, it is crucial to have extended singularities which reach the boundary, giving rise to a singular locus. The latter now have an interpretation in terms of higher-dimensional bulk theories. This discussion extends to the 4D edge theories arising from orbifolds of $G_2$ cones.

\paragraph{Symmetry Inheritance for 4D Edge Modes:} We expect that in our systems with 4D edge modes that the symmetries of the building blocks of the bulk system may also act in a natural way in the full theory. In this context, symmetry inheritance relates the symmetries of the SCFT localized along $\Sigma$ to those of the lower-dimensional modes at the tip of the asymptotically conical geometry $X$. In physical terms, one simply considers the topological symmetry operators of the bulk theory, and pushes them fully into the edge. Geometrically, symmetry inheritance is quantified by the boundary maps
\begin{equation}\label{eq:symminherit}
    \partial_n: H_{n}(\partial X)\rightarrow H_{n-1}(\partial T(\Sigma))
\end{equation}
which are taken from the Mayer-Vietoris long exact sequence with respect to the covering $\partial X=(\partial X\setminus \Sigma)\cup \;\!T(\Sigma)$. Since the $k$-form symmetries take values in the Pontryagin duals of these groups and the map $(\partial_n)^\vee$ (induced from the contravariant functor specified by Pontryagin duality) is in the reverse direction of \eqref{eq:symminherit}, this tells us how symmetries of the 5D/7D theories map onto symmetries of the 4D theory. In general this map can have a non-trivial kernel and cokernel, so our study of symmetry inheritance will take note of which symmetries of the 5D/7D theory do not become symmetries in 4D, and which 4D symmetries do not originate from the higher-dimensional bulk.

To illustrate how this works in more detail, let us specialize further to the electric polarization, and consider the inheritance of 5D/7D lines to 4D local operators. For this we need to consider the geometric realizations of the 5D/7D theories as subsets of the orbifolded $G_2$-holonomy space. The radial direction of the $G_2$ cone is part of the 5D/7D world volume and a line can therefore be oriented radially stretching from the tip of the cone to infinity. In the $G_2$-holonomy space, the cycles associated to the line in the local model of the 5D/7D theory no longer make sense globally. Rather, the relevant cones over 1-cycles (in a given radial shell) must be restrictions of 2-cycles. Such 2-cycles can be stretched between different 5D/7D loci. An M2-brane wrapped on such a 2-cycle gives a local operator in 4D which looks like a collection of line operators in the 5D/7D systems. Such local operators are inherited from the higher-dimensional lines by ``restriction'' to the tip of the $G_2$ cone. Modding out all 4D local operators by the inherited local operators gives the intrinsic local operators of the 4D system.

All of this is quantified by the boundary map
\be
\partial_2\,: H_2(\partial X_0)\rightarrow H_1(\partial T(\Sigma))
\ee
of the Mayer-Vietoris sequence. Here $X_0$ denotes the $G_2$ cone (with collapsed zero section),
and $T(\Sigma)$ the tubular neighborhood of its singularities in the asymptotic boundary $\partial X_0$. The map $\partial_2$ sends a 2-cycle, at a given radius, to (the cone over) a 1-cycle.
The image $\textnormal{Im}\,\partial_2$ is therefore isomorphic to the inherited symmetries, while the kernel $\textnormal{Ker}\,\partial_2$ is isomorphic to the intrinsic symmetries. The former is extended by the latter
\be
0~\rightarrow~\textnormal{Ker}\,\partial_2 ~\rightarrow~ H_2(\partial X_0) ~\rightarrow~\textnormal{Im}\,\partial_2 ~\rightarrow~0\,.
\ee
In the examples below we find this short exact sequence splits.

By a similar token, we can consider 0-form symmetries of the 5D theory, which are often associated with 7D SYM sectors (viewed as another ``bulk theory''). As we already saw in section \ref{sec:EXAMP}, these symmetries directly descend, along with some 5D gauge symmetries, to flavor symmetries of the 4D theory. An interesting feature of inheritance is that in some cases, the contributions from the defect group end up trivializing. This happens because once we perform a consistent 5D gauging, additional states are often included which end up changing the global form of the flavor group / flavor symmetry in the 4D edge mode theory compared with the isolated 5D theories used to construct the bulk.

In some of our 5D theories, there can also be more intricate structures such as an entwinement between 0-form and 1-form symmetries via a 2-group structure (see e.g., \cite{Apruzzi:2021vcu, Cvetic:2022imb, DelZotto:2022joo}). Precisely because the structure of the bulk 0-form symmetry is often modified by 5D gauging effects, this also impacts the structure of such higher-group structures as well. That being said, we leave a full treatment of such phenomena for future work.

\subsection{Spontaneous Symmetry Breaking}

One of the important features of our orbifolds of $G_2$ cones is that resolving the tip of the cone often ends up breaking some of the symmetries of the system. We already explained in section \ref{sec:EXAMP} that this phenomenon is generic when we have non-Abelian flavor symmetries, and indicates that there are localized degrees of freedom at the tip of the cone charged under these symmetries.

It is natural to ask whether we can perform a similar analysis for the various generalized symmetries of our system. One way to approach this
question is to calculate the defect group both before and after resolution of the tip of the cone. At this point it is worth remarking
that in many previous studies of the defect group, a special class of $k$-form symmetries was considered in which the process of resolution does not alter this structure
(as in \cite{DelZotto:2015isa, Albertini:2020mdx, Morrison:2020ool}). For example, in 5D SCFTs realized via orbifolds of the form $\mathbb{C}^3 / \Gamma_{SU(3)}$, one method for extracting the defect group is to explicitly resolve the geometry \cite{Tian:2021cif}, but one can alternatively work directly in terms of singular homology of the bulk Calabi-Yau and its boundary $S^5 / \Gamma_{SU(3)}$. The issue we face in some of the cases considered here is that in general, blowups and smoothing deformations of a space can have different Betti numbers, and this in turn means that the structure of the defect group is in principle sensitive to such changes. So, a priori, we can expect the structure of the defect group to be different depending on whether we have resolved our $G_2$ cone or not. To emphasize this point, we shall sometimes write $\mathbb{D}(X_0)$ and $\mathbb{D}(X)$ to indicate the defect group of the singular and resolved cases, respectively. Similarly we denote by $\mathcal{E}(X_0)$ and $\mathcal{E}(X)$ the electric subgroups of the defects group for the geometries $X_0$ and $X$, respectively. For technical reasons outlined earlier, we mainly confine our attention to the electric polarization. In all the examples we consider, the part of the defect group which is broken is indeed just the contribution to the 0-form symmetry. For additional discussion of the various continuous $k$-form symmetries, see Appendix \ref{sec:appContSym}.

From the perspective of the boundary, however, the topology of $\partial X$ is unchanged by such localized deformations (be they blowups or smoothings). Consequently, the analysis if \cite{Heckman:2022muc} implies that wrapping branes ``at infinity'' will be insensitive to such operations, and as such, the topological symmetry operators will remain the same in both the singular and resolved phase. This is all compatible with \textit{spontaneous} symmetry breaking: The theory still retains the original symmetry, but the spectrum of localized degrees of freedom which are charged under the symmetry can indeed change.\footnote{For example, consider the vev of a scalar bifundamental of $(SU(N) \times SU(N)) / \mathbb{Z}_N$.}

\subsection{Examples}

Having specified some general features of higher-symmetries, we now turn to an analysis in some explicit examples. The core mathematical tool we use to read off the various higher-form symmetries and how they are inherited in the edge mode system is a calculation of the relevant homology groups. In the special case of the homology groups of $\partial X_{0}^{\circ}$, namely the boundary geometry with singularities excised, there is an additional subtlety because this geometry sometimes deformation retracts to a lower-dimensional space. To implicitly indicate this feature, when we present the generators of $H_{\ast}(\partial X_{0}^{\circ})$, we shall therefore only list the highest degree term contribution which is non-trivial, i.e., we do not go ``all the way to $6$'' in all cases.

\subsubsection{Quadrion Theories}

As a first example, consider again the quadrion theories.
The quotient on the North and South pole patches are, with $\Gamma=\mathbb{Z}_K$ and $\zeta=\exp(2\pi i/K)$,
\begin{eqnarray}
(v_{1},v_{2},v_{3},t) &\sim &(\zeta ^{c}v_{1},\zeta ^{d}v_{2},\zeta
^{-c-d}v_{3},t) \\
(v_{1}^{\prime },v_{2}^{\prime },v_{3}^{\prime },t^{\prime }) &\sim
&(\zeta ^{-c}v_{1}^{\prime },\zeta ^{d}v_{2}^{\prime },\zeta
^{c-d}v_{3}^{\prime },t^{\prime }).
\end{eqnarray}%
and with constraints on exponents as
\begin{equation}
    (\left\vert c\right\vert,K)=(\left\vert d\right\vert, K)=(\left\vert c+d\right\vert, K)=(\left\vert c-d\right\vert, K)=1.
\end{equation}
The geometry $X_0$ is characterized by the homology groups
\begin{equation}\label{eq:Homologies2}
    \begin{aligned}
     H_*(X_0,\partial X_0)/H_*(X_0)&\cong\{0,0,0,\mathbb{Z}\oplus \mathbb{Z}_K^2,0, \mathbb{Z}\oplus \mathbb{Z}_K,0,\mathbb{Z}  \} \\
        H_*(\partial X_0)&\cong \{\mathbb{Z},0,\mathbb{Z}\oplus \mathbb{Z}_K^2,0, \mathbb{Z}\oplus \mathbb{Z}_K,0,\mathbb{Z}  \}  \\
         H_*(\partial X_0^\circ)&\cong \{\mathbb{Z},\mathbb{Z}_K,\mathbb{Z}, \mathbb{Z}_K^2,\mathbb{Z},\mathbb{Z}^3  \} \\
         H_*(X,\partial X)/H_*(X)&\cong\{0,0,0,\mathbb{Z}\oplus \mathbb{Z}_K,0,  \mathbb{Z}_K,0,\mathbb{Z}  \}
    \end{aligned}
\end{equation}
and the Pontryagin dual of the defect group, interpreted as the possible higher-form symmetry group $\mathcal{E} = \mathbb{D}_{\mathrm{M2}}^{\vee}$ (in an electric polarization). We therefore find that in the limit where we have a collapsed zero section
(via $ H_*(X_0,\partial X_0)/H_*(X_0)$)
\begin{equation}\label{}\begin{aligned}
   \mathcal{E}(X_0)&=
   U(1)^{(0)}\times ( \mathbb{Z}_K^{(0)})^2  \,.
   \end{aligned}
\end{equation}

With this, the 4D 0-form symmetry contributed from the defect group is $U(1)\times \mathbb{Z}^2_K$. We interpret the torsional elements
as the center of the non-Abelian flavor symmetry group. We can determine which contributions are inherited from 5D by
studying the boundary map
\begin{equation}\label{eq:symminherit2}
    \partial_2: H_{2}(\partial X_0)=\mathbb{Z} \times \mathbb{Z}^2_K\rightarrow H_{1}(\partial T(\Sigma))=\mathbb{Z}^4_K
\end{equation}
of the Mayer-Vietoris sequence for the covering $\partial X_0=\partial X_0^\circ \cup T(\Sigma)$. This boundary map specifies how the internal support of 5D defects given in \eqref{eq:Inherit5D} glue together to 4D defects, which are therefore inherited from 5D. The 5D theories from which the symmetries are inherited are four copies of $\mathcal{T}^{\textnormal{\tiny\!  (M)}}_{\mathbb{C}^3/\Gamma}$ and we have
\begin{align*}
     \mathrm{Ker}(\partial_2)&=\left \{\begin{array}{c} \textnormal{4D 0-form symmetries that do not come from} \\ \textnormal{ 5D bulk 1-form symmetries} \end{array}\right\}\\[2mm]
     \mathrm{Coker}(\partial_2)&=\left \{\begin{array}{c} \textnormal{5D 1-form symmetries that are transparent} \\ \textnormal{to the 4D edge modes} \end{array}\right\}\\[2mm]
     \mathrm{Im}(\partial_2)&=\left \{\begin{array}{c} \textnormal{5D 1-form symmetries that are faithful 0-form} \\ \textnormal{symmetries acting on the 4D edge modes} \end{array}\right\}\,.
\end{align*}

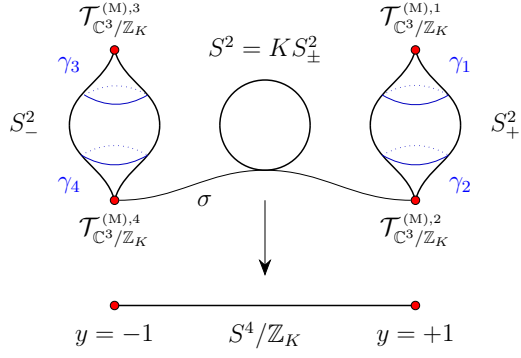
\begin{figure}
    \centering
    \scalebox{0.8}{\begin{tikzpicture}
	\begin{pgfonlayer}{nodelayer}
		\node [style=none] (0) at (-2.5, 1.25) {};
		\node [style=none] (1) at (-2.5, -1.25) {};
		\node [style=none] (2) at (-1.75, 0) {};
		\node [style=none] (3) at (-3.25, 0) {};
		\node [style=none] (4) at (2.5, 1.25) {};
		\node [style=none] (5) at (2.5, -1.25) {};
		\node [style=none] (6) at (3.25, 0) {};
		\node [style=none] (7) at (1.75, 0) {};
		\node [style=none] (8) at (0, 0.75) {};
		\node [style=none] (9) at (0, -0.75) {};
		\node [style=none] (10) at (0.75, 0) {};
		\node [style=none] (11) at (-0.75, 0) {};
		\node [style=none] (12) at (-4, 0) {$S^2_-$};
		\node [style=none] (13) at (4, 0) {$S^2_+$};
		\node [style=none] (14) at (0, 1.25) {$S^2=KS^2_\pm$};
		\node [style=none] (15) at (-2.5, -3) {};
		\node [style=none] (16) at (2.5, -3) {};
		\node [style=none] (17) at (0, -3.5) {$S^4/\mathbb{Z}_K$};
		\node [style=none] (18) at (-2.5, -3.5) {$y=-1$};
		\node [style=none] (19) at (2.5, -3.5) {$y=+1$};
		\node [style=none] (20) at (0, -1.25) {};
		\node [style=none] (21) at (0, -2.5) {};
		\node [style=CircleRed] (22) at (-2.5, -3) {};
		\node [style=CircleRed] (23) at (-2.5, -1.25) {};
		\node [style=CircleRed] (24) at (-2.5, 1.25) {};
		\node [style=CircleRed] (25) at (2.5, -1.25) {};
		\node [style=CircleRed] (26) at (2.5, 1.25) {};
		\node [style=CircleRed] (27) at (2.5, -3) {};
		\node [style=none] (28) at (2.5, 1.75) {$\mathcal{T}^{\textnormal{\tiny\!  (M),1}}_{\mathbb{C}^3/\mathbb{Z}_K}$};
		\node [style=none] (29) at (2.5, -1.75) {$\mathcal{T}^{\textnormal{\tiny\!  (M),2}}_{\mathbb{C}^3/\mathbb{Z}_K}$};
		\node [style=none] (30) at (-2.5, 1.75) {$\mathcal{T}^{\textnormal{\tiny\!  (M),3}}_{\mathbb{C}^3/\mathbb{Z}_K}$};
		\node [style=none] (31) at (-2.5, -1.75) {$\mathcal{T}^{\textnormal{\tiny\!  (M),4}}_{\mathbb{C}^3/\mathbb{Z}_K}$};
  \node [style=none] (32) at (-1, -1.25) {$\sigma$};
  \node [style=none] (33) at (-3, 0.5) {};
		\node [style=none] (34) at (-1.95, 0.5) {};
		\node [style=none] (35) at (1.95, 0.5) {};
		\node [style=none] (36) at (3.05, 0.5) {};
		\node [style=none] (37) at (-3.05, -0.5) {};
		\node [style=none] (38) at (-1.95, -0.5) {};
		\node [style=none] (39) at (1.95, -0.5) {};
		\node [style=none] (40) at (3.05, -0.5) {};
		\node [style=none] (41) at (3.25, 1) {$\color{blue}\gamma_1$};
		\node [style=none] (42) at (3.25, -1) {$\color{blue}\gamma_2$};
		\node [style=none] (43) at (-3.25, 1) {$\color{blue}\gamma_3$};
		\node [style=none] (44) at (-3.25, -1) {$\color{blue}\gamma_4$};
	\end{pgfonlayer}
	\begin{pgfonlayer}{edgelayer}
		\draw [style=ThickLine, in=-105, out=90] (3.center) to (0.center);
		\draw [style=ThickLine, in=90, out=-75] (0.center) to (2.center);
		\draw [style=ThickLine, in=75, out=-90] (2.center) to (1.center);
		\draw [style=ThickLine, in=-90, out=105] (1.center) to (3.center);
		\draw [style=ThickLine, in=-105, out=90] (7.center) to (4.center);
		\draw [style=ThickLine, in=90, out=-75] (4.center) to (6.center);
		\draw [style=ThickLine, in=75, out=-90] (6.center) to (5.center);
		\draw [style=ThickLine, in=-90, out=105] (5.center) to (7.center);
		\draw [style=ThickLine, in=-180, out=90] (11.center) to (8.center);
		\draw [style=ThickLine, in=90, out=0] (8.center) to (10.center);
		\draw [style=ThickLine, in=0, out=-90] (10.center) to (9.center);
		\draw [style=ThickLine, in=-90, out=180] (9.center) to (11.center);
		\draw [style=ThickLine] (15.center) to (16.center);
		\draw [style=ArrowLineRight] (20.center) to (21.center);
  \draw [in=-180, out=0] (23) to (9.center);
		\draw [in=-180, out=0] (9.center) to (25);
  \draw [style=LightBlue, bend right] (33.center) to (34.center);
		\draw [style=BlueDottedLight, bend left] (33.center) to (34.center);
		\draw [style=LightBlue, bend right] (35.center) to (36.center);
		\draw [style=BlueDottedLight, bend left] (35.center) to (36.center);
		\draw [style=LightBlue, bend right] (37.center) to (38.center);
		\draw [style=BlueDottedLight, bend left] (37.center) to (38.center);
		\draw [style=LightBlue, bend right] (39.center) to (40.center);
		\draw [style=BlueDottedLight, bend left] (39.center) to (40.center);
	\end{pgfonlayer}
\end{tikzpicture}}
    \caption{Sketch of the twistor space associated with the Quadrion theory. The fibration over $B=S^4/\mathbb{Z}_K$ exhibits two exceptional fibers projecting to $y=\pm1$ and running between pairs of codimension-6 singularities $\Sigma$ modeled on $\mathbb{C}^3/\mathbb{Z}_K$. $\Sigma$ consists of the four red dots. The horizontal 2-cycle $\sigma$ projects to the generator of $H_2(S^4/\Gamma)$. The fractional fibers $S_\pm^2$ intersect the boundaries of the tubular neighborhoods of each singularity along a torsional 1-cycle $\gamma_i$ (blue). The 2-cycle $\sigma$ has a similar intersection pattern (not depicted). }
    \label{fig:fibers}
\end{figure}

In order to describe these groups, we consider the cycles generating
\begin{equation}\label{eq:2ndHomo}
H_{2}(\partial X_0)\cong \mathbb{Z}\times \mathbb{Z}^2_K\cong ((\mathbb{Z}\times \mathbb{Z})/K\mathbb{Z}) \times \mathbb{Z}_K\,.
\end{equation}
For this, let us introduce a coordinate $y$ as a height coordinate on the $S^4$ with North and South pole at $y=+1$ and $y=-1$ respectively. Then, the factor $\mathbb{Z}\times \mathbb{Z}$ in \eqref{eq:2ndHomo} is generated by vertical cycles, they are the two fiber classes $S^2_\pm$ projecting to the North and South pole of the base $S^4/\Gamma$ with coordinates $y=\pm1$ respectively. These fibers are rigid, however, the multiple $KS^{2}_\pm$ is homologous to the generic 2-sphere fiber $S^2$ of the twistor space $\partial X_0$ and therefore $KS^{2}_\pm$ are homologous. Accounting for this equivalence we find a contribution of $(\mathbb{Z}\times \mathbb{Z})/K\mathbb{Z}$ to $H_2(\partial X_0)$ where $K\mathbb{Z}$ is embedded diagonally. The generator $\sigma$ of the remaining $\mathbb{Z}_K$ factor is horizontal, it is the lift of the torsional generator of $H_2(S^4/\Gamma)$ to the total space, see Figure \ref{fig:fibers}.

In order to describe the image of $\partial_2$ we pick the natural basis of $\mathbb{Z}^4_K$ where each factor is associated to the torsional 1-cycle $\gamma_i$ in the $S^5/\mathbb{Z}_K$ linking each codimension-6 singularity. With these conventions in place we have
\begin{equation}\begin{aligned}
    \partial_2\,:\quad ((\mathbb{Z}\times \mathbb{Z})/K\mathbb{Z}) \times \mathbb{Z}_K &\rightarrow \mathbb{Z}_K^4 \\
    (n,m,h) &\mapsto (n+h,n,m+h,m)
    \end{aligned}
\end{equation}
and we readily compute
\begin{align}\label{eq:syminhertet2}
     \mathrm{Ker}(\partial_2)&=K\mathbb{Z} \\
     \mathrm{Coker}(\partial_2)&=\mathbb{Z}_K \\
     \mathrm{Im}(\partial_2)&=\mathbb{Z}^3_K\,.
\end{align}

Let us now discuss the inherited symmetries at the level of defect operators. Consider an M2-brane wrapped on the cone over $S^{2}_\pm$. The fibers $S^2_\pm$ run between pairs of singularities modeled on $\mathbb{C}^3/\mathbb{Z}_K$ as shown in Figure \ref{fig:fibers}. Restricting to the local models $T(\Sigma)$ of the 5D theory we find that $S^2_\pm$ decompose into four cones over the torsional 1-cycles $\gamma_i$ generating $H_1(\partial T(\Sigma))\cong \mathbb{Z}_K^4$. These are the relative 2-cycles in $T(K)$ wrapped by M2-branes to engineer line defects in the 5D theory. Similarly, we find the horizontal 2-cycle $\sigma$ to restrict to two cones. Overall, therefore we have $\partial_2^\vee$ pushing forward the 1-form symmetry subgroup $\mathbb{Z}_K^3$ of the 5D theory, embedding these into the 4D 0-form symmetry group $U(1) \times \mathbb{Z}_K^2$.

Upon giving a finite volume to $S^4/\mathbb{Z}_K$, the symmetry breaks as follows:\footnote{We read this off from the relative homology groups $ H_*(X,\partial X)/H_*(X)$.}
\begin{align}
\mathcal{E}(X_0) & \rightarrow \mathcal{E}(X) \\
U(1)^{(0)} \times \mathbb{Z}_{K}^{(0)} \times \mathbb{Z}_{K}^{(0)} & \rightarrow  U(1)^{(0)} \times \mathbb{Z}_{K}^{(0)}.
\end{align}

\subsubsection{Quadrions Coupled to Bifundamental Matter}

Consider next our example of quadrions coupled to bifundamental matter.
The quotient on the North and South pole patches are, with $\Gamma=\mathbb{Z}_K$ and $\zeta=\exp(2\pi i/K)$,
\begin{eqnarray}
(v_{1},v_{2},v_{3},t) &\rightarrow &(\zeta ^{c}v_{1},\zeta ^{d}v_{2},\zeta
^{-c-d}v_{3},t) \\
(v_{1}^{\prime },v_{2}^{\prime },v_{3}^{\prime },t^{\prime }) &\rightarrow
&(\zeta ^{-c}v_{1}^{\prime },\zeta ^{d}v_{2}^{\prime },\zeta
^{c-d}v_{3}^{\prime },t^{\prime }).
\end{eqnarray}%
and with constraints on exponents as
\begin{equation}
    (\left\vert c\right\vert,K)=(\left\vert c+d\right\vert, K)=(\left\vert c-d\right\vert, K)=1, \quad K=md,
\end{equation}
for some integers $m,d>1$. The relevant homology groups are:
\begin{equation}\label{eq:Homologies3}
    \begin{aligned}
     H_*(X_0,\partial X_0)/H_*(X_0)&\cong\{0,0,0,\mathbb{Z}\oplus \mathbb{Z}_m^2,0, \mathbb{Z}\oplus \mathbb{Z}_K,0,\mathbb{Z}  \} \\
        H_*(\partial X_0)&\cong\{\mathbb{Z},0,\mathbb{Z}\oplus \mathbb{Z}_m^2,0, \mathbb{Z}\oplus \mathbb{Z}_K,0,\mathbb{Z}  \}  \\
         H_*(\partial X_0^\circ)&\cong \{\mathbb{Z}, \mathbb{Z}_K,\mathbb{Z},\mathbb{Z}^2\oplus \mathbb{Z}_m^2, \mathbb{Z},\mathbb{Z}  \} \\
         H_*(X,\partial X)/H_*(X)&\cong\{0,0,0,\mathbb{Z}\oplus \mathbb{Z}_m,0,  \mathbb{Z}_K,0,\mathbb{Z}  \}
    \end{aligned}
\end{equation}
and the electric polarization of the defect group is therefore
\begin{equation}\label{}\begin{aligned}
   \mathcal{E}(X_0)&=
   U(1)^{(0)}\times ( \mathbb{Z}_m^{(0)})^2.
   \end{aligned}
\end{equation}

\begin{figure}
    \centering
   \scalebox{0.8}{
\begin{tikzpicture}
	\begin{pgfonlayer}{nodelayer}
		\node [style=none] (15) at (-3, -1.75) {};
		\node [style=none] (16) at (2.5, -1.75) {};
		\node [style=none] (17) at (-0.25, -2.25) {$S^2/\mathbb{Z}_m$};
		\node [style=none] (18) at (-3, -2.25) {$y=-1$};
		\node [style=none] (19) at (2.5, -2.25) {$y=+1$};
		\node [style=none] (20) at (-0.25, -0.5) {};
		\node [style=none] (21) at (-0.25, -1.25) {};
		\node [style=CircleRed] (22) at (-3, -1.75) {};
		\node [style=CircleRed] (27) at (2.5, -1.75) {};
		\node [style=none] (45) at (-0.25, 2.25) {};
		\node [style=none] (46) at (-0.25, 0.75) {};
		\node [style=none] (47) at (0.5, 1.5) {};
		\node [style=none] (48) at (-1, 1.5) {};
		\node [style=none] (49) at (-0.25, 2.75) {$S^3/\mathbb{Z}_d=m(S^3/\mathbb{Z}_K)$};
		\node [style=none] (50) at (-3, 2.25) {};
		\node [style=none] (51) at (-3, 0.75) {};
		\node [style=none] (52) at (-2.25, 1.5) {};
		\node [style=none] (53) at (-3.75, 1.5) {};
		\node [style=none] (54) at (2.5, 2.25) {};
		\node [style=none] (55) at (2.5, 0.75) {};
		\node [style=none] (56) at (3.25, 1.5) {};
		\node [style=none] (57) at (1.75, 1.5) {};
		\node [style=none] (58) at (2.5, 2.75) {~~$(S^3/\mathbb{Z}_K)_+$};
		\node [style=none] (59) at (-3, 2.75) {$(S^3/\mathbb{Z}_K)_-$~~};
		\node [style=none] (60) at (-3.25, 1.25) {};
		\node [style=none] (61) at (-2.75, 1.75) {};
		\node [style=none] (62) at (-0.5, 1.25) {};
		\node [style=none] (63) at (0, 1.75) {};
		\node [style=none] (64) at (2.25, 1.25) {};
		\node [style=none] (65) at (2.75, 1.75) {};
		\node [style=none] (66) at (-3, 0.25) {\color{blue}$\gamma_{-}$};
		\node [style=none] (67) at (2.5, 0.25) {\color{blue}$\gamma_{+}$};
		\node [style=none] (68) at (-0.25, 0.25) {\color{blue}$\gamma_{d}=m\gamma_\pm$};
	\end{pgfonlayer}
	\begin{pgfonlayer}{edgelayer}
		\draw [style=ThickLine] (15.center) to (16.center);
		\draw [style=ArrowLineRight] (20.center) to (21.center);
		\draw [style=ThickLine, in=-180, out=90] (48.center) to (45.center);
		\draw [style=ThickLine, in=90, out=0] (45.center) to (47.center);
		\draw [style=ThickLine, in=0, out=-90] (47.center) to (46.center);
		\draw [style=ThickLine, in=-90, out=180] (46.center) to (48.center);
		\draw [style=ThickLine, in=-180, out=90] (53.center) to (50.center);
		\draw [style=ThickLine, in=90, out=0] (50.center) to (52.center);
		\draw [style=ThickLine, in=0, out=-90] (52.center) to (51.center);
		\draw [style=ThickLine, in=-90, out=180] (51.center) to (53.center);
		\draw [style=ThickLine, in=-180, out=90] (57.center) to (54.center);
		\draw [style=ThickLine, in=90, out=0] (54.center) to (56.center);
		\draw [style=ThickLine, in=0, out=-90] (56.center) to (55.center);
		\draw [style=ThickLine, in=-90, out=180] (55.center) to (57.center);
		\draw [style=BlueLine, bend right] (60.center) to (61.center);
		\draw [style=BlueLine, bend right] (61.center) to (60.center);
		\draw [style=BlueLine, bend right] (62.center) to (63.center);
		\draw [style=BlueLine, bend right] (63.center) to (62.center);
		\draw [style=BlueLine, bend right] (64.center) to (65.center);
		\draw [style=BlueLine, bend right] (65.center) to (64.center);
	\end{pgfonlayer}
\end{tikzpicture}
   }
    \caption{Geometry of Hopf fibration in the case of quadrions coupled to bifundamental matter. The boundary $\partial T(\Sigma)$ has two connected components, the picture shows one. The boundary component is fibered over the tear drop $S^2/\mathbb{Z}_m$ with generic lens space fibers $S^3/\mathbb{Z}_d$ and two exceptional fibers $S^3/\mathbb{Z}_K$ projecting to its North and South pole at $y=\pm1$. The Hopf circles $\gamma_\pm$ are rigid, however $m\gamma_\pm$ is homolgous to the Hopf circle of the generic lens space fiber, and therefore $\gamma_\pm$ generate a $(\mathbb{Z}_K\oplus\mathbb{Z}_K)/\mathbb{Z}_d\cong \mathbb{Z}_K\oplus \mathbb{Z}_m $ contribution to the first homology group. }
    \label{fig:HopfCircles}
\end{figure}
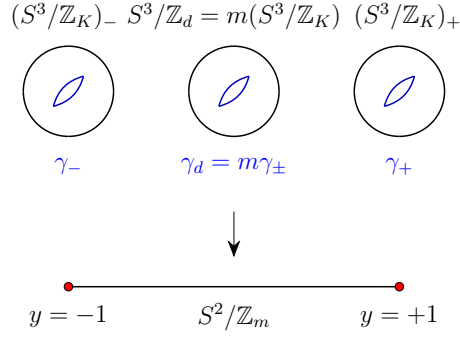

Next, recall that the flavor symmetry algebra is $\mathfrak{g}=\mathfrak{su}(d)\times \mathfrak{su}(d)\times \mathfrak{u}(1)$.
The flavor symmetry group is determined by the boundary map
\begin{equation}\label{eq:boundmapcase3}
    \partial_2: H_2(\partial X_0)\cong \mathbb{Z}\oplus \mathbb{Z}^2_m\quad \rightarrow \quad H_1(\partial T(\Sigma))\cong ((\mathbb{Z}_K\oplus \mathbb{Z}_K)/\mathbb{Z}_d)^2\cong (\mathbb{Z}_K\oplus \mathbb{Z}_m)^2
\end{equation}
with domain generated by the fibers $S^2_\pm$, the two fiber classes projecting to the North and South pole of the base, and $\sigma$ and codomain generated by four Hopf circles (see Figure \ref{fig:HopfCircles}). Similar to the case of Quadrions we can associate four 1-cycles $\gamma_i$ to the four codimension-6 singularities contained in $K$. These generate the $\mathbb{Z}_K^4$ and the quotient by $\mathbb{Z}_d^2$ identifies linear multiples in pairs according to the flavor branes stretching between these. We find
\begin{align}
     \mathrm{Ker}(\partial_2)&=K\mathbb{Z} \\
     \mathrm{Coker}(\partial_2)&=\mathbb{Z}_K \\
     \mathrm{Im}(\partial_2)&=\mathbb{Z}_K\oplus \mathbb{Z}^2_m\,.
\end{align}
In this example the 0-form symmetry is inherited from the 1-form symmetries of two 5D/7D systems, each with electric line defect group isomorphic to $\mathbb{Z}_K \times \mathbb{Z}_m$. A 5D/7D system is given by two 5D SCFTs engineered via $\mathbb{C}^3/\mathbb{Z}_{md}$ glued together such that the two $\mathfrak{su}(d)$ loci compactify resulting in a diagonal gauging of the flavor symmetries. Via symmetry inheritance,
we find the 4D flavor symmetry group in the $X_0$ phase (i.e., the quasi-SCFT phase) is:
\be
G_{F}= \frac{SU(d)\times SU(d)\times U(1)}{\mathbb{Z}_K(m,0,1)\times \mathbb{Z}_K(0,m,1)}\times \mathbb{Z}_m^2
\ee
 with the $\mathbb{Z}_K\times \mathbb{Z}_m^2$ subgroup inherited from the two 5D/7D systems.

Upon giving a finite volume to $S^4/\mathbb{Z}_K$, some of the zero-form symmetries break:
\begin{align}
\mathcal{E}(X_0) & \rightarrow \mathcal{E}(X) \\
U(1)^{(0)} \times \mathbb{Z}_{m}^{(0)} \times \mathbb{Z}_{m}^{(0)} & \rightarrow  U(1)^{(0)} \times \mathbb{Z}_{m}^{(0)}.
\end{align}

\subsubsection{5D SCFT Intersecting a Flavor Stack}

Consider next the system given by a 5D SCFT intersecting a flavor stack. The quotient on the North and South pole patches are, with $\Gamma=\mathbb{Z}_K$ and $\zeta=\exp(2\pi i/K)$,
\begin{eqnarray}
(v_{1},v_{2},v_{3},t) &\rightarrow &(\zeta v_{1},\zeta ^{-1}v_{2},v_{3},t) \\
(v_{1}^{\prime },v_{2}^{\prime },v_{3}^{\prime },t^{\prime }) &\rightarrow
&(\zeta ^{-1}v_{1}^{\prime },\zeta ^{-1}v_{2}^{\prime },\zeta
^{2}v_{3}^{\prime },t^{\prime })\,.
\end{eqnarray}
The relevant homology groups are:
\begin{equation}
    \begin{aligned}
     H_*(X_0,\partial X_0)/H_*(X_0)&\cong\{0,0,0,\mathbb{Z}\oplus \mathbb{Z}_K,0, \mathbb{Z}\oplus \mathbb{Z}_K,0,\mathbb{Z}  \} \\
        H_*(\partial X_0)&\cong \{\mathbb{Z},0,\mathbb{Z}\oplus \mathbb{Z}_K,0, \mathbb{Z}\oplus \mathbb{Z}_K,0,\mathbb{Z}  \} \\
         H_*(\partial X_0^\circ)&\cong \begin{cases} \{\mathbb{Z},\mathbb{Z}_K,\mathbb{Z},\mathbb{Z}\oplus \mathbb{Z}_K,\mathbb{Z},\mathbb{Z}^2  \} \qquad K\textnormal{ odd}\\ \{ \mathbb{Z}, \mathbb{Z}_K, \mathbb{Z}, \mathbb{Z}\oplus \mathbb{Z}_K, 0, \mathbb{Z} \} \qquad K\textnormal{ even}
\end{cases}  \\
H_*(X,\partial X)/H_*(X)&\cong\{0,0,0,\mathbb{Z},0, \mathbb{Z}_K,0,\mathbb{Z}  \}
    \end{aligned}
\end{equation}
and the Pontryagin dual of the electric polarization of the defect group is therefore
\begin{equation}\label{}\begin{aligned}
   \mathcal{E}(X_0) &=
   U(1)^{(0)}\times  \mathbb{Z}_K^{(0)}.
   \end{aligned}
\end{equation}

Returning to section \ref{sec:EXAMP}, the flavor symmetry algebra $\mathfrak{g}=\mathfrak{su}(K)\times \mathfrak{su}(2)\times \mathfrak{u}(1)$ and $\mathfrak{g}=\mathfrak{su}(K)\times \mathfrak{u}(1)$ when $K$ is even and odd respectively.
The flavor symmetry group in the quasi-SCFT phase is thus given by:
\be \begin{aligned}
G_{F}&= \frac{SU(K)\times SU(2)}{\mathbb{Z}_2(K/2,1)}\times U(1)\qquad\qquad (K\textnormal{ even})\\[2mm]
G_{F}&= \frac{SU(K)\times U(1)}{\mathbb{Z}_K(1,1)}\qquad\qquad\qquad\qquad (K\textnormal{ odd})\\
\end{aligned} \ee

Upon giving a finite volume to $S^4/\mathbb{Z}_K$, the Pontryagin dual of the electric polarization of the defect group breaks to
\begin{equation}\begin{aligned}
   \mathcal{E}(X)&=
   U(1)^{(0)}
   \end{aligned}
\end{equation}

\subsubsection{Trinion-Like Pairs}

As a final example, we consider the trinion-like pairs, where we quotient $\Lambda^{2}_{\mathrm{ASD}}(S^4)$ by $\Gamma = \mathbb{Z}_K \times \mathbb{Z}_L$. Letting $\omega=\exp(2\pi i/K)$ and $\eta=\exp(2\pi i/L)$, the group action on the North and South poles is:
\begin{eqnarray}
(v_{1},v_{2},v_{3},t) &\rightarrow &(\omega ^{2b}v_{1},\eta
^{-2a}v_{2},\eta ^{2a}\omega ^{-2b}v_{3},t) \\
(v_{1}^{\prime },v_{2}^{\prime },v_{3}^{\prime },t^{\prime }) &\rightarrow
&(\omega ^{-2b}v_{1}^{\prime },\eta ^{-2a}v_{2}^{\prime },\eta ^{2a}\omega ^{2b}v_{3}^{\prime },t^{\prime }).
\end{eqnarray}
Define $K'=\textnormal{gcd}(K,2a)$ and $L'=\textnormal{gcd}(L,2b)$. The relevant homology groups are:
\begin{equation}\label{eq:Homologies4}
    \begin{aligned}
     H_*(X_0,\partial X_0)/H_*(X_0)&\cong\{0,0,0,\mathbb{Z},\mathbb{Z}_{\textnormal{gcd}(K',L')}, \mathbb{Z}\oplus \mathbb{Z}_{K'}\oplus \mathbb{Z}_{L'}0,\mathbb{Z}  \} \\
        H_*(\partial X_0)&\cong\{\mathbb{Z},0,\mathbb{Z},\mathbb{Z}_{\textnormal{gcd}(K',L')}, \mathbb{Z}\oplus \mathbb{Z}_{K'}\oplus \mathbb{Z}_{L'},0,\mathbb{Z}  \}  \\
         H_*(\partial X_0^\circ)&\cong \{\mathbb{Z},\mathbb{Z}_{K'}\oplus \mathbb{Z}_{L'},\mathbb{Z}\oplus \mathbb{Z}_{\textnormal{gcd}(K',L')} ,\mathbb{Z}^5,\mathbb{Z}^3\} \\
         H_*(X,\partial X)/H_*(X)&\cong\{0,0,0,\mathbb{Z},\mathbb{Z}_{\textnormal{gcd}(K',L')}, \mathbb{Z}\oplus \mathbb{Z}_{K'}\oplus \mathbb{Z}_{L'}0,\mathbb{Z}  \},
    \end{aligned}
\end{equation}
and the electric symmetry read off from the defect group is simply
\begin{equation}\label{}\begin{aligned}
   \mathcal{E}_{\textnormal{SCFT}}&=
   U(1)^{(0)}.
   \end{aligned}
\end{equation}
The flavor symmetry algebra is $\mathfrak{g}=\mathfrak{u}(1)\times \mathfrak{su}(K') \times \mathfrak{su}(L')  \times\mathfrak{su}(\textnormal{gcd}(K',L')) $. We can again determine the global structure of the flavor group studying the map
\begin{equation}
    \partial_2: H_2(\partial X) \cong \mathbb{Z}~\rightarrow ~H_1(\partial T(\Sigma))\,.
\end{equation}
The group $H_2(\partial X)$ is generated by the fibers projecting to either $y,y'=1$ which are fixed by $\Gamma$ and we denote these respectively by
$S_\pm^2$. They are an $\textnormal{lcm}(K',L')$-folding of the generic twistor space fiber. The two choices are homologous. Whenever $\textnormal{gcd}(K',L')\neq 1$ we find these to be contained in the singular locus $\Sigma$ and, in particular, they do not intersect $\partial T(\Sigma)$. Therefore $\partial_2=0$. In other words, the flavor symmetry group is:
\begin{equation}
    G_F= PSU((K',L')) \times PSU(K')^2\times PSU(L')^2\times  PSU((K',L')) \times U(1)\,.
\end{equation}

\section{Further Examples}\label{sec:Rest}

In the previous sections we primarily focused on the space $\Lambda^{2}_{\mathrm{ASD}}(S^4)$ and its quotients by a finite subgroup of isometries. Many of the structures encountered in this class of geometries naturally generalize to other $G_2$-holonomy spaces with complete AC metrics. In this regard, a natural class of examples are obtained from Bryant-Salamon metrics on $\Lambda^{2}_{\mathrm{ASD}}(M)$, the bundle of anti-self-dual 2-forms over a self-dual Einstein 4-orbifold $M$ and quotients thereof.
We will first discuss $M=\mathbb{CP}^2$ and discrete quotients thereof and
we also briefly revisit the case of the bundle of anti-self-dual 2-forms of a weighted projective space $\Lambda_{\mathrm{ASD}}^{2}(\mathbb{WCP}^2)$ considered in in \cite{Acharya:2001gy}  and how it fits with the considerations of the present work. We also show that starting from an ambient eight-dimensional geometry, all of these different cases involving orbifold singularities generated by Abelian group actions can be unified in a single construction.

\subsection{Physics and Geometry of $\Lambda^{2}_{\mathrm{ASD}}(\mathbb{CP}^2)$}

To set the stage for our analysis, we begin by briefly reviewing the physics and geometry of $X = \Lambda^{2}_{\mathrm{ASD}}(\mathbb{CP}^2)$ with its complete AC Bryant-Salamon metric.
In this case, $\partial X$ is the six-manifold $\mathbb{F}$, the twistor space of $\mathbb{CP}^2$. This is also referred to as a flag manifold and can be realized as the quotient $SU(3) / U(1) \times U(1)$.

Since we will be interested in taking finite quotients of $\Lambda^{2}_{\mathrm{ASD}}(\mathbb{CP}^2)$, it will be helpful to give a more uniform characterization of the boundary geometry $\partial X = \mathbb{F}$. This space is obtained as a quadric in $\mathbb{CP}^2 \times \mathbb{CP}^2$. Consider two copies of $\mathbb{CP}^2$, which we denote as $\mathbb{CP}^{2}_{V}$ and $\mathbb{CP}^2_W$, with respective homogeneous coordinates $V_i$ and $W_i$ for $i=1,2,3$. The flag manifold is parameterized by the hypersurface swept out by the zero set $V \cdot W = V_1 W_1 + V_2 W_2 + V_3 W_3 = 0$ (see e.g., \cite{Altavilla2021TwistorGO} for a helpful exposition):
\be
\mathbb{F}=\{ ([V_1:V_2:V_3],[W_1:W_2:W_3])\in \mathbb{CP}^2_V \times \mathbb{CP}^2_W\,\big|\, V \cdot W = 0  \}.
\ee
The flag manifold $\mathbb{F}$ admits the twistor fibration $\pi:\mathbb{F}\rightarrow \mathbb{CP}^2_{U}$ given by the cross product:
\be \begin{aligned}
\pi([V_1,V_2,V_3],[W_1,W_2,W_3])&= [ V_2^*W_3- V_3^*W_2, V_3^*W_1- V_1^*W_3, V_1^*W_2- V_2^*W_1]\\
&\equiv [U_1,U_2,U_3].
\end{aligned} \ee
Note that this map is non-holomorphic. So, all told, we actually have three different $\mathbb{CP}^2$'s, $\mathbb{CP}^{2}_{U}$, $\mathbb{CP}^{2}_{V}$ and $\mathbb{CP}^{2}_{W}$. As explained in \cite{Atiyah:2001qf}, the role of these three $\mathbb{CP}^2$'s can be permuted in the limit where the zero-section collapses to zero size, and this is interpreted as a $\mathbb{Z}_3$ symmetry which is broken once we resolve the tip of the cone $X_0$ to finite size. Indeed, the minimal field content describing this would be three 4D $\mathcal{N} = 1$ chiral superfields $\Phi_i$ which are coupled by a cubic superpotential $\Phi_1 \Phi_2 \Phi_3$, and the vev of one $\Phi_i$ parameterizes the volume modulus of the $\mathbb{CP}^2$ for a given geometry.\footnote{One might ask whether this is already evidence for an interacting fixed point in 4D. For example, if we were to consider IIA on the same singular geometry the so-called ``XYZ model'' would lead to a non-trivial fixed point. The order of limits is somewhat different here, in particular the same cubic coupling in 4D is marginal irrelevant.}

\paragraph{Symmetries of $\Lambda^{2}_{\mathrm{ASD}}(\mathbb{CP}^2)$}

Let us now turn to the symmetries of $X = \Lambda^{2}_{\mathrm{ASD}}(\mathbb{CP}^2)$, including the limit $X_0$ where the zero-section has collapsed to zero size. To begin, we observe that because $\mathbb{F} = SU(3) / U(1) \times U(1)$ is a coset space, it clearly admits an $SU(3)$ group action. In terms of the parameterization as a hypersurface in $\mathbb{CP}^2_{V} \times \mathbb{CP}^2_{W}$, we have, for $A\in SU(3)$:
\be
A \cdot (V, W) = ( A^* V,A W)\,,
\ee
where here, $A$ acts as left-multiplication on a column vector of $W_i$'s, and $A^{*}$ denotes the complex conjugate (but not transpose) matrix acting by left-multiplication on a column vector of the $V_{i}$'s. The center of $SU(3)$ acts trivially and therefore we are actually describing a $PSU(3)$ automorphism on $\mathbb{F}$. The $G_2$-metric on $\Lambda_{\textnormal{ASD}}^2(\mathbb{CP}^2)$ exhibits these symmetries asymptotically as isometries \cite{Atiyah:2001qf}. In addition to these continuous symmetries, there are also some discrete groups. These act on the flag manifold coordinates as follows:
\begin{align}
   & \mathbb{Z}_2\subset S_3: \quad \sigma_2\cdot (V,W)=(W^*,V^*)  \\
   & \mathbb{Z}_3\subset S_3: \quad \sigma_3 \cdot (U,V,W)=(W,V,U),
\end{align}
where in the second line we have also included a permutation involving the base $\mathbb{CP}^2_{U}$ of the twistor space.
The full isometry symmetry group for $\mathbb{F}$ is therefore $PSU(3)\times S_3$. This is retained as a symmetry for $X_0$, but the $\mathbb{Z}_3$ subgroup of $S_3$ is broken in the resolved phase with geometry $X$.

\subsubsection{Abelian Group Actions}

We now specialize further, restricting attention to Abelian group actions. The maximal torus of $PSU(3)$ is the quotient $U(1)_1\times U(1)_2/\mathbb{Z}_3$. Let $\alpha$ and $\beta$ denote elements in $U(1)_1$ and $U(1)_2$ respectively.
Then, we have the action on the flag manifold $\mathbb{F}$:\footnote{Here we have used the homogeneity of the coordinates to pick a convenient presentation of the $PSU(3)$ group action.}
\be
([V_1,V_2,V_3],[W_1,W_2,W_3]) ~\mapsto ~ ([\alpha^{} V_1, \beta V_2, V_3],[\alpha^{-1} W_1,\beta^{-1} W_2, W_3]).
\ee

To get a handle on the fixed loci, consider the fibration $\mathbb{F}\rightarrow \mathbb{CP}_V^2$. This favors one of the three branches of the geometric moduli space but we will make clear how the fixed loci is modified after the flop transitions. Codimension-6 fixed points can potentially occur at three locations at in the base: $\{V_1=V_2=0\}, \; \{ V_2=V_3=0\},$ and/or $\{V_3=V_1=0\}$. Note that the condition $V_1=V_2=0$ implies $W_3=0$ from the condition $V\cdot W = V_1 W_1 + V_2 W_2 + V_3 W_3 =0$ used in the defining equation for $\mathbb{F}$ in
$\mathbb{CP}^{2}_{V} \times \mathbb{CP}^{2}_{W}$.

We can now cover $\mathbb{CP}^2_{V}$ by three distinct patches and in each the group actions read
\be \begin{aligned}\label{eq:local}
[V_1/V_3,V_2/V_3,1] ~&\mapsto ~ [\alpha V_1/V_3, \beta V_2/V_3, 1] \\
[V_1/Y_2,1,V_3/V_2] ~&\mapsto ~ [\alpha\beta^{-1} V_1/V_2, 1, \beta^{-1} V_3/V_2] \\
[1,V_2/V_1,V_3/V_1] ~&\mapsto ~ [1,\alpha^{-1}\beta V_2/V_1, \alpha^{-1} V_3/V_1].
\end{aligned}\ee
Focusing on the $(V_3 \neq 0)$ patch, this can be locally extended in the fiber direction via the coordinate $W_1/W_2$ or $W_2/W_1$ depending on whether we are at the North pole, $W_2=0$, or South pole, $W_1=0$, of the fibral $\mathbb{CP}^1$. This doubling is familiar from the $\Lambda^2_{\textnormal{ASD}}(S^4)/\Gamma$ geometries where two singular loci can project to the same locus in the base $S^4$ which signifies a connected singular locus (codimension-6 or 4) starting from asymptotic infinity, passing through the zero-section to another loci of asymptotic infinity. From flop transitions, which ``out-going" singularities are matched with each ``in-coming" singularities can be permuted, so we now list the maximal torus action on six $\mathbb{C}^3$ patches of $SU(3)/U(1)^2$:
\be \begin{aligned}
(x_1,x_2,x_3)~&\mapsto ~ (\alpha x_1,\beta^{-1} x_2, \alpha^{-1}\beta^{}  x_3) \\
(x'_1,x'_2,x'_3)~&\mapsto ~ (\alpha^{-1} x'_1,\beta^{} x'_2, \alpha^{}\beta^{-1}  x'_3) \\
(y_1,y_2,y_3)~&\mapsto ~ (\alpha^{-1}\beta^{} y_1,\beta^{-1} y_2, \alpha^{}  y_3) \\
(y'_1,y'_2,y'_3)~&\mapsto ~ (\alpha^{}\beta^{-1} y'_1,\beta^{} y'_2, \alpha^{-1} y'_3) \\
(z_1,z_2,z_3)~&\mapsto ~ (\alpha^{-1}\beta z_1,\alpha^{} z_2, \beta^{-1}  z_3) \\
(z'_1,z'_2,z'_3)~&\mapsto ~ (\alpha^{}\beta^{-1} z'_1,\alpha^{-1} z'_2, \beta^{}  z'_3).
\end{aligned}\ee
The $x_i$ and $x'_i$ coordinates parameterize the $\mathbb{C}^3$ patch centered at the North and South poles in the $\mathbb{P}^1$ fiber above $V_1=V_2=0$. They are given by:
\begin{align}
    & x_1=V_1/V_3, \quad x_2=(V_2/V_3)^*, \quad x_3=W_1/W_2\\
    & x'_1=(V_1/V_3)^*, \quad x'_2=V_2/V_3, \quad x'_3=W_2/W_1
\end{align}
where we see that the complex structure of the $\mathbb{C}^3$ patch does not necessarily align with that of the $\mathbb{CP}^2_V$ base. The coordinates $y_i$ and $y'_i$ are similarly associated with $V_3=V_1=0$, and $z_i$ and $z'_i$ with $V_2=V_3=0$ where their definitions follow from suitable cyclic permutations.

\subsection{Examples of $\Lambda_{\mathrm{ASD}}^{2}(\mathbb{CP}^2) / \Gamma$}

Let us now turn to some explicit examples of quotients $X = \Lambda_{\mathrm{ASD}}^{2}(\mathbb{CP}^2) / \Gamma$ with $\Gamma$ a finite Abelian group. Our aim here is not to give an exhaustive treatment, but rather, to illustrate some of the qualitative features. As a first general comment, we remark that quotients by finite Abelian groups do not generate an electric 1-form symmetry. Indeed, this is because all of the singularities we generate stretch to infinity, so we find that every element of $\Gamma$ has fixed points on the boundary of $\Lambda_{\textnormal{ASD}}^2(\mathbb{CP}^2)$. By Armstrong's theorem \cite{armstrong_1968} we therefore have $\pi_1(\partial X/\Gamma)=0$ and these models do not have discrete 1-form symmetries as in the examples based on quotients of $S^4$.
Their main difference to those examples therefore lies in the structure of their singular loci, which is the main focus of our analysis.

\subsubsection{Single Cyclic Factor with Generic $\Gamma=\mathbb{Z}_K$}

\begin{figure}
    \centering
    \scalebox{0.8}{\begin{tikzpicture}
	\begin{pgfonlayer}{nodelayer}
		\node [style=none] (14) at (0, 3.575) {\large  $\mathcal{T}^{\textnormal{\tiny\!  (M)}}_{\mathbb{C}^3/\Gamma_u}$};
		\node [style=none] (15) at (0, 4.25) {};
		\node [style=none] (16) at (0, 2.75) {};
		\node [style=none] (18) at (-0.75, 3.5) {};
		\node [style=none] (20) at (0.75, 3.5) {};
		\node [style=none] (21) at (2.875, 0) { \large $\mathcal{T}^{\textnormal{\tiny\!  (M)}}_{\mathbb{C}^3/\Gamma_v}$};
		\node [style=none] (22) at (2.75, 0.75) {};
		\node [style=none] (23) at (2.75, -0.75) {};
		\node [style=none] (24) at (2, 0) {};
		\node [style=none] (25) at (3.5, 0) {};
		\node [style=none] (26) at (-2.875, 0) { \large $\mathcal{T}^{\textnormal{\tiny\!  (M)}}_{\mathbb{C}^3/\Gamma_w}$};
		\node [style=none] (27) at (-2.75, 0.75) {};
		\node [style=none] (28) at (-2.75, -0.75) {};
		\node [style=none] (29) at (-3.5, 0) {};
		\node [style=none] (30) at (-2, 0) {};
		\node [style=none] (31) at (-0.5, 5.5) {};
		\node [style=none] (32) at (-0.5, 6.5) {};
		\node [style=none] (33) at (0.5, 5.5) {};
		\node [style=none] (34) at (0.5, 6.5) {};
		\node [style=none] (35) at (0, 6) {$\mathfrak{su}_{N_3}$};
		\node [style=none] (36) at (4.5, -2.5) {};
		\node [style=none] (37) at (4.5, -1.5) {};
		\node [style=none] (38) at (5.5, -2.5) {};
		\node [style=none] (39) at (5.5, -1.5) {};
		\node [style=none] (40) at (5, -2) {$\mathfrak{su}_{N_1}$};
		\node [style=none] (41) at (-5.5, -2.5) {};
		\node [style=none] (42) at (-5.5, -1.5) {};
		\node [style=none] (43) at (-4.5, -2.5) {};
		\node [style=none] (44) at (-4.5, -1.5) {};
		\node [style=none] (45) at (-5, -2) {$\mathfrak{su}_{N_2}$};
		\node [style=none] (46) at (1.375, 1.75) {$\mathfrak{su}_{N_2}$};
		\node [style=none] (47) at (1.375, 2.25) {};
		\node [style=none] (48) at (1.375, 1.25) {};
		\node [style=none] (49) at (0.875, 1.75) {};
		\node [style=none] (50) at (1.875, 1.75) {};
		\node [style=none] (51) at (0, 0) {$\mathfrak{su}_{N_3}$};
		\node [style=none] (52) at (0, 0.5) {};
		\node [style=none] (53) at (0, -0.5) {};
		\node [style=none] (54) at (-0.5, 0) {};
		\node [style=none] (55) at (0.5, 0) {};
		\node [style=none] (56) at (-1.375, 1.75) {$\mathfrak{su}_{N_1}$};
		\node [style=none] (57) at (-1.375, 2.25) {};
		\node [style=none] (58) at (-1.375, 1.25) {};
		\node [style=none] (59) at (-1.875, 1.75) {};
		\node [style=none] (60) at (-0.875, 1.75) {};
		\node [style=none] (61) at (0, 5.5) {};
		\node [style=none] (62) at (-3.3, -0.5) {};
		\node [style=none] (63) at (3.3, -0.5) {};
		\node [style=none] (64) at (-2.3, 0.58) {};
		\node [style=none] (65) at (-1.675, 1.375) {};
		\node [style=none] (66) at (-0.45, 2.92) {};
		\node [style=none] (67) at (-1.08, 2.13) {};
		\node [style=none] (68) at (1.08, 2.13) {};
		\node [style=none] (69) at (0.45, 2.92) {};
		\node [style=none] (70) at (1.675, 1.375) {};
		\node [style=none] (71) at (2.3, 0.58) {};
		\node [style=none] (72) at (-2, 0.375) {};
		\node [style=none] (73) at (2, 0.375) {};
		\node [style=none] (74) at (0, 2.75) {};
		\node [style=none] (75) at (0, 1.375) {};		\node [style=none] (76) at (0, -1) {};
  \node [style=none] (77) at (0.4, -0.3) {};
  \node [style=none] (78) at (-0.4, -0.3) {};
  \node [style=none] (79) at (2.125, 2.5) {};
  \node [style=none] (80) at (1.875, 1.75) {};
  \node [style=none] (81) at (1.375, 2.25) {};
  \node [style=none] (82) at (-2.125, 2.5) {};
  \node [style=none] (83) at (-1.875, 1.75) {};
  \node [style=none] (84) at (-1.375, 2.25) {};
	\end{pgfonlayer}
	\begin{pgfonlayer}{edgelayer}
		\draw [style=ThickLine] (31.center) to (33.center);
		\draw [style=ThickLine] (34.center) to (33.center);
		\draw [style=ThickLine] (34.center) to (32.center);
		\draw [style=ThickLine] (32.center) to (31.center);
		\draw [style=ThickLine] (36.center) to (38.center);
		\draw [style=ThickLine] (39.center) to (38.center);
		\draw [style=ThickLine] (39.center) to (37.center);
		\draw [style=ThickLine] (37.center) to (36.center);
		\draw [style=ThickLine] (41.center) to (43.center);
		\draw [style=ThickLine] (44.center) to (43.center);
		\draw [style=ThickLine] (44.center) to (42.center);
		\draw [style=ThickLine] (42.center) to (41.center);
		\draw [style=ThickLine, bend left=45] (49.center) to (47.center);
		\draw [style=ThickLine, bend left=45] (47.center) to (50.center);
		\draw [style=ThickLine, bend left=45] (50.center) to (48.center);
		\draw [style=ThickLine, bend left=45] (48.center) to (49.center);
		\draw [style=ThickLine, bend left=45] (54.center) to (52.center);
		\draw [style=ThickLine, bend left=45] (52.center) to (55.center);
		\draw [style=ThickLine, bend left=45] (55.center) to (53.center);
		\draw [style=ThickLine, bend left=45] (53.center) to (54.center);
		\draw [style=ThickLine, bend left=45] (59.center) to (57.center);
		\draw [style=ThickLine, bend left=45] (57.center) to (60.center);
		\draw [style=ThickLine, bend left=45] (60.center) to (58.center);
		\draw [style=ThickLine, bend left=45] (58.center) to (59.center);
		\draw [style=ThickLine] (61.center) to (15.center);
		\draw [style=ThickLine] (67.center) to (66.center);
		\draw [style=ThickLine] (69.center) to (68.center);
		\draw [style=ThickLine] (70.center) to (71.center);
		\draw [style=ThickLine] (65.center) to (64.center);
		\draw [style=ThickLine] (62.center) to (44.center);
		\draw [style=ThickLine] (63.center) to (37.center);
		\draw [style=ThickLine] (30.center) to (54.center);
		\draw [style=ThickLine] (55.center) to (24.center);
  \draw [style=ThickLine] (76.center) to (77.center);
  \draw [style=ThickLine] (76.center) to (78.center);
  \draw [style=ThickLine] (79.center) to (80.center);
  \draw [style=ThickLine] (79.center) to (81.center);
  \draw [style=ThickLine] (82.center) to (83.center);
  \draw [style=ThickLine] (82.center) to (84.center);
	\end{pgfonlayer}
\end{tikzpicture}
}
    \caption{5D theory associated to $\Lambda^2_{\textnormal{ASD}}(\mathbb{CP}^2)/\Gamma$ with group action $\Gamma=\mathbb{Z}_K$. The cones of the gauge nodes denote a 5D gauging with gauge coupling depending on the radial coordinate $r$ of the $G_2$-holonomy space such that 4D $\mathcal{N} = 1$ supersymmetry is preserved.}
    \label{fig:5DCp2}
\end{figure}
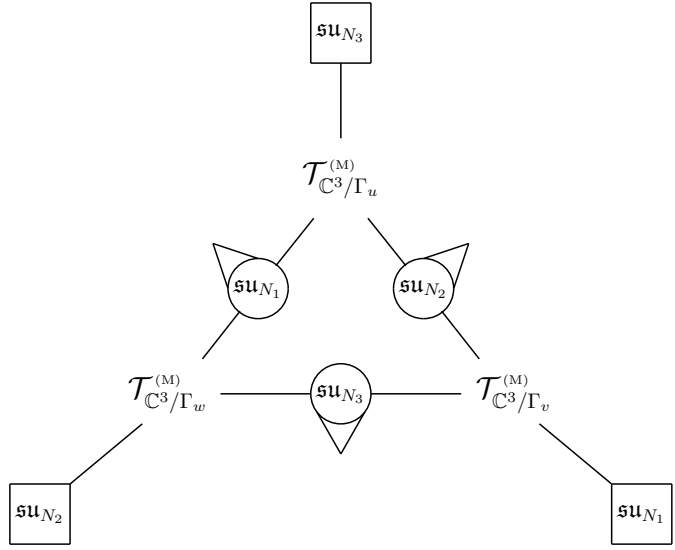

When considering $\Gamma=\mathbb{Z}_K$ we need to specify how its generator embeds within $U(1)_1\times U(1)_2$. This embedding is parameterized by integers $a,b$ and setting $\alpha=\zeta^a$ and $\beta=\zeta^b$ with $\zeta=\exp(2\pi i/K)$. With this we find the group action on various (unprimed) patches to take the form
\be \begin{aligned}\label{eq:local2}
(x_1,x_2,x_3,t_x)~&\mapsto ~ (\zeta^{a} x_1,\zeta^{-b} x_2, \zeta^{-a+b}   x_3,t_x) \\
(y_1,y_2,y_3,t_y)~&\mapsto ~ (\zeta^{-a+b} y_1,\zeta^{-b}  y_2,\zeta^{a}  y_3,t_y) \\
(z_1,z_2,z_3,t_z)~&\mapsto ~(\zeta^{-a+b}   z_1,\zeta^{a}  z_2, \zeta^{-b}  z_3,t_z)
\end{aligned}\ee
where in each patch of the cone, we have specified the action on $\mathbb{C}^3\times \mathbb{R}$ with group action purely on $\mathbb{C}^3$. Each patch behaves much as in the case of quotients of $\Lambda_{\mathrm{ASD}}^{2}(S^4)$. Let us define
\be\begin{aligned}
N_1&=\textnormal{gcd}(K,|a|)\,, \quad N_2=\textnormal{gcd}(K,|b|)\,, \quad N_3=\textnormal{gcd}(K,|a-b|).
\end{aligned}\ee
We can then present the resulting network of 5D theories as shown in Figure \ref{fig:5DCp2}. All of these factors are interpreted as flavor symmetries in the 4D edge mode theory.

Observe that in the 5D theory, the volumes of the compact $S^2$'s supporting the 5D gaugings are related. At a fixed radial slice, we can compare the volume of each $S^2$ before and after the group action. This is reduced by a factor of $N_i / K$ for each $S^2$, and results in the following relation between the three gauge couplings:
\be
\frac{1}{N_1}\frac{1}{g_1^{\textnormal{5D}}(r)^2}=\frac{1}{N_2}\frac{1}{g_2^{\textnormal{5D}}(r)^2}=\frac{1}{N_3}\frac{1}{g_3^{\textnormal{5D}}(r)^2}
\ee
where $g_i^{\textnormal{5D}}(r)$ is the gauge coupling of $\mathfrak{su}(N_i)$ at radius $r$ in the $G_2$ cone.

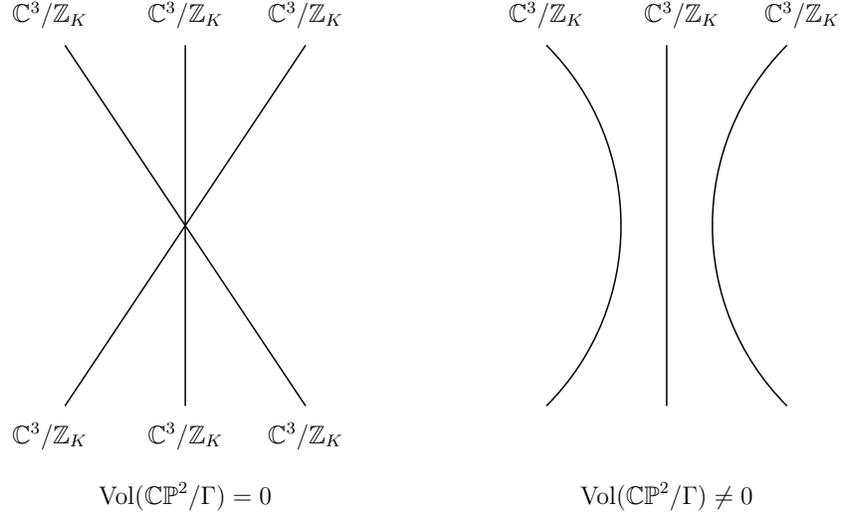
\begin{figure}
\centering
\scalebox{0.8}{
\begin{tikzpicture}
	\begin{pgfonlayer}{nodelayer}
		\node [style=none] (0) at (-4, 3) {};
		\node [style=none] (1) at (-4, -3) {};
		\node [style=none] (2) at (-2, -3) {};
		\node [style=none] (3) at (-6, -3) {};
		\node [style=none] (4) at (-6, 3) {};
		\node [style=none] (5) at (-2, 3) {};
		\node [style=none] (6) at (-4, 0) {};
		\node [style=none] (7) at (4, 3) {};
		\node [style=none] (8) at (4, -3) {};
		\node [style=none] (9) at (6, -3) {};
		\node [style=none] (10) at (2, -3) {};
		\node [style=none] (11) at (2, 3) {};
		\node [style=none] (12) at (6, 3) {};
		\node [style=none] (13) at (4, 0) {};
		\node [style=none] (14) at (-4, -4.5) {$\textnormal{Vol}(\mathbb{CP}^2/\Gamma)=0$};
		\node [style=none] (15) at (4, -4.5) {$\textnormal{Vol}(\mathbb{CP}^2/\Gamma)\neq 0$};
		\node [style=none] (16) at (-4, -3.5) {$\mathbb{C}^3/\mathbb{Z}_K$};
		\node [style=none] (17) at (-2, -3.5) {$\mathbb{C}^3/\mathbb{Z}_K$};
		\node [style=none] (18) at (-6.25, -3.5) {$\mathbb{C}^3/\mathbb{Z}_K$};
		\node [style=none] (19) at (-4, 3.5) {$\mathbb{C}^3/\mathbb{Z}_K$};
		\node [style=none] (20) at (-2, 3.5) {$\mathbb{C}^3/\mathbb{Z}_K$};
		\node [style=none] (21) at (-6.25, 3.5) {$\mathbb{C}^3/\mathbb{Z}_K$};
		\node [style=none] (22) at (4.25, 3.5) {$\mathbb{C}^3/\mathbb{Z}_K$};
		\node [style=none] (23) at (6.25, 3.5) {$\mathbb{C}^3/\mathbb{Z}_K$};
		\node [style=none] (24) at (2, 3.5) {$\mathbb{C}^3/\mathbb{Z}_K$};
	\end{pgfonlayer}
	\begin{pgfonlayer}{edgelayer}
		\draw [style=ThickLine] (4.center) to (2.center);
		\draw [style=ThickLine] (5.center) to (3.center);
		\draw [style=ThickLine] (0.center) to (1.center);
		\draw [style=ThickLine] (7.center) to (8.center);
		\draw [style=ThickLine, bend left=45] (11.center) to (10.center);
		\draw [style=ThickLine, bend right=45] (12.center) to (9.center);
	\end{pgfonlayer}
\end{tikzpicture}
}
\caption{Breaking the Sexion. \textsc{left:} Singularity corresponding to a Sexion theory, a sixvalent junction of 5D SCFTs. \textsc{right:} Resolving the tip of the cone triggers a breaking pattern to a triple of 5D SCFTs. Note that this is one of three branches of the geometric moduli space, when flopping to the other two, the 5D SCFT loci will consist of pairing different combinations of top line segments with bottom line segments on the left figure. }
\label{Fig:Hexion}
\end{figure}

\paragraph{Example: Sexion Theories.} These are characterized by $N_i=1$ which have no codimension-4 loci, and thus is the analog of the Quadrion orbifolds we considered previously. In this case there are only codimension-6 loci modelled on $\mathbb{C}^3/\Gamma\times \mathbb{R}$. An explicit example is $K=7$ and $a=1$ and $b=2$. When the $\mathbb{CP}^2$ is collapsed we obtain a codimension-7 singularity at the location where three copies of $\mathbb{C}^3/\Gamma\times \mathbb{R}_+$ and $\mathbb{C}^3/\Gamma\times \mathbb{R}_-$ meet. See Figure \ref{Fig:Hexion}.

\begin{figure}[t!]
\centering
\scalebox{0.8}{
\begin{tikzpicture}
	\begin{pgfonlayer}{nodelayer}
		\node [style=none] (0) at (-9, 6) {};
		\node [style=none] (1) at (-9, 4) {};
		\node [style=none] (2) at (-7, 4) {};
		\node [style=none] (3) at (-7, 6) {};
		\node [style=none] (4) at (-3, 6) {};
		\node [style=none] (5) at (-3, 4) {};
		\node [style=none] (6) at (-1, 4) {};
		\node [style=none] (7) at (-1, 6) {};
		\node [style=none] (8) at (-6, 2) {};
		\node [style=none] (9) at (-4, 2) {};
		\node [style=none] (10) at (-6, 0) {};
		\node [style=none] (11) at (-4, 0) {};
		\node [style=none] (12) at (-5, 1) {\Large{ $\mathfrak{su}_K$}};
		\node [style=none] (13) at (-8, 5) {\Large{$\mathfrak{su}_K$}};
		\node [style=none] (14) at (-2, 5) {\Large{$\mathfrak{su}_K$}};
		\node [style=none] (15) at (-7, 5) {};
		\node [style=none] (16) at (-3, 5) {};
		\node [style=none] (17) at (-8, 4) {};
		\node [style=none] (18) at (-2, 4) {};
		\node [style=none] (19) at (-4.5, 5.5) {};
		\node [style=none] (20) at (-5, 5) {};
		\node [style=none] (21) at (-4.5, 4.5) {};
		\node [style=none] (22) at (-7, 3) {};
		\node [style=none] (23) at (-7, 3.5) {};
		\node [style=none] (24) at (-7.5, 3) {};
		\node [style=none] (25) at (-3, 3) {};
		\node [style=none] (26) at (-3.5, 3) {};
		\node [style=none] (27) at (-3, 2.5) {};
	\end{pgfonlayer}
	\begin{pgfonlayer}{edgelayer}
		\draw (0.center) to (3.center);
		\draw (3.center) to (2.center);
		\draw (0.center) to (1.center);
		\draw (1.center) to (2.center);
		\draw (4.center) to (5.center);
		\draw (5.center) to (6.center);
		\draw (6.center) to (7.center);
		\draw (7.center) to (4.center);
		\draw (8.center) to (10.center);
		\draw (10.center) to (11.center);
		\draw (11.center) to (9.center);
		\draw (8.center) to (9.center);
		\draw (17.center) to (8.center);
		\draw (9.center) to (18.center);
		\draw (15.center) to (16.center);
		\draw (19.center) to (20.center);
		\draw (21.center) to (20.center);
		\draw (23.center) to (22.center);
		\draw (24.center) to (22.center);
		\draw (26.center) to (25.center);
		\draw (25.center) to (27.center);
	\end{pgfonlayer}
\end{tikzpicture}
}
\caption{Quiver diagram of low energy 4D $\mathcal{N}=1$ matter associated to the pure flavor orbifold geometry. Here the lines denote bifundamental chiral multiplets, and the loop signifies a cubic superpotential.}
\label{fig:PureFlavcp2}
\end{figure}
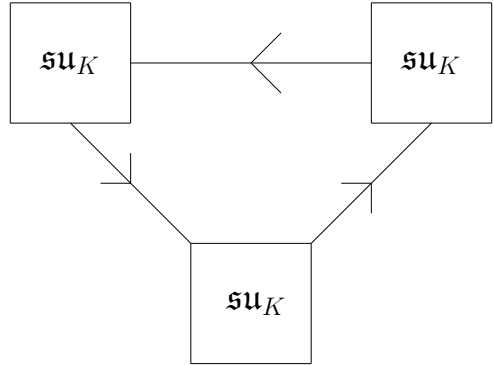

\paragraph{Example: Pure Flavor / Coupled Bifundamentals}
At the other extreme, we can consider orbifold geometries that only contain codimension-4 singularities. These are given by taking either $a=0$, $b=0$, or $a=b$, which each imply\footnote{We also assume $\mathbb{Z}_K$ acts effectively on the geometry so, for instance, if $a=0$, we assume $\mathrm{gcd}(K,|b|)=1$.} $\mathbb{Z}_K\subset SU(2)$. These $\mathbb{Z}_K$ quotients are identical in form to the $U(1)$ quotient of $\Lambda^2_{ASD}(\mathbb{CP}^2)$ to IIA given in \cite{Atiyah:2001qf}, which corresponds to three D6-branes in $\mathbb{R}^6$ intersecting at a point. The 4D theory generated from this example involves 4D $\mathcal{N} = 1$ chiral multiplets in bifundamental representations, with a flavor invariant superpotential coupling generated by a closed loop which passes through each symmetry factor once (see Figure \ref{fig:PureFlavcp2}).

As in the discussion of the cases with base $S^4 / \Gamma$ we can now impose different constraints on the $N_i$ realizing different ranks and connectedness properties of the quiver \ref{fig:5DCp2}. However, the discussion parallels the already analyzed cases and we leave it for the enthusiastic reader.

\subsection{Generalizations with $\mathbb{WCP}^2$ Base}

In the previous sections we first considered smooth $G_2$-holonomy spaces and then constructed large classes of orbifolds from these taking quotients by discrete isometries. In this section, we skip straight to the singular geometry by considering the bundle of anti-self-dual 2-forms over weighted projective space $\mathbb{WCP}^2_{q_1,q_2,q_3}$ first analyzed in \cite{Acharya:2001gy}. The isometry groups of such spaces is $U(1)^2$ and in principle we could take further quotients by discrete subgroups thereof. However, even without such quotients, we already find  singular structures similar to the ones already discussed here, so we elect to focus on these.

To frame the discussion to follow, let us denote the projective coordinates of the weighted projective space by $Z_i$ and introduce the integers
\begin{align}
&n_1=\textnormal{gcd}(q_2,q_3)\,, &n_2&=\textnormal{gcd}(q_3,q_1)\,,  &n_3&=\textnormal{gcd}(q_1,q_2)\,, \\
&r_1=\textnormal{gcd}(q_1,|q_2-q_3|)\,, &r_2&=\textnormal{gcd}(q_2,|q_3-q_1|)\,,  &r_3&=\textnormal{gcd}(q_3,|q_1-q_2|)\,.
\end{align}
These are such that there are $A_{n_i-1}$ ADE singularities supported at 3 two-spheres $Z_i=0$ and $A_{r_k-1}$ ADE singularities filling the fibers projecting to the point with $Z_k=1$ with all other coordinates vanishing.

Overall the singular loci are of similar structure as those in $\Lambda_{\textnormal{ASD}}^2(\mathbb{CP}^2)/\Gamma$ and we can again represent the 5D setup as shown in Figure \ref{Fig:Hexion}, with identifications $N_i=n_i$ and $N_i'=r_i$.


The homology groups of weighted projective space were determined by Kawasaki in \cite{Kawasaki1973} to be isomorphic to those of unweighted projective space, in particular we have
\be H_2(\mathbb{WCP}^2_{q_1,q_2,q_3})=\mathbb{Z}\ee
which is torsion free. In comparison to the case with base $S^4$ discussed at length in earlier sections we therefore find no additional massive modes from branes wrapped on torsional 2-cycles. However, there are additional modes from wrapping the sphere which generates this homology group. When the volume of the projective space goes to zero both electric particles and magnetic strings become light in the semi-classical limit so we again expect interesting physics at the conical singularity.

\begin{figure}
\centering
\scalebox{0.8}{
\begin{tikzpicture}
	\begin{pgfonlayer}{nodelayer}
		\node [style=none] (0) at (0, 4.25) {\Large $\mathcal{T}^{\textnormal{\tiny\!  (M)}}_{\mathbb{C}^3/\Gamma_2}$};
		\node [style=none] (1) at (-4, 2) { \Large $\mathcal{T}^{\textnormal{\tiny\!  (M)}}_{\mathbb{C}^3/\Gamma_1}$};
		\node [style=none] (2) at (-4, -2) {\Large $\mathcal{T}^{\textnormal{\tiny\!  (M)}}_{\mathbb{C}^3/\Gamma_{m+1}}$};
		\node [style=none] (3) at (0.125, -4.25) { \Large $\mathcal{T}^{\textnormal{\tiny\!  (M)}}_{\mathbb{C}^3/\Gamma_{m}}$};
		\node [style=none] (4) at (4, -2) { \Large $\mathcal{T}^{\textnormal{\tiny\!  (M)}}_{\mathbb{C}^3/\Gamma_{m-1}}$};
		\node [style=none] (5) at (4, 2) {\Large  $\mathcal{T}^{\textnormal{\tiny\!  (M)}}_{\mathbb{C}^3/\Gamma_{3}}$};
		\node [style=none] (10) at (1.75, 3.5) {};
		\node [style=none] (11) at (2.75, 3.5) {};
		\node [style=none] (12) at (2.25, 4) {};
		\node [style=none] (13) at (2.25, 3) {};
		\node [style=none] (14) at (-2.75, 3.5) {};
		\node [style=none] (15) at (-1.75, 3.5) {};
		\node [style=none] (16) at (-2.25, 4) {};
		\node [style=none] (17) at (-2.25, 3) {};
		\node [style=none] (18) at (-2.75, -3.5) {};
		\node [style=none] (19) at (-1.75, -3.5) {};
		\node [style=none] (20) at (-2.25, -3) {};
		\node [style=none] (21) at (-2.25, -4) {};
		\node [style=none] (22) at (1.75, -3.5) {};
		\node [style=none] (23) at (2.75, -3.5) {};
		\node [style=none] (24) at (2.25, -3) {};
		\node [style=none] (25) at (2.25, -4) {};
		\node [style=none] (26) at (4, 1) {};
		\node [style=none] (27) at (4, -1) {};
		\node [style=none] (28) at (-4, 1) {};
		\node [style=none] (29) at (-4, -1) {};
		\node [style=none] (30) at (-6.5, 4.5) {};
		\node [style=none] (31) at (-5.5, 4.5) {};
		\node [style=none] (32) at (-5.5, 3.5) {};
		\node [style=none] (33) at (-6.5, 3.5) {};
		\node [style=none] (34) at (-0.5, 7) {};
		\node [style=none] (35) at (0.5, 7) {};
		\node [style=none] (36) at (0.5, 6) {};
		\node [style=none] (37) at (-0.5, 6) {};
		\node [style=none] (38) at (5.5, 4.5) {};
		\node [style=none] (39) at (6.5, 4.5) {};
		\node [style=none] (40) at (6.5, 3.5) {};
		\node [style=none] (41) at (5.5, 3.5) {};
		\node [style=none] (42) at (5.5, -3.5) {};
		\node [style=none] (43) at (6.5, -3.5) {};
		\node [style=none] (44) at (6.5, -4.5) {};
		\node [style=none] (45) at (5.5, -4.5) {};
		\node [style=none] (46) at (-6.5, -3.5) {};
		\node [style=none] (47) at (-5.5, -3.5) {};
		\node [style=none] (48) at (-5.5, -4.5) {};
		\node [style=none] (49) at (-6.5, -4.5) {};
		\node [style=none] (50) at (-0.5, -6) {};
		\node [style=none] (51) at (0.5, -6) {};
		\node [style=none] (52) at (0.5, -7) {};
		\node [style=none] (53) at (-0.5, -7) {};
		\node [style=none] (54) at (-4.75, 2.75) {};
		\node [style=none] (55) at (0, 5) {};
		\node [style=none] (56) at (4.75, 2.75) {};
		\node [style=none] (57) at (4.75, -2.75) {};
		\node [style=none] (58) at (0, -5) {};
		\node [style=none] (59) at (-4.75, -2.75) {};
		\node [style=none] (60) at (0, 6) {};
		\node [style=none] (61) at (0, -6) {};
		\node [style=none] (62) at (0, 6.5) {$\mathfrak{su}$};
		\node [style=none] (63) at (6, 4) {$\mathfrak{su}$};
		\node [style=none] (64) at (6, -4) {$\mathfrak{su}$};
		\node [style=none] (65) at (0, -6.5) {$\mathfrak{su}$};
		\node [style=none] (66) at (-6, 4) {$\mathfrak{su}$};
		\node [style=none] (67) at (-6, -4) {$\mathfrak{su}$};
		\node [style=none] (68) at (2.25, 3.5) {$\mathfrak{su}$};
		\node [style=none] (69) at (-2.25, 3.5) {$\mathfrak{su}$};
		\node [style=none] (70) at (2.25, -3.5) {$\mathfrak{su}$};
		\node [style=none] (71) at (-2.25, -3.5) {$\mathfrak{su}$};
		\node [style=none] (73) at (-3.25, 2.75) {};
		\node [style=none] (74) at (1, 4) {};
		\node [style=none] (75) at (-1, 4) {};
		\node [style=none] (76) at (-3.25, -2.75) {};
		\node [style=none] (77) at (3.25, -2.75) {};
		\node [style=none] (78) at (-1, -4) {};
		\node [style=none] (79) at (1, -4) {};
		\node [style=none] (80) at (3.25, 2.75) {};
		\node [style=none] (81) at (-2.625, 3.125) {};
		\node [style=none] (82) at (-1.75, 3.625) {};
		\node [style=none] (83) at (1.75, 3.625) {};
		\node [style=none] (84) at (2.625, 3.125) {};
		\node [style=none] (85) at (2.625, -3.125) {};
		\node [style=none] (86) at (1.75, -3.625) {};
		\node [style=none] (87) at (-1.75, -3.625) {};
		\node [style=none] (88) at (-2.625, -3.125) {};
		\node [style=none] (89) at (0, -7.75) {};
        \node [style=none] (90) at (3, 4.25) {};
		\node [style=none] (91) at (-3, 4.25) {};
		\node [style=none] (92) at (3, -4.25) {};
		\node [style=none] (93) at (-3, -4.25) {};
	\end{pgfonlayer}
	\begin{pgfonlayer}{edgelayer}
		\draw [style=ThickLine, bend right=45] (13.center) to (11.center);
		\draw [style=ThickLine, bend right=45] (11.center) to (12.center);
		\draw [style=ThickLine, bend right=45] (12.center) to (10.center);
		\draw [style=ThickLine, bend right=45] (10.center) to (13.center);
		\draw [style=ThickLine, bend right=45] (17.center) to (15.center);
		\draw [style=ThickLine, bend right=45] (15.center) to (16.center);
		\draw [style=ThickLine, bend right=45] (16.center) to (14.center);
		\draw [style=ThickLine, bend right=45] (14.center) to (17.center);
		\draw [style=ThickLine, bend right=45] (21.center) to (19.center);
		\draw [style=ThickLine, bend right=45] (19.center) to (20.center);
		\draw [style=ThickLine, bend right=45] (20.center) to (18.center);
		\draw [style=ThickLine, bend right=45] (18.center) to (21.center);
		\draw [style=ThickLine, bend right=45] (25.center) to (23.center);
		\draw [style=ThickLine, bend right=45] (23.center) to (24.center);
		\draw [style=ThickLine, bend right=45] (24.center) to (22.center);
		\draw [style=ThickLine, bend right=45] (22.center) to (25.center);
		\draw [style=DashedLine] (26.center) to (27.center);
		\draw [style=DashedLine] (28.center) to (29.center);
		\draw [style=ThickLine] (31.center) to (32.center);
		\draw [style=ThickLine] (33.center) to (32.center);
		\draw [style=ThickLine] (33.center) to (30.center);
		\draw [style=ThickLine] (30.center) to (31.center);
		\draw [style=ThickLine] (35.center) to (36.center);
		\draw [style=ThickLine] (37.center) to (36.center);
		\draw [style=ThickLine] (37.center) to (34.center);
		\draw [style=ThickLine] (34.center) to (35.center);
		\draw [style=ThickLine] (39.center) to (40.center);
		\draw [style=ThickLine] (41.center) to (40.center);
		\draw [style=ThickLine] (41.center) to (38.center);
		\draw [style=ThickLine] (38.center) to (39.center);
		\draw [style=ThickLine] (43.center) to (44.center);
		\draw [style=ThickLine] (45.center) to (44.center);
		\draw [style=ThickLine] (45.center) to (42.center);
		\draw [style=ThickLine] (42.center) to (43.center);
		\draw [style=ThickLine] (47.center) to (48.center);
		\draw [style=ThickLine] (49.center) to (48.center);
		\draw [style=ThickLine] (49.center) to (46.center);
		\draw [style=ThickLine] (46.center) to (47.center);
		\draw [style=ThickLine] (51.center) to (52.center);
		\draw [style=ThickLine] (53.center) to (52.center);
		\draw [style=ThickLine] (53.center) to (50.center);
		\draw [style=ThickLine] (50.center) to (51.center);
		\draw [style=ThickLine] (32.center) to (54.center);
		\draw [style=ThickLine] (60.center) to (55.center);
		\draw [style=ThickLine] (41.center) to (56.center);
		\draw [style=ThickLine] (57.center) to (42.center);
		\draw [style=ThickLine] (58.center) to (61.center);
		\draw [style=ThickLine] (59.center) to (47.center);
		\draw [style=ThickLine] (73.center) to (81.center);
		\draw [style=ThickLine] (82.center) to (75.center);
		\draw [style=ThickLine] (74.center) to (83.center);
		\draw [style=ThickLine] (84.center) to (80.center);
		\draw [style=ThickLine] (77.center) to (85.center);
		\draw [style=ThickLine] (86.center) to (79.center);
		\draw [style=ThickLine] (78.center) to (87.center);
		\draw [style=ThickLine] (88.center) to (76.center);
        \draw [style=ThickLine] (88.center) to (76.center);
		\draw [style=ThickLine] (12.center) to (90.center);
		\draw [style=ThickLine] (90.center) to (11.center);
		\draw [style=ThickLine] (14.center) to (91.center);
		\draw [style=ThickLine] (91.center) to (16.center);
		\draw [style=ThickLine] (93.center) to (18.center);
		\draw [style=ThickLine] (21.center) to (93.center);
		\draw [style=ThickLine] (25.center) to (92.center);
		\draw [style=ThickLine] (23.center) to (92.center);
	\end{pgfonlayer}
\end{tikzpicture}
}
\caption{The quiver associated to the $n$-zene geometry (a generalization of benzene). We leave the ranks of the 5D gauge and flavor symmetries implicit. The 5D gaugings preserve 4D $\mathcal{N}=1$ supersymmetry, but break 5D Lorentz symmetry due to a radially dependent gauge coupling. There could be further massive modes from M2-branes wrapped on torsional cycles of the zero-section. Generically $\Gamma_i\neq \Gamma_j$. The respective codimension-4 and -6 singular loci are non-compact. The dashed lines indicate a necklace. The cones of the gauge nodes denote a 5D gauging with gauge coupling depending on the radial coordinate $r$ of the $G_2$-holonomy space such that 4D $\mathcal{N} = 1$
supersymmetry is preserved.}
\label{fig:Polyon}
\end{figure}
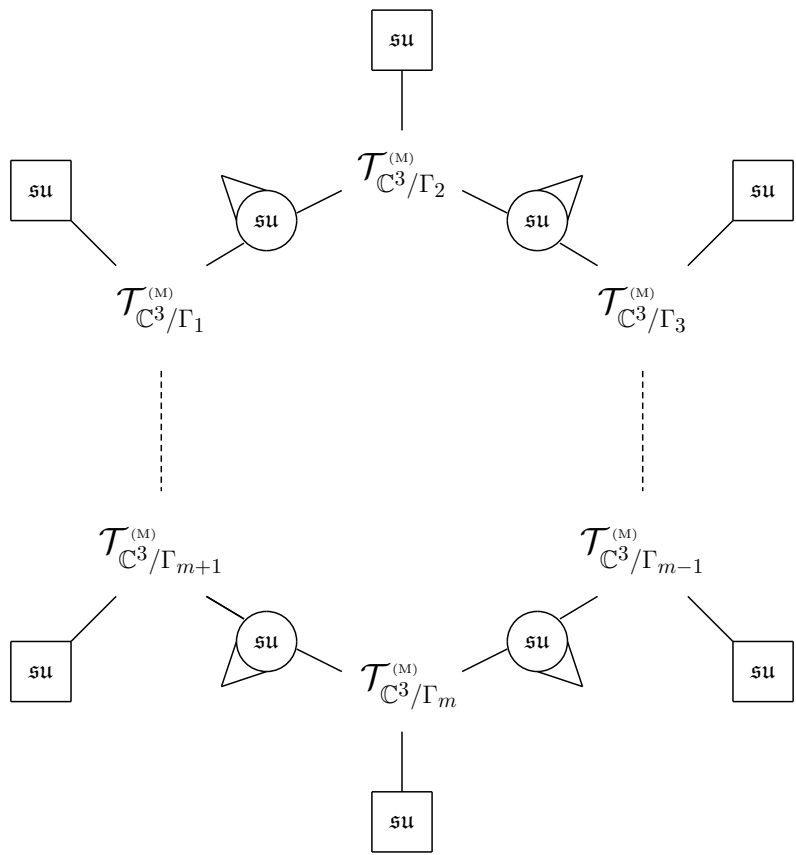

\subsection{Unification and Further Generalizations}

Having presented a number of different examples, it is also helpful to provide a unifying perspective on the various spaces considered in this paper. We largely follow the discussion given in \cite{Acharya:2001gy}.

The relevant construction relies on the hyperk\"ahler quotient construction applied to the flat space $\mathbb{H}^{n+2}$ which is acted on by $SU(2)\times Sp(n+2)$ where $Sp(n+2)$ acts from the right preserving the three complex structures while $SU(2)$ acts from the left rotating the complex structures. Let $H\subset Sp(n+2)$ denote a subgroup of dimension $n$. The hyperk\"ahler quotient $\mathscr{X}=\mathbb{H}^{n+2}// H$ is of dimension 8 and inherits an $SU(2)$ action with torus $U(1)$. The quotient space $X_0=\mathscr{X}/U(1)$ is a $G_2$ cone and admits a deformation to $X$ which is the bundle of anti-self-dual 2-forms over the link of the quotient $\mathscr{X}/SU(2)$, which in general is a compact 4-orbifold with self-dual Einstein metric.

Let us consider the example $H=U(1)^n\subset Sp(n+2)$ in closer detail. We have quaternionic coordinates $Q_m=a_m+b_mj$ for the $m$th quaternionic plane in $\mathbb{H}^{m+2}$ and consider an action with hyperk\"ahler moment maps
\be\begin{aligned}
\mu^\alpha_{\mathbb{R}}(p)&=\sum_{m=1}^{n+2}p^\alpha_m(|a_m|^2-|b_m|^2)\\
\mu^\alpha_{\mathbb{C}}(p)&=\sum_{m=1}^{n+2}p^\alpha_m a_m b_m^*\,.\\
\end{aligned}\ee
Here $\alpha=1,\dots,k$ and $p$ abbreviates the $n(n+2)$ integers $p_i^\alpha$ which are organized into $n$ $U(1)$ charge vectors $p^\alpha$ of length $n+2$. We then have
\be
\mathscr{X}= \left(\,\bigcap_{\alpha=1}^k (\mu^\alpha_{\mathbb{R}})^{-1}(0) \cap  (\mu^\alpha_{\mathbb{C}})^{-1}(0)\right) /\,U(1)^n
\ee
which we quotient by an additional $U(1)$ acting as $(a_i,b_i) \mapsto (\lambda a_i, \lambda b_i) $ where $\lambda=\exp(i\phi)$, this gives the cone $X_0=\mathscr{X}/U(1)$. The bolt $M$ resolving $X_0$ to $X$ is
\be
M=\left[ S^{4n+7}\cap \left(\,\bigcap_{\alpha=1}^k (\mu^\alpha_{\mathbb{R}})^{-1}(0) \cap  (\mu^\alpha_{\mathbb{C}})^{-1}(0)\right) \right]/\,U(1)^n\times SU(2)\,.
\ee
For example, when $n=0$ we have $M=S^7/SU(2)=S^4$ and when $n=1$ with charge vector $(p_1,p_2,p_3)$ one finds $M=\mathbb{WCP}$ with weights $(p_2+p_3,p_1+p_3,p_1+p_2)$ or half of that, when all $p_i$ are even \cite{Acharya:2001gy}. Taking further finite quotients from here covers all of the previous examples.

For this example the base $M$ can be visualized by noting that $U(1)^2\subset Sp(n+2)$ acts on the base and consequently we have a toric fibration $T^2\hookrightarrow B\rightarrow P$ with real 2-dimensional base $P$. By the hyperk\"ahler quotient construction $P$ must be simply connected and therefore takes the general form of a polygon with boundary edges and vertices. At the edges of $P$ the $T^2$ fiber degenerates to a circle, while at the vertices the fiber collapses completely. Fibers therefore trace out a necklace of 2-spheres over the edges which meet at the vertices. With this, when $n\geq 2$, we find the natural generalization of Figures \ref{fig:QFT} and \ref{fig:5DCp2} as schematically depicted in Figure \ref{fig:Polyon}. More precisely, we have $n+2$ 5D SCFTs $\mathcal{T}^{\textnormal{\tiny\!  (M)}}_{\mathbb{C}^3/\Gamma_m}$, each of which comes with up to three local flavor loci. In the $G_2$-holonomy orbifold, some of these flavor loci now reside on compact subspaces, and this generically results in two of the three local flavor symmetries being gauged in 5D (with a position dependent gauge coupling). This gauging involves pairing up neighboring 5D SCFTs and gauging a common diagonal subgroup, and so each 5D theory is generically left with a single ``ungauged'' flavor symmetry (7D SYM on a non-compact 2-cycle). Indeed, the third remains a flavor symmetry in 5D and fills a full fiber of the bundle of ASD 2-forms. With this we generically expect to have $n+2$ singular loci of topology $S^2\times \mathbb{R}$ and $n+2$ loci of topology $\mathbb{R}^3$. In 4D we therefore find a flavor symmetry with $2n+4$ simple Lie algebra flavor factors.

When we collapse the bolt $X\rightarrow X_0$ the 5D gauge loci deform as  $S^2\times \mathbb{R}\rightarrow \mathbb{R}^3\cup \mathbb{R}^3 $ while the topology of the 5D flavor loci is unaltered. The breathing mode therefore spontaneously breaks $n+2$ pairs of 5D $\mathfrak{su}$ gauge symmetries to their diagonal.

Clearly it would be extremely interesting to consider the above construction for non-Abelian groups $H$, to quotient $X$ further by discrete isometries introducing additional singularities or to make contact with the constructions presented in \cite{Braun:2023fqa}. We leave such questions for future work.

\section{Conclusions} \label{sec:CONC}

Junctions and intersections of SQFTs provide a general method for engineering couplings between bulk degrees of freedom and modes localized on lower-dimensional defects. In this paper we have shown how to engineer examples of this sort where the bulk itself is a strongly coupled system. The key geometric ingredient in our construction is the asymptotically conical $G_2$-holonomy orbifold. When the zero-section of this space is of finite size, we can interpret the geometry as 5D SCFTs which are coupled via a diagonal gauging of flavor symmetries. In the limit where the zero-section collapses to zero size, this results in a 4D quasi-SCFT edge mode with non-trivial coupling to the 5D bulk modes. We have used this geometric perspective to extract the non-Abelian flavor symmetries, as well as various discrete higher-form symmetries of the 5D theories and their descent to the 4D system. In the rest of this section we discuss some natural avenues for generalization.

With the structure of the bulk theories in place, it would be quite natural to study the anomaly inflow to the quasi-SCFT degrees of freedom. One approach to extracting this data would be to determine the topological terms of the 5D / 7D bulk theories, and to use this to extract quantities such as various continuous and discrete anomalies.

For the most part, our analysis has centered on a special class of quotients $X / \Gamma$ where $\Gamma$ is an Abelian group with a particularly simple group action on $X$. It would be interesting to study the resulting geometries generated by more general Abelian group actions, as well as genuinely non-Abelian group actions. These will lead to additional novel structures and non-trivial bulk / boundary couplings. A first glance at these is given in Appendix \ref{sec:ExtraQuotients}.

It is also natural to consider IIA string theory on the same quotient space $X / \Gamma$ to engineer 3D $\mathcal{N} = 2$ quasi-SCFTs.
An interesting feature of such examples is that compared with their 4D counterparts, the IR dynamics of 3D $\mathcal{N} = 2$ theories can already exhibit new strong coupling phenomena. A related comment is that type IIB string theory on such backgrounds will likely exhibit different strong coupling dynamics, simply because IIA and IIB on an ADE singularity can result in rather different IR behavior.

While we have primarily focused on $G_2$-holonomy spaces, one could contemplate carrying out a related analysis for M-theory and IIA / IIB on $Spin(7)$-holonomy spaces. This would lead to examples of quasi-SCFTs in 3D $\mathcal{N} = 1$ backgrounds (in the M-theory case) and quasi-SCFTs in 2D with $\mathcal{N} = (1,1)$ and $\mathcal{N} = (0,2)$ supersymmetry for IIA and IIB, respectively.

The condition that we have a complete $G_2$-holonomy space imposes non-trivial restrictions on the ways in which 5D bulk theories can combine to couple to 4D edge modes. For example, in the $G_2$ space, 5D gauging of the flavor symmetries of the 5D SCFTs is also often accompanied by a gauge coupling which depends on the radial position in the AC $G_2$-holonomy metric. This suggests constraints on self-consistent junctions of $D > 4$ theories which can support edge modes. It is tempting to speculate that just as the Swampland program (see \cite{Vafa:2005ui}) imposes non-trivial constraints on when an effective field theory can consistently couple to quantum gravity, there may be ``hard to spot'' bottom up consistency conditions which are straightforward to identify from a top down perspective.\footnote{One precise version of such a conjecture is that any consistent $d$-dimensional QFT must admit a consistent coupling to a $D \geq d$-dimensional theory of gravity \cite{Heckman:2018jxk}, for some choice of $D$. Observe that any string realization of a QFT automatically satisfies this requirement.} To carry this out in the present context would require a general classification of possible singularities in non-compact special holonomy spaces, a challenging topic which would no doubt be interesting for many reasons.


\newpage

\section*{Acknowledgements}

We thank D. Xie for initial collaboration when this project was in a nascent phase
of development. MDZ thanks C. Closset and I. Garc\'{i}a-Etxebarria for several discussions on related topics.
This work was initiated at the ``Physics and Special Holonomy'' meeting
held at KITP in 2019, and BSA, MDZ and JJH thank the organizers for kind hospitality during this meeting.
Some of this work was finalized at the ``Physics and Special Holonomy'' meeting held at KITP in 2023, and BSA and MH thank the
organizers for kind hospitality during this meeting. The work of BSA is supported by a grant from the Simons Foundation (\#488569, Bobby Acharya). The work of MDZ has received funding from the European Research Council
(ERC) under the European Union’s Horizon 2020 research and innovation programme
(grant agreement No. 851931). MDZ also acknowledges support from the Simons Foundation Grant \#888984
(Simons Collaboration on Global Categorical Symmetries).
The work of JJH and ET is supported by DOE (HEP) Award DE-SC0013528. The work of JJH and MH is supported in
part by a University Research Foundation grant at the University of Pennsylvania. The
work of MH is also supported by the Simons Foundation Collaboration grant
\#724069 on ``Special Holonomy in Geometry, Analysis and Physics''.


\appendix
\addtocontents{toc}{\protect\setcounter{tocdepth}{0}}

\section{Continuous $k$-Form Symmetries} \label{sec:appContSym}
In this Appendix we focus on the collection of continuous higher form symmetries of the form $U(1)^{(k)}$ that appear in all of our quotients of the $G_2$-holonomy manifold $\Lambda^2_{\mathrm{ASD}}(S^4)$.\footnote{There may be additional continuous non-Abelian 0-form symmetries after the quotient, but we restrict ourselves to Abelian symmetries in this section. } Specifically, we analyze what are the charged objects under these symmetries, determine whether these symmetries spontaneously break in the resolved phase $\mathrm{Vol}(S^4)\neq 0$, and the fate of these charged objects in the resolved phase.

Recall that we denote by $X_0=\mathrm{Cone}(\mathbb{CP}^3)$ the unresolved $G_2$ cone, and $X = X=\Lambda^2_{\mathrm{ASD}}(S^4)$ the resolved phase of the geometry. We emphasize that while this unquotiented geometry engineers a free chiral multiplet $\Phi$ \cite{Atiyah:2001qf}, the remarks in this subsection generalize to quotients of $X_0$ because the (co)homology classes of $\mathbb{CP}^3$ survive these quotients.

Let $\mathbb{M}_4$ be the 4D spacetime transverse to the $G_2$-holonomy geometry, then we can KK expand the 11D supergravity forms $C_3$ and $C_6$ along $\mathbb{M}_4\times \mathbb{CP}^3$ as:\footnote{We ignore the expansion component $C_3\supset a_3$ because the would-be defect operator is just an M2-brane which would not wrap the radial direction.}
\begin{align}\label{eq:kku1exp}
  & C_3=a_1\wedge \omega_2   \\
  & C_6=b_4\wedge\omega_2 +b_2\wedge \omega_4+b_0\wedge \mathrm{Vol}_{\mathbb{CP}^3}
\end{align}
where $\omega_2\in H^2(\mathbb{CP}^2,\mathbb{Z})$ is the hyperplane class, $\omega_4=\omega^2_2$, and $\mathrm{Vol}_{\mathbb{CP}^3}=\omega^3_2$.

We see that the fields $a_i$ and $b_i$ are background fields for the higher form-symmetries
\begin{equation}\label{eq:abelianksyms}
   U(1)^{(-1)}_{b_0}\times U(1)^{(0)}_{a_1}\times U(1)^{(1)}_{b_2} \times U(1)^{(3)}_{b_4}
\end{equation}
of the 4D SQFT on $M_4$ where here we denote the background fields by the subscripts. Generalizing the argument from \cite{Atiyah:2001qf} to higher-form symmetries, the symmetries that are spontaneously broken in $X_0$ or $X$ are those associated to $\mathbb{CP}^3$ cohomology classes that cannot be extended into the bulk. In other words, the symmetry is spontaneously broken if the pullback of the embedding map
\begin{align}\label{eq:embedmap}
   \iota^*: \; H^i(X)\rightarrow H^i(\mathbb{CP}^3)
\end{align}
is trivial. To understand why, let us specialize to $U(1)^{(0)}$, then $H^2(X)=0$ since $\Lambda^2_{\mathrm{ASD}}(S^4)$ retracts to $S^4$, and we see that we can only extend $C_3=a_1\wedge \omega_2$ from \eqref{eq:kku1exp} into $X$ as $C_3=a_1\wedge \widetilde{\omega}_2$ where $d\widetilde{\omega}_2\neq 0$. The $U(1)^{(0)}$ gauge transformations, $a_1\rightarrow a_1+d\lambda_0$ are no longer symmetries because, $dC_3=da_1\wedge \widetilde{\omega}-a_1\wedge d\widetilde{\omega}$, is no longer invariant. One subtlety that arises for $X_0$ is that while $H^j(X_0)=0$ for $j>0$, we can only extend the expansions \eqref{eq:kku1exp} into a dense subset of $X_0$ by extending $\omega_2$ (and its powers) to a closed form on $X_0^\circ\simeq \mathbb{CP}^3\times \mathbb{R}_{r>\epsilon}$ by demanding that $\omega_2$ be constant along $r$ so as to satisfy $(d_{\mathbb{CP}^3}+d_r)\omega_2=0$. This allows us to conclude that all the symmetries in \eqref{eq:abelianksyms} are indeed symmetries of the unresolved phase.

Now having a criteria for when one of the $U(1)^{(k)}$ symmetries are broken \eqref{eq:embedmap}, we claim that the spontaneously broken symmetries upon resolving to $X$ are
\begin{equation}\label{eq:broksyms}
  \textnormal{Broken Symmetries:}\; \; U(1)^{(0)}.
\end{equation}
This follows from observing that
\begin{equation}\label{eq:homologywidetilde}
  H^*(X)=\{\mathbb{Z}, 0, 0, 0, \mathbb{Z}, 0 ,0, 0\}
\end{equation}
which immediately tells us that the pullback of the embedding map \eqref{eq:embedmap} is trivial, in particular, for $i=2$ and $i=6$. Additionally, the fact that $\iota^*: H^4(X)\rightarrow H^4(\partial
X)$ is a bijection tells us that the $U(1)^{(1)}$ symmetry is unbroken. This latter claim follows from considering the long exact sequence in relative cohomology
\begin{equation}\label{eq:les}
 ... \rightarrow  H^4(X,\partial X) \rightarrow H^4(X)\xrightarrow{\iota^*} H^4(\partial X)\rightarrow H^5(X,\partial X)\rightarrow ...
\end{equation}
and observing that from Poincar\'e-Lefchetz duality $H^4(X,\partial X)=H_3(X)=0$ and $H^5(X,\partial X)=H_2(X)=0$. Notice, importantly, that we excluded $U(1)^{(3)}$ from the list \eqref{eq:broksyms} despite the fact that $b_4\wedge \omega_2 \subset C_6$ and $\omega_2$ could not be expanded into to bulk allowing us to conclude that $U(1)^{(0)}$ is broken. The reason is that the putative NG-boson is a massless 3-form field which does not have a consistent kinetic term in 4D. This is consistent with a higher-form generalization of the Coleman-Mermin-Wagner (CMW) theorem \cite{Lake:2018dqm}.

Briefly commenting on the Nambu-Goldstone (NG) bosons associated to \eqref{eq:broksyms}, first note that $U(1)^{(-1)}$ will not have NG bosons since this would be a $(-1)$-form field and the NG boson of $U(1)^{(3)}$ is non-propagating since there is no kinetic term one can write down for a 3-form in 4D.\footnote{This may induce some topological effects on the theory whose precise effects we leave to future work.} So we are left with the NG boson for $U(1)^{(0)}$ which was already addressed in \cite{Atiyah:2001qf}. Unsurprisingly, it is the phase of the scalar $\Phi$ (or $\mathcal{O}$ for the case of 4D (Quasi)-SCFTs). This arises because a gauge transformation of $C_3$ that is constant along $M_4$ (and setting $a_3$=0) acts as $\delta C_3=d(a_1\wedge \widetilde{\omega}_2)=-a_1\wedge d\widetilde{\omega}_2$, which then transforms the phase of $\Phi$ as
\begin{equation}
\delta(e^{i\int_{\mathbb{R}^3}C_3})=e^{i\int_{\mathbb{R}^3}d\widetilde{\omega}_2}=e^{i\int_{S^2}\widetilde{\omega}_2}\neq 0.
\end{equation}

\begin{figure}
\begin{center}
\includegraphics[scale=0.3]{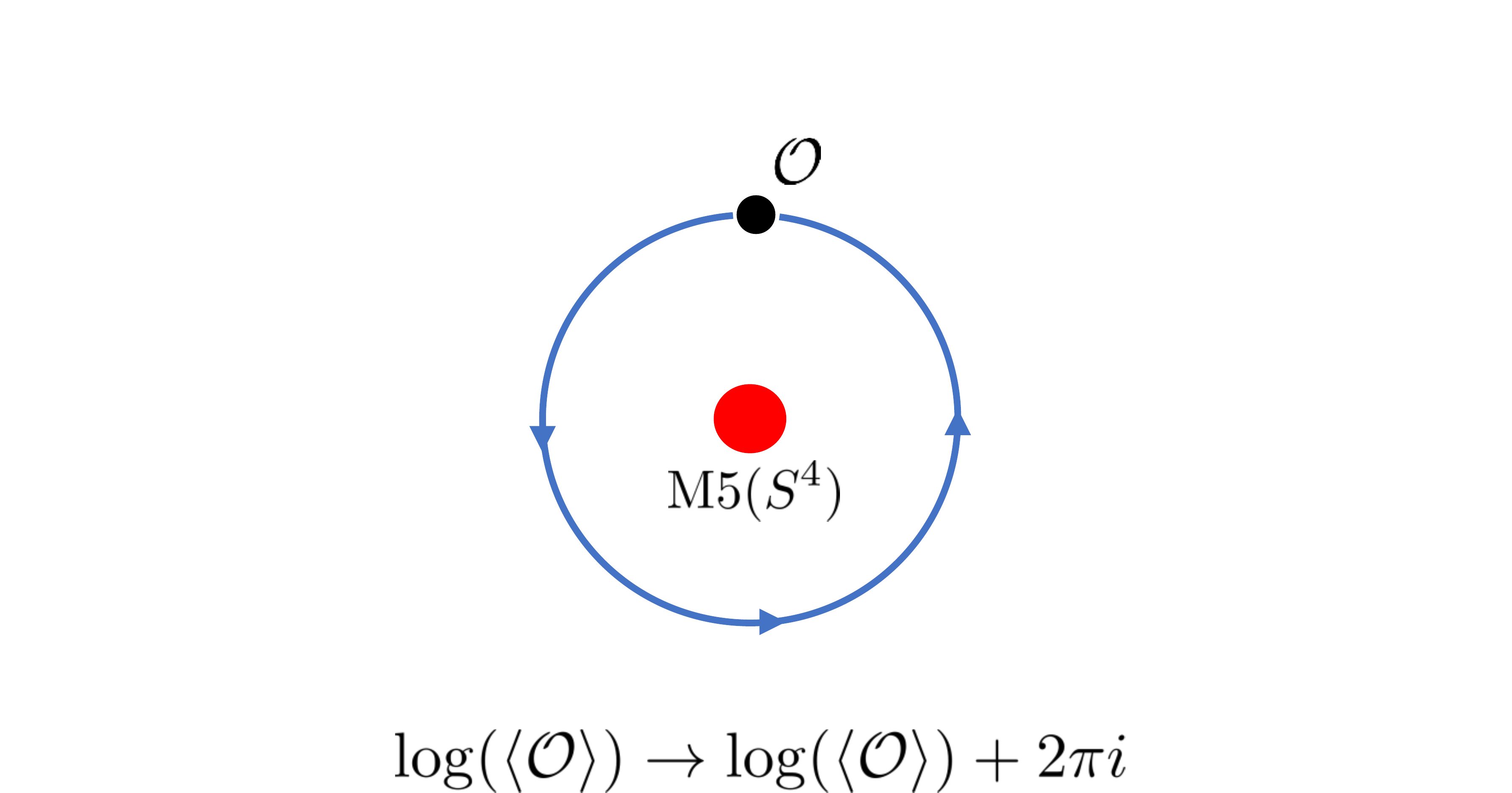}
\end{center}
\caption{Illustration of transverse directions of string defect in the 4D theory that arises from an M5 wrapped on the zero-section of $\Lambda^2_{ASD} (S^4)/\Gamma$. This object has tension $T \sim \langle \mathcal{O}\rangle$, is charged under $U(1)^{(1)}$ symmetry, and induces a monodromy in the phase of $\langle \mathcal{O}\rangle$ in the plane transverse to the string. }\label{fig:m5stringdefect}
\end{figure}

We now move on to discuss the charged objects of the $U(1)^{(k)}$ symmetries. We restrict ourselves to $k=0$ and $k=1$ in order to avoid the subtley with the CMW theorem, and by the fact that $(-1)$-form symmetries do not have charged defect operators. For the unresolved geometry, $X_0$, the charged objects associated with the unbroken symmetries are then
\begin{align}
     U(1)^{(0)}:   \mathrm{M2}(\mathrm{Cone}(\gamma_2)) \quad U(1)^{(1)}:  \mathrm{M5}(\mathrm{Cone}(\gamma_4))
\end{align}
where $\gamma_2$ and $\gamma_4$ are generators of $H_2(\mathbb{CP}^3,\mathbb{Z})$ and $H_4(\mathbb{CP}^3,\mathbb{Z})$ respectively. Meanwhile in the resolved geometry, $X$, there are dynamical objects charged under the unbroken 1-form symmetry:
\begin{align}
    U(1)^{(1)}: \; \; \mathrm{M5}(S^4).
\end{align}
Notice that because the $\mathrm{M5}$ brane wraps a finite size $S^4$, this creates a finite tension string-like excitation in the 4D theory proportional to $\langle \Phi \rangle$ (or $\langle \mathcal{O} \rangle$ for 4D quasi-SCFTs). We can see that this M5 induces a monodromy in the phase of $\Phi$ similar to the Nielsen-Olesen vortex for Abelian Higgs theory\footnote{Of course, the key difference here is that the free chiral multiplet is still gapless after $\Phi$ gains a vev.} by the fact that $\int_{N_4} G_4=N_{M5}=1$ for any four-manifold $N_4$ that links the worldvolume in the M5 in question. We see that $N_4=S^1\times \mathbb{R}^3$ where $S^1$ links the M5 worldvolume inside the 4D spacetime (see Figure \ref{fig:m5stringdefect}) and the $\mathbb{R}^3$ is a fiber of $\Lambda^2_{\mathrm{ASD}}(S^4)$, (which intersects the zero-section once) is one such example and thus causes the monodromy.

\section{Further Quotients}
\label{sec:ExtraQuotients}

In the main text we focussed on isometric quotients of asymptotically conical $G_2$-holonomy manifolds, with no compact 3-cycles, by finite Abelian groups. In this Appendix we consider more general types of quotients. We given an example of a finite non-Abelian quotient of $\Lambda_{\textnormal{ASD}}^2(S^4)$ together with examples of Abelian quotients of asymptotically conical $G_2$-holonomy manifolds with compact 3-cycles, more precisely finite Abelian quotients of the spinor bundle $\mathbb{S}(S^3)$.

\subsection{Non-Abelian Quotients of $\Lambda^{2}_{\mathrm{ASD}}(S^4)$}
\label{sec:NonAb}

In sections \ref{sec:CONEHEAD} and \ref{sec:EXAMP} we considered the case of finite, Abelian group actions $\Gamma\subset Sp(2)/\mathbb{Z}_2 \simeq SO(5)$ on the $G_2$-holonomy space $\Lambda_{\textnormal{ASD}}^2(S^4)$. One common feature of these constructions, and also in those considered in previous subsection of this section, were non-compact codimension-6 fixed point loci. These run radially, as in the Quadrion and Sexion, and were fundamentally what enabled an interpretation of these geometries as 4D edge modes coupled to a 5D bulk.

When considering finite, non-Abelian groups $\Gamma\subset SO(5)$ codimension-6 fixed point loci are not necessarily non-compact. In this case we first consider the fixed point loci of individual elements $\Gamma$ separately, generating a finite Abelian subgroup of $\Gamma$ and therefore the previously considered characterization of fixed point loci applies. In particular, the codimension-4 fixed point loci are topologically either $\mathbb{R}^3$ or $S^2\times \mathbb{R}$. Compact codimension-6 loci now arise at intersections of the latter.

Locally, the compact codimension-6 loci therefore take the form $S^1\times \mathbb{C}^3/\Gamma'$ for some finite $\Gamma'\subset SU(3)$. The circle $S^1$ is contained in the base $S^4$ and therefore collapses in the conical limit. The interface theory is therefore some 5D $\mathcal{N}=1$ theory compactified on a circle with some configuration of twisted background profiles such that the resulting theory preserves half of the supersymmetry, yielding a 4D $\mathcal{N}=1$ theory.

It would be very interesting to exhaust the finite subgroups of $SO(5)$ and map out the full list of physical systems
which can be constructed in this way. However, we leave a more complete treatment for future work and
concentrate on an example with many of the salient features of non-Abelian quotients.

\paragraph{Example: }

Consider the finite group $\Gamma_{D_{K+2}}\subset Sp(1) \simeq SU(2)$, where $\Gamma_{D_{K+2}}$ is the binary dihedral group of order $4K$ generated by the matrices
\begin{equation}
   \lambda =\left(\begin{array}{cc}
        \exp\!\left(\frac{i\pi}{K}\right) & 0  \\
        0 & \exp\!\left(-\frac{i\pi}{K}\right)
    \end{array}\right)\,, \qquad \mu =\left(\begin{array}{cc}
        0 & i  \\
        i & 0
    \end{array}\right)\,,
\end{equation}
acting on the column vector with entries $(z_1,z_2)$ via left multiplication. In terms of the quaternion $Q=z_1+z_2j$, this action is obtained via quaternionic right multiplication by setting $\lambda=\exp\!\left(i\pi/K\right)$ and $\mu=ij=k$.

We take a diagonal embedding of $\Gamma_{D_{K+2}}$ in $Sp(1)_{(1)}\times Sp(1)_{(2)} \subset Sp(2)$.
With this, the action on the base coordinates of the North pole patch as
\begin{equation}
    \begin{aligned}
            \lambda:~~&Q_2^{-1}Q_1=v_1+v_2j \quad \rightarrow \quad\lambda^{-1}Q_2^{-1}Q_1 \lambda= \lambda^2 v_1 +v_2j\\
    \mu:~~&Q_2^{-1}Q_1 =  v_1+ v_2j\quad\rightarrow\quad\mu^{-1} Q_2^{-1}Q_1\mu =\bar  v_1+\bar v_2j
     \end{aligned}
\end{equation}
The group action on the boundary $\mathbb{CP}^3$ is:
\begin{equation}\begin{aligned}
    \lambda\,: &\qquad [Z_1 , Z_2 , Z_3 , Z_4  ]~\rightarrow~[\lambda Z_1 , \lambda^{-1}Z_2 , \lambda Z_3 , \lambda^{-1}Z_4  ] \\
    \mu\,: &\qquad [Z_1 , Z_2 , Z_3 , Z_4  ]~\rightarrow~[ i Z_2 , i Z_1 , i Z_4 , i Z_3  ]\,.
    \end{aligned}
\end{equation}
Here we should note that since $\lambda^{K}=\mu^2=-1$ does not act, the geometry is only acted on by the quotient group $\Gamma\equiv \Gamma_{D_{k+2}}/\mathbb{Z}_2$. We also see that the generators satisfy $\mu \lambda=\lambda^{-1}\mu$ and therefore every element can be represented as $\mu^{n}\lambda^m$ which thus fully parameterize the action. On the North and South pole patch of $\Lambda_{\textnormal{ASD}}^2(S^4)$ the full action is
\begin{eqnarray}
\lambda\,:&\quad (v_{1},v_{2},v_{3},t) &\rightarrow~~ (\lambda^2 v_{1},v_{2},\lambda^{-2}v_{3},t) \\
\mu\,:&\quad (v_{1},v_{2},v_{3},t) &\rightarrow~~ ( \bar v_{1}, \bar v_{2},\bar v_{3},-t)\\
\lambda\,:&\quad (v_{1}',v_{2}',v_{3}',t') &\rightarrow~~ (\lambda^{-2} v_{1}',v_{2}',\lambda^{2}v_{3}',t') \\
\mu\,:&\quad (v_{1}',v_{2}',v_{3}',t') &\rightarrow~~ (  \bar v_{1}', \bar v_{2}', \bar v_{3}',-t')\,.
\end{eqnarray}
To extract the fixed point loci, we focus on the loci fixed by individual elements of the group action:
\begin{itemize}
\item $ \lambda^m\,:$ The Abelian subgroup generated by this element is of order $K/\textnormal{gcd}(K,m)$. For different values of $m$ these elements commute and their fixed point loci therefore coincide. The fixed point locus on the North patch of the base $S^4$ is $v_1=0$. Within the $G_2$-holonomy orbifold the fixed point locus is extended radially to a copy of $S^2\times \mathbb{R}$.
\item $ \mu \lambda^m\,:$ The Abelian subgroup generated by this element is of order $2$. The fixed point locus on the North patch of the base $S^4$ is $\textnormal{Im}\,v_2=0$ and $v_1=c\times \lambda^{-m}$ with real constant $c$. Within the $G_2$-holonomy orbifold the fixed point locus is extended radially to a copy of $S^2\times \mathbb{R}$.

\item $\mu\,:$ The Abelian subgroup generated by this element is of order $2$. The fixed point locus on the North patch of the base $S^4$ is $\textnormal{Im}\,v_1=\textnormal{Im}\,v_2=0$. Within the $G_2$-holonomy orbifold the fixed point locus is extended radially to a copy of $S^2\times \mathbb{R}$.
\end{itemize}
These fixed point loci give the following singularities:
\begin{itemize}
\item $\mathfrak{su}(n)$ locus on $\mathbb{R}_{\geq 0}\times S^2/\Gamma$\,: Note that all order 2 elements act as $(v_2,t)\rightarrow (v_2,-t)$ and therefore the fixed point locus is folded in half to an $\mathfrak{su}(n)$ locus in the $G_2$-holonomy orbifold with a single asymptotic boundary.
\item $\mathfrak{su}(2)$ locus on $\mathbb{R}\times S^2/\Gamma$\,: Note that $\Gamma$ groups all fixed point loci of order 2 into an orbit. In the $G_2$-holonomy orbifold there is a single $\mathfrak{su}(2)$ locus.
\end{itemize}
Here $ S^2/\Gamma \subset S^4/\Gamma$. As written above the action is not manifestly holomorphic. However, if one defines new coordinates as
\begin{align}
(w_1, w_2, w_3) &\equiv \Big(\textnormal{Re}(v_1)+i\textnormal{Re}(v_3) , \textnormal{Im}(v_1)+i\textnormal{Im}(v_3)  , \textnormal{Im}(v_2)  + it\Big) \\
(w_1', w_2', w_3') &\equiv \Big(\textnormal{Re}(v_1')+i\textnormal{Re}(v_3') , \textnormal{Im}(v_1')+i\textnormal{Im}(v_3')  , \textnormal{Im}(v_2')  + it'\Big)
\end{align}
then the group acts holomorphically, with $\theta=2\pi i/K$,
\be\label{eq:Action}
\lambda =
\begin{pmatrix}
\cos\theta & \sin\theta & 0\\
-\sin\theta & \cos\theta & 0\\
0 & 0 & 1
\end{pmatrix}
, \quad \mu =
\begin{pmatrix}
1 & 0 & 0\\
0 & -1 & 0\\
0 & 0 & - 1
\end{pmatrix}\,.
\ee

Let us discuss the physics of the setup. For this purpose first note that the fixed points intersect along the circle
 \be\label{eq:circle}
S^1=\{ v_1,v_1'=0\,;~ \textnormal{Im}\,v_2,\textnormal{Im}\,v_2'=0  \}\,.
\ee
As such, $\Lambda_{\textnormal{ASD}}^2(S^4)/\Gamma$ contains a compact locus of codimension-6 singularities modelled on $S^1\times \mathbb{C}^3/\Gamma'$ with $\Gamma'$ as in $\eqref{eq:Action}$. We now consider the Betti numbers of the crepant resolution of $\mathbb{C}^3/\Gamma'$. All elements of $\Gamma'$ are junior, they have age 1, and therefore $b_4=0$. In this example we therefore do not find mutually non-local massless degrees of freedom localized along $S^1$.

Let us now consider symmetry enhancement / breaking as we go from the conical limit $X_0$ to the resolved geometry $X$. In going from the
singular cone to the resolved phase, we find the breaking pattern:
\be
\mathfrak{su}(2)\times \mathfrak{su}(2) \times \mathfrak{su}(n)~\rightarrow~ \mathfrak{su}(2)_{\textnormal{diag}}\times \mathfrak{su}(n)\,,
\ee
the $\mathfrak{su}(n)$ factor is unbroken due to the boundary containing only a single $S^2/\Gamma$ worth of $\mathfrak{su}(n)$ singularities, or equivalently, due to the singular locus extending along the half line $\mathbb{R}_{\geq 0}$.

\subsection{Quotients of the Spinor Bundle $\mathbb{S}(S^3)$}

We now give an example of a quotient space with a compact 3-cycle.
The spinor bundle $X=\mathbb{S}(S^3)$ over the three-sphere has isometries Isom$(X)=SU(2)^3$. The boundary $\partial X=S^3\times S^3$ can be parameterized by to quaternions $(x,y)$ individually of unit norm. In this parameterization the isometries act as
\be
(x,y)\mapsto (pxr,qyr)
\ee
where $p,q,r$ are unit quaternions parameterizing Isom$(X)=SU(2)_p\times SU(2)_q\times SU(2)_r$. We take $x$ to parameterize the $S^3$ in the fibers of the spinor bundles, such three-spheres are topologically trivial in $X$. We let $y$ parameterize the base $S^3$ of the spinor bundle.

In general finite Abelian subgroups take the form
\be
\Gamma\cong \mathbb{Z}_{K_1}\times  \mathbb{Z}_{K_2}\times  \mathbb{Z}_{K_3}
\ee
and are further specified by the embedding
\be
\Gamma ~\hookrightarrow~ SU(2)_p\times SU(2)_q\times SU(2)_r
\ee
in which the generator of $ \mathbb{Z}_{K_i}$ is mapped onto the triple $(\omega^{l_i},\omega^{m_i},\omega^{n_i})$ where $\omega$ generates a finite Abelian subgroup of $SU(2)$ of order $K_i$ and $l_i,m_i,n_i\in\mathbb{Z}$ and $i=1,2,3$.

Let us specialize to the subclass of finite Abelian quotients with $\Gamma\cong \mathbb{Z}_P\times \mathbb{Z}_Q\times \mathbb{Z}_R$ of order $PQR$ where $\mathbb{Z}_P\subset SU(2)_p$ and $ \mathbb{Z}_Q\subset SU(2)_q$ and $ \mathbb{Z}_R\subset SU(2)_r$. Let us further assume that $R=\textnormal{gcd}(P,Q)$ and denote the generators respectively by $\gamma_p,\gamma_q,\gamma_r$ in obvious notation. We have the following fixed point loci:

\begin{itemize}
\item $ (1,\gamma_q^{Q/R},\gamma_r)$ has fixed points along a circle $S^1_1$ of the zero section, the corresponding subgroup of order $R$ gives a fixed points of order $R$ and codimension-6.
\item $  (\gamma_p^{P/R} ,\gamma_q^{Q/R},\gamma_r)$ fixes a circle $S^1_1$ in the base of the spinor bundle and above every point on that circle a plane $F_1$ in the fiber direction, the subgroup of order $R$ generated by this element gives fixed points of order $R$ and codimension-4.
\item $  (-\gamma_p^{P/R} ,\gamma_q^{Q/R},\gamma_r)$ fixes a circle $S^1_1$ in the base of the spinor bundle and above every point on that circle a plane $F_2$ in the fiber direction, the subgroup of order $R$ generated by this element gives fixed points of order $R$ and codimension-4.

\item $ (1,-\gamma_q^{Q/R},\gamma_r)$ has fixed points along a circle $S^1_2$ of the zero section, the corresponding subgroup of order $R$ gives a fixed points of order $R$ and codimension-6.
\item $  (\gamma_p^{P/R} ,-\gamma_q^{Q/R},\gamma_r)$ fixes a circle $S^1_2$ in the base of the spinor bundle and above every point on that circle a plane $F_2$ in the fiber direction, the subgroup of order $R$ generated by this element gives fixed points of order $R$ and codimension-4.
\item $  (-\gamma_p^{P/R} ,-\gamma_q^{Q/R},\gamma_r)$ fixes a circle $S^1_2$ in the base of the spinor bundle and above every point on that circle a plane $F_1$ in the fiber direction, the subgroup of order $R$ generated by this element gives fixed points of order $R$ and codimension-4.
\item $\gamma_p\in SU(2)_p$ fixes the zero section of the spinor bundle, the subgroup $\mathbb{Z}_P$ gives a fixed points of order $P$ and codimension-4 along the zero section.
\item $ \gamma_q\in SU(2)_q$ acts fixed point free.
\item $ \gamma_r\in SU(2)_r$ acts fixed point free.
\end{itemize}
It therefore follows that we have the singularities
\begin{itemize}
\item Codimension-6 modelled on $\mathbb{C}^3/\mathbb{Z}_P\times \mathbb{Z}_R$ along two linking, disjoint circles in the zero section mod $\Gamma$.
\item Codimension-4 modelled on $\mathbb{C}^2/\mathbb{Z}_P$ along the zero section mod $\Gamma$, this locus intersects both the codimension-6 circles.
\item Codimension-4 modelled on $\mathbb{C}^2/\mathbb{Z}_R$ with one compact circular direction and two non-compact direction. In total there are four of these loci, pairs of these intersect in the codimension-6 circles, when they do not intersect at the zero section they are otherwise disjoint.
\end{itemize}

Crucially, the locus of $A_{P-1}$ singularities is compact and we therefore have a gauge symmetry in 4D (not 5D, as in previous examples). Further, the codimension-6 loci supporting 5D SCFTs are compact! However, we now have a compact 3-cycle which can be wrapped by Euclidean M2-branes, so there are quantum corrections to the classical geometry.
Clearly, these examples exhibit much interesting physics, we leave a detailed analysis to future work.

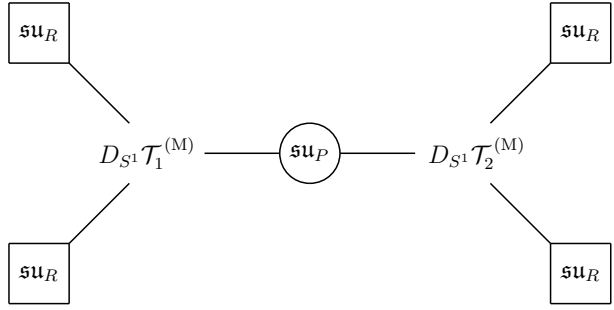
\begin{figure}[t!]
\centering
\scalebox{0.8}{\begin{tikzpicture}
	\begin{pgfonlayer}{nodelayer}
		\node [style=none] (0) at (-0.5, 0) {};
		\node [style=none] (1) at (0.5, 0) {};
		\node [style=none] (2) at (0, 0.5) {};
		\node [style=none] (3) at (0, -0.5) {};
		\node [style=none] (4) at (0, 0) {$\mathfrak{su}_P$};
		\node [style=none] (5) at (-2.5, 0) {$D_{S^1}\mathcal{T}_1^{\tiny\textnormal{\scriptsize\,(M)}}$~~~~};
		\node [style=none] (6) at (2.5, 0) {~~~~$D_{S^1}\mathcal{T}_2^{\tiny \textnormal{\,(M)}}$};
		\node [style=none] (7) at (-5, 2.5) {};
		\node [style=none] (8) at (-4, 2.5) {};
		\node [style=none] (9) at (-4, 1.5) {};
		\node [style=none] (10) at (-5, 1.5) {};
		\node [style=none] (11) at (-4.5, 1.5) {};
		\node [style=none] (12) at (-4.5, 2) {$\mathfrak{su}_R$};
		\node [style=none] (13) at (-5, -1.5) {};
		\node [style=none] (14) at (-4, -1.5) {};
		\node [style=none] (15) at (-4, -2.5) {};
		\node [style=none] (16) at (-5, -2.5) {};
		\node [style=none] (17) at (-4.5, -2.5) {};
		\node [style=none] (18) at (-4.5, -2) {$\mathfrak{su}_R$};
		\node [style=none] (19) at (4, 2.5) {};
		\node [style=none] (20) at (5, 2.5) {};
		\node [style=none] (21) at (5, 1.5) {};
		\node [style=none] (22) at (4, 1.5) {};
		\node [style=none] (23) at (4.5, 1.5) {};
		\node [style=none] (24) at (4.5, 2) {$\mathfrak{su}_R$};
		\node [style=none] (25) at (4, -1.5) {};
		\node [style=none] (26) at (5, -1.5) {};
		\node [style=none] (27) at (5, -2.5) {};
		\node [style=none] (28) at (4, -2.5) {};
		\node [style=none] (29) at (4.5, -2.5) {};
		\node [style=none] (30) at (4.5, -2) {$\mathfrak{su}_R$};
		\node [style=none] (31) at (-3, 0.5) {};
		\node [style=none] (32) at (-3, -0.5) {};
		\node [style=none] (33) at (3, -0.5) {};
		\node [style=none] (34) at (3, 0.5) {};
		\node [style=none] (35) at (-1.75, 0) {};
		\node [style=none] (36) at (1.75, 0) {};
	\end{pgfonlayer}
	\begin{pgfonlayer}{edgelayer}
		\draw [style=ThickLine, bend right=45] (3.center) to (1.center);
		\draw [style=ThickLine, bend right=45] (1.center) to (2.center);
		\draw [style=ThickLine, bend right=45] (2.center) to (0.center);
		\draw [style=ThickLine, bend right=45] (0.center) to (3.center);
		\draw [style=ThickLine] (8.center) to (9.center);
		\draw [style=ThickLine] (10.center) to (9.center);
		\draw [style=ThickLine] (10.center) to (7.center);
		\draw [style=ThickLine] (7.center) to (8.center);
		\draw [style=ThickLine] (14.center) to (15.center);
		\draw [style=ThickLine] (16.center) to (15.center);
		\draw [style=ThickLine] (16.center) to (13.center);
		\draw [style=ThickLine] (13.center) to (14.center);
		\draw [style=ThickLine] (20.center) to (21.center);
		\draw [style=ThickLine] (22.center) to (21.center);
		\draw [style=ThickLine] (22.center) to (19.center);
		\draw [style=ThickLine] (19.center) to (20.center);
		\draw [style=ThickLine] (26.center) to (27.center);
		\draw [style=ThickLine] (28.center) to (27.center);
		\draw [style=ThickLine] (28.center) to (25.center);
		\draw [style=ThickLine] (25.center) to (26.center);
		\draw [style=ThickLine] (9.center) to (31.center);
		\draw [style=ThickLine] (32.center) to (14.center);
		\draw [style=ThickLine] (35.center) to (0.center);
		\draw [style=ThickLine] (1.center) to (36.center);
		\draw [style=ThickLine] (22.center) to (34.center);
		\draw [style=ThickLine] (33.center) to (25.center);
	\end{pgfonlayer}
\end{tikzpicture}
}
\caption{4D Quiver for $\mathbb{S}(S^3)/\Gamma$ with $\Gamma\cong \mathbb{Z}_P\times \mathbb{Z}_Q\times \mathbb{Z}_R$ and $R=\textnormal{gcd}(P,Q)$, here $D_{S^1}$ denotes circle reduction. The 5D SCFTs $\mathcal{T}_{1,2}^{\scriptsize (\textnormal{M})}$ are modelled on $\mathbb{C}^3/\Gamma'_{1,2}$ where $\Gamma'_{1,2}\cong \mathbb{Z}_P\times \mathbb{Z}_R$ are isomorphic but realize distinct subgroups of $\Gamma$. The circular (rectangular) node indicates a (non-)compact codimension-4 singular locus in the geometry.}
\label{fig:Last}

\end{figure}

To close this section we give a 4D (not 5D) quiver picture of the geometry, Figure \ref{fig:Last}. Note that the gauging, as described by how the compact codimension-4 locus stretches between the codimension-6 loci only preserves 4D $\mathcal{N}=1$ supersymmetry.

\section{Homology Group Computations}\label{app:homologygrps}

In this Appendix we collect the various homology group computations used to calculate the generalized symmetries in an electric polarization. Additional care is needed in interpreting the magnetic polarization, a topic we defer to future work.

To frame the analysis to follow, we recall the various geometric objects whose topological properties are under consideration in this Appendix. We consider in detail the bundle of anti-self-dual 2-forms
\begin{equation}
   X=\Lambda^2_{\textnormal{ASD}}(M)/\Gamma\,,\qquad  M=S^4 \,,
\end{equation}
with asymptotic boundary
$
 \partial X=\mathbb{CP}^3/\Gamma
$
which has generic 2-sphere fiber. The boundary is singular with singular locus
\begin{equation}
    \textnormal{Sing}(\partial X)=\Sigma\,.
\end{equation}
The singularities $\Sigma$ consist of codimension-4 and -6 singularities supported along wedge sums of 2-spheres and at points respectively. The complement
\begin{equation}
    \partial X^\circ= \partial X\setminus \Sigma
\end{equation}
is the smooth boundary. In this Appendix we compute the homology groups
\begin{equation}
  H_*(X)\cong H_*(M/\Gamma)\,, \qquad H_*(X,\partial X)/H_*(X)\,,\qquad H_*(\partial X)\,, \qquad H_*(\partial X^\circ)\,.
\end{equation}
All homology groups have integer coefficients.\medskip

Ultimately all homology groups will be computed via an application of the Mayer-Vietoris
sequence with covering given by the North and South pole patch with coordinates:
\begin{eqnarray}
    \textnormal{North Patch }W_N\,:&\qquad (v_1,v_2,v_3,t)\\
    \textnormal{South Patch } W_S\,:&\qquad (v_1',v_2',v_3',t')
\end{eqnarray}
The zero section $v_3=v_3'=t=t'=0$ is parameterized by coordinates $v_1,v_2$ and $v_1',v_2'$. Let us introduce the coordinate $1-y^2=|v_1|^2+|v_2|^2$ and similarly $y'$ which parameterize the radius of the unit ball in each copy of $\mathbb{C}^2$. Then the coordinates on the zero section are related as
\begin{equation}\label{eq:basetransition}
    v_1'=\frac{\bar v_1}{1-y^2}\,, \qquad  v_2'=- \frac{v_2}{1-y^2}\,,
\end{equation}
along the 3-spheres $y,y'=0$. Alternatively, $(v_1+v_2j)^{-1}\equiv v_1'+v_2'j$ in quaternionic formulation. The total space is then glued together at $y=y'=0$ from two halves which consist of the fibers of the bundle $\Lambda_{\textnormal{ASD}}^-(S^4)$ projecting to the two unit balls above. The transition function on the fiber coordinates follows form considering anti-self-dual 2-forms. The gluing region are the fibers projecting to $y=y'=0$ and from \eqref{eq:basetransition} it follows that the gluing is such that it realizes a degree $-1$ map on the radial $S^2$ shells of the identified $\mathbb{C}\times \mathbb{R}$ fibers.

Let us recall the Abelian quotients, with $\Gamma=\mathbb{Z}_K$ and $\zeta=\exp(2\pi i/K)$,
\begin{eqnarray}
W_N\,:&\quad (v_{1},v_{2},v_{3},t) &\rightarrow ~~(\zeta ^{c}v_{1},\zeta ^{d}v_{2},\zeta
^{-c-d}v_{3},t) \\
W_S\,:&\quad  (v_{1}^{\prime },v_{2}^{\prime },v_{3}^{\prime },t^{\prime }) &\rightarrow
~~(\zeta ^{-c}v_{1}^{\prime },\zeta ^{d}v_{2}^{\prime },\zeta
^{c-d}v_{3}^{\prime },t^{\prime }).
\end{eqnarray}
for which we considered three cases in the main text,
\begin{equation}\label{eq:BaseHomo}
\begin{aligned}
  \textnormal{Case 1}:&\quad   (c,K)=(d,K)=(c+d,K)=(c-d,K)=1\\
  \textnormal{Case 2}:&\quad  (c,K)=(c+d,K)=(c-d,K)=1\,, \quad K=md \\
  \textnormal{Case 3}:& \quad   c=-d=1\,,
\end{aligned}
\end{equation}
and the Abelian quotients, with $\Gamma=\mathbb{Z}_K\times \mathbb{Z}_L$ and $\omega=\exp(2\pi i/K),\eta=\exp(2\pi i/L)$,
\begin{eqnarray}
(v_{1},v_{2},v_{3},t) &\rightarrow &(\eta ^{a-c}\omega ^{b-d}v_{1},\eta
^{-a-c}\omega ^{-b-d}v_{2},\eta ^{2c}\omega ^{2d}v_{3},t) \\
(v_{1}^{\prime },v_{2}^{\prime },v_{3}^{\prime },t^{\prime }) &\rightarrow
&(\eta ^{c-a}\omega ^{d-b}v_{1}^{\prime },\eta ^{-a-c}\omega
^{-b-d}v_{2}^{\prime },\eta ^{2a}\omega ^{2b}v_{3}^{\prime },t^{\prime }).
\end{eqnarray}
for which we considered one example in the main text,
\begin{equation}\label{eq:BaseHomo2}
\begin{aligned}
  \textnormal{Case 4}:& \quad   a=c\,, ~b=-d\,.
\end{aligned}
\end{equation}

Let us compute the homology groups of the folded zero section $S^4/\Gamma$ first. By noting that $S^4/\Gamma$ is a suspension of $S^3/\Gamma$ along the coordinates $y,y'$ we find
\begin{equation}
\begin{aligned}
  \textnormal{Case 1}:&\quad   H_*(S^4/\Gamma)\cong \{\mathbb{Z},0,\mathbb{Z}_K,0,\mathbb{Z} \} \\
  \textnormal{Case 2}:&\quad   H_*(S^4/\Gamma)\cong \{\mathbb{Z},0,\mathbb{Z}_m,0,\mathbb{Z} \} \\
  \textnormal{Case 3}:& \quad   H_*(S^4/\Gamma)\cong \{\mathbb{Z},0,\mathbb{Z}_K,0,\mathbb{Z} \} \\
   \textnormal{Case 4}:& \quad   H_*(S^4/\Gamma)\cong \{\mathbb{Z},0,0,0,\mathbb{Z} \}
\end{aligned}
\end{equation}
and therefore in all but the last case we find massive modes from M2-branes wrapped on torsional 2-cycles which become massless when the $S^4/\Gamma$ is contracted to a point.

\begin{figure}
    \centering
    \scalebox{0.8}{
    \begin{tikzpicture}
	\begin{pgfonlayer}{nodelayer}
		\node [style=none] (0) at (0, 2) {};
		\node [style=none] (1) at (0, -2) {};
		\node [style=none] (2) at (-3, 0) {};
		\node [style=none] (3) at (3, 0) {};
		\node [style=none] (4) at (-0.75, 0) {};
		\node [style=none] (5) at (0.75, 0) {};
		\node [style=none] (6) at (-0.75, 3.75) {};
		\node [style=none] (7) at (0.75, 3.75) {};
		\node [style=none] (8) at (0, 4.5) {};
		\node [style=none] (9) at (0, 3) {};
		\node [style=none] (10) at (0, 2.75) {};
		\node [style=none] (11) at (0, 2.25) {};
		\node [style=none] (12) at (0, -2.75) {$S^3/\Gamma$};
		\node [style=none] (13) at (-4.5, 0) {$S^4/\Gamma\,:$};
		\node [style=none] (14) at (-3, -3.25) {};
		\node [style=none] (15) at (3, -3.25) {};
		\node [style=none] (16) at (-3, -3.75) {$y'=1$};
		\node [style=Circle] (17) at (3, -3.25) {};
		\node [style=Circle] (18) at (-3, -3.25) {};
		\node [style=none] (19) at (3, -3.75) {$y=1$};
		\node [style=Circle] (20) at (0, -3.25) {};
		\node [style=none] (21) at (-1.5, -1) {$W_S$};
		\node [style=none] (22) at (1.5, 1) {$W_N$};
		\node [style=none] (23) at (0, 1) {$V_1$};
		\node [style=none] (24) at (0, -1) {$V_2$};
		\node [style=none] (25) at (-1.375, 3.75) {$S^2$};
		\node [style=none] (26) at (-0.25, 0) {};
		\node [style=none] (27) at (0.25, 0) {};
		\node [style=Star] (28) at (3, 0) {};
		\node [style=Star] (29) at (-3, 0) {};
		\node [style=none] (30) at (1.5, 0) {$T^2/\Gamma$};
		\node [style=none] (31) at (5, 0) {};
	\end{pgfonlayer}
	\begin{pgfonlayer}{edgelayer}
		\draw [style=ThickLine, in=90, out=0] (0.center) to (3.center);
		\draw [style=ThickLine, in=0, out=-90] (3.center) to (1.center);
		\draw [style=ThickLine, in=-90, out=180] (1.center) to (2.center);
		\draw [style=ThickLine, in=-180, out=90] (2.center) to (0.center);
		\draw [in=-135, out=90] (4.center) to (0.center);
		\draw [in=90, out=-45] (0.center) to (5.center);
		\draw [in=45, out=-90] (5.center) to (1.center);
		\draw [in=-90, out=135] (1.center) to (4.center);
		\draw [bend right=45] (4.center) to (5.center);
		\draw [bend left=45] (4.center) to (5.center);
		\draw [style=ThickLine, bend left=45] (6.center) to (8.center);
		\draw [style=ThickLine, bend left=45] (8.center) to (7.center);
		\draw [style=ThickLine, bend left=45] (7.center) to (9.center);
		\draw [style=ThickLine, bend left=45] (9.center) to (6.center);
		\draw [bend right] (6.center) to (7.center);
		\draw [style=DottedLine, bend left] (6.center) to (7.center);
		\draw [style=ArrowLineRight] (10.center) to (11.center);
		\draw (14.center) to (15.center);
		\draw [bend left] (27.center) to (26.center);
		\draw [bend left] (26.center) to (27.center);
	\end{pgfonlayer}
\end{tikzpicture}
}
    \caption{Sketch of the decomposition used for various Mayer-Vietoris exact sequences. We show $\partial X$ fibered by $Y_\Gamma$ over the interval parameterized by $y,y'$.}
    \label{fig:Decomposition}
\end{figure}
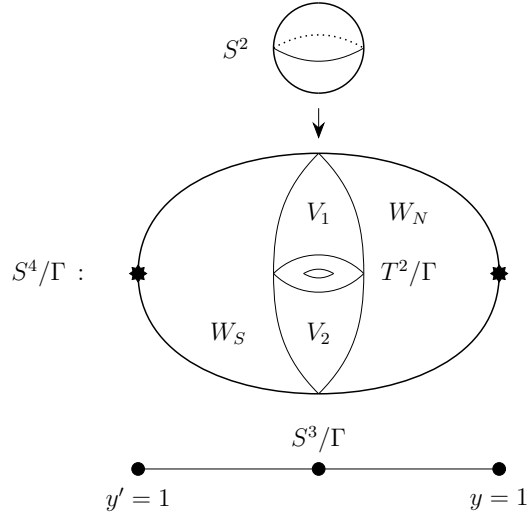

We now compute the homology groups $H_n(\partial X)$ via the Mayer-Vietoris long exact sequence
\begin{equation}
   \dots ~\rightarrow~ H_n(W_N\cap W_S)  ~ \xrightarrow[]{\,\iota_n \,}~ H_n(W_N)\oplus H_n(W_S) ~\rightarrow~ H_n(\partial X)~\xrightarrow[]{\,\partial_n \,}~  \dots
\end{equation}
where we choose the North and South pole patches $W_{N,S}$ such that their intersection is $Y/\Gamma\equiv Y_\Gamma$. Here $Y$ is the collection of all fiberes with base coordinate $y,y'=0$, topologically $Y=S^3\times S^2$. We therefore begin by computing the homology groups $H_n(Y_\Gamma)$ via the Mayer-Vietoris long exact sequence
\begin{equation}
   \dots ~\rightarrow~ H_n( V_1\cap V_2)  ~ \xrightarrow[]{\,\iota_n \,}~ H_n(V_1)\oplus H_n(V_2) ~\rightarrow~ H_n(Y_\Gamma)~\xrightarrow[]{\,\partial_n \,}~  \dots\,.
\end{equation}
Here, the covering $Y_\Gamma=V_1\cup V_2$ descends from a covering $Y=U_1\cup U_2$, and $U_i$ are topologically solid tori, from the canonical decomposition of $S^3$ as two solid tori, times the fiber $S^2$. See Figure \ref{fig:Decomposition} for a sketch of the decompositions.
With this we compute:
\begin{equation}
\label{eq:Identifications2}
\begin{aligned}
  \textnormal{Case 1}:&\quad  H_*(Y_\Gamma)=\{ \mathbb{Z}, \mathbb{Z}_K, \mathbb{Z}, \mathbb{Z}\oplus \mathbb{Z}_K, 0, \mathbb{Z} \}   \\[1mm]
  \textnormal{Case 2}:& \quad H_*(Y_\Gamma)=\{ \mathbb{Z}, \mathbb{Z}_m, \mathbb{Z}, \mathbb{Z}\oplus \mathbb{Z}_K, 0, \mathbb{Z} \}  \\[1mm]
  \textnormal{Case 3}:&\quad  H_*(Y_\Gamma)=\{ \mathbb{Z}, \mathbb{Z}_K, \mathbb{Z}, \mathbb{Z}\oplus \mathbb{Z}_K, 0, \mathbb{Z} \}  \\[1mm]
   \textnormal{Case 4}:& \quad  H_*(Y_\Gamma)= \{ \mathbb{Z}, 0, \mathbb{Z}\oplus \mathbb{Z}_{\textnormal{gcd}(K',L')}, \mathbb{Z}\oplus \mathbb{Z}_{K'}\oplus \mathbb{Z}_{L'} , 0, \mathbb{Z} \}
\end{aligned}
\end{equation}
where $K'=\textnormal{gcd}(K,2a)$ and $L'=\textnormal{gcd}(L,2b)$. Feeding these results into the next Mayer-Vietoris we find
\begin{equation}
\label{eq:IdentificationsHZ}
\begin{aligned}
   \textnormal{Case 1}:&\quad   H_*(\partial X)=\{\mathbb{Z},0,\mathbb{Z}\oplus \mathbb{Z}_K^2,0, \mathbb{Z}\oplus \mathbb{Z}_K,0,\mathbb{Z}  \}  \\[1mm]
    \textnormal{Case 2}:& \quad  H_*(\partial X)=\{\mathbb{Z},0,\mathbb{Z}\oplus \mathbb{Z}_m^2,0, \mathbb{Z}\oplus \mathbb{Z}_K,0,\mathbb{Z}  \} \\[1mm]
  \textnormal{Case 3}:& \quad  H_*(\partial X)=\{\mathbb{Z},0,\mathbb{Z}\oplus \mathbb{Z}_K,0, \mathbb{Z}\oplus \mathbb{Z}_K,0,\mathbb{Z}  \}
  \\[1mm]
   \textnormal{Case 4}:& \quad  H_*(\partial X)=\{\mathbb{Z},0,\mathbb{Z},\mathbb{Z}_{\textnormal{gcd}(K',L')}, \mathbb{Z}\oplus \mathbb{Z}_{K'}\oplus \mathbb{Z}_{L'},0,\mathbb{Z}  \}\,.
\end{aligned}
\end{equation}

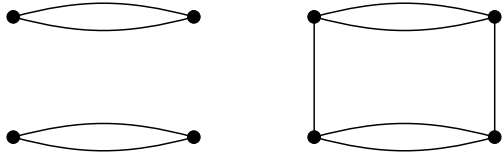
\begin{figure}
\centering
\scalebox{0.8}{
\begin{tikzpicture}
	\begin{pgfonlayer}{nodelayer}
		\node [style=none] (0) at (-4, 1) {};
		\node [style=none] (1) at (-1, 1) {};
		\node [style=none] (2) at (-4, -1) {};
		\node [style=none] (3) at (-1, -1) {};
		\node [style=none] (4) at (1, -1) {};
		\node [style=none] (5) at (4, -1) {};
		\node [style=none] (6) at (4, 1) {};
		\node [style=none] (7) at (1, 1) {};
		\node [style=Circle] (8) at (-4, 1) {};
		\node [style=Circle] (9) at (-1, 1) {};
		\node [style=Circle] (10) at (-1, -1) {};
		\node [style=Circle] (11) at (-4, -1) {};
		\node [style=Circle] (12) at (1, 1) {};
		\node [style=Circle] (13) at (4, 1) {};
		\node [style=Circle] (14) at (4, -1) {};
		\node [style=Circle] (15) at (1, -1) {};
	\end{pgfonlayer}
	\begin{pgfonlayer}{edgelayer}
		\draw [style=ThickLine, bend left=15] (7.center) to (6.center);
		\draw [style=ThickLine, bend right=15] (4.center) to (5.center);
		\draw [style=ThickLine, bend left=15] (0.center) to (1.center);
		\draw [style=ThickLine, bend right=15] (2.center) to (3.center);
		\draw [style=ThickLine] (7.center) to (4.center);
		\draw [style=ThickLine] (6.center) to (5.center);
		\draw [style=ThickLine, bend right=15] (0.center) to (1.center);
		\draw [style=ThickLine, bend left=345] (3.center) to (2.center);
		\draw [style=ThickLine, bend right=15] (7.center) to (6.center);
		\draw [style=ThickLine, bend left=345] (5.center) to (4.center);
	\end{pgfonlayer}
\end{tikzpicture}

}
\caption{Fixed point loci for Case 4 in the asymptotic $\mathbb{CP}^3$, presented as wedge sums. Lines represents 2-spheres. Left:  $\textnormal{gcd}(K',L')=1$. Right: $\textnormal{gcd}(K',L')\neq1$.   }
\label{fig:singularlociwedge}
\end{figure}

Next we compute the homology groups $H_n(\partial X^\circ)$. The computation runs via Lefschetz-duality and deformation equivalence which give
\be
H_n(\partial X^\circ)\cong H^{6-n}(\partial X^\circ,\partial(\partial X^\circ))\cong  H^{6-n}(\partial X,\Sigma)
\ee
allowing us to employ the long exact sequence in relative cohomology of the pair $(\partial X, \Sigma)$. These isomorphisms allow us to make use of the simple topology of the comparably singular loci $\Sigma$ which are:
\begin{equation}
\label{eq:SingLoci}
\begin{aligned}
   \textnormal{Case 1}:&\quad   \Sigma= *_1 \,\dot \cup *_2  \,\dot \cup *_3  \,\dot\cup *_4 \\[1mm]
    \textnormal{Case 2}:& \quad \Sigma= S^2  \,\dot \cup\, S^2 \\[1mm]
  \textnormal{Case 3}:& \quad  \Sigma= \begin{cases} S^2  \,\dot \cup\, S^2\,, \qquad~~\,  K \textnormal{~even}  \\ S^2  \,\dot\cup *_1  \dot\cup\, *_2 \,, \quad K \textnormal{~odd}   \end{cases}
  \\[1mm]
   \textnormal{Case 4}:& \quad \Sigma= \begin{cases}\vee_{i=1}^4 S^2_i \,, \qquad~~~\,  \textnormal{gcd}(K',L')= 1  \\ \vee_{i=1}^6 S^2_i  \,,  \qquad~~~\,  \textnormal{gcd}(K',L')\neq 1   \end{cases}
\end{aligned}
\end{equation}
where $\dot \cup$ denotes disjoint union, $*_i$ indicates a point, and in the final line the wedge sums are as shown in figure \ref{fig:singularlociwedge}. With this we compute
\begin{equation}
\label{eq:IdentificationsHZcirc}
\begin{aligned}
   \textnormal{Case 1}:&\quad   H_*(\partial X^\circ)=\{\mathbb{Z},\mathbb{Z}_K,\mathbb{Z}, \mathbb{Z}_K^2,\mathbb{Z},\mathbb{Z}^3  \}  \\[1mm]
  \textnormal{Case 2}:& \quad  H_*(\partial X^\circ)=\{\mathbb{Z}, \mathbb{Z}_K,\mathbb{Z},\mathbb{Z}^2\oplus \mathbb{Z}_m^2, \mathbb{Z},\mathbb{Z}  \}  \\[1mm]
   \textnormal{Case 3}:& \quad  H_*(\partial X^\circ)=\begin{cases} \{\mathbb{Z},\mathbb{Z}_K,\mathbb{Z},\mathbb{Z}\oplus \mathbb{Z}_K,\mathbb{Z},\mathbb{Z}^2  \} \qquad K\textnormal{ odd}\\ \{ \mathbb{Z}, \mathbb{Z}_K, \mathbb{Z}, \mathbb{Z}\oplus \mathbb{Z}_K, 0, \mathbb{Z} \} \qquad K\textnormal{ even}
\end{cases}  \\[1mm]
   \textnormal{Case 4}:& \quad  H_*(\partial X^\circ)=\{\mathbb{Z},\mathbb{Z}_{K'}\oplus \mathbb{Z}_{L'},\mathbb{Z}\oplus \mathbb{Z}_{\textnormal{gcd}(K',L')} ,\mathbb{Z}^5,\mathbb{Z}^3\}
\end{aligned}
\end{equation}
where for case 4 we have only given the result for the generic case with $ \textnormal{gcd}(K',L')\neq 1$. Here the universal coefficient theorem can be used to check the ranks of the homology groups in lower degrees by considering $H_n(\partial X,\Sigma)$. In higher degrees the full groups follow by exactness as $\Sigma$ is low-dimensional.

Finally we compute the homology quotients $H_n(X,\partial X)/H_n(X)$ from the long exact sequence in relative homology for the pair $(X,\partial X)$. Here the only non-trivial computation lies in degree $n=4$ where the free cycles in $\partial X$ are quotients of the hyperplane class in $\mathbb{CP}^2$ while in $X$ they are quotients of $S^4$. Here we have $H_4(Z,\partial Z)\cong H^3(Z)\cong H^3(S^4)=0$ with the smooth space $Z=\Lambda_{\textnormal{ASD}}^{2}(S^4)$, and similarly $H_5(Z,\partial Z)=0$. We see that  $\mathbb{CP}^2$ does not trivialize upon inclusion into the bulk, rather $[\mathbb{CP}^2]\mapsto [S^4]$ in homology. We find this to be preserved in the quotient and overall compute:
\begin{equation}
\label{eq:IdentificationsHZrel}
\begin{aligned}
   \textnormal{Case 1}:&\quad   H_*(X,\partial X)/H_*(X)=\{{0},0,0,\mathbb{Z}\oplus \mathbb{Z}_K,0,  \mathbb{Z}_K,0,\mathbb{Z}  \}  \\[1mm]
  \textnormal{Case 2}:& \quad  H_*(X,\partial X)/H_*(X)=\{{0},0,0,\mathbb{Z}\oplus \mathbb{Z}_m,0, \mathbb{Z}_K,0,\mathbb{Z}  \} \\[1mm]
    \textnormal{Case 3}:&\quad   H_*(X,\partial X)/H_*(X)=\{{0},0,0,\mathbb{Z},0, \mathbb{Z}_K,0,\mathbb{Z}  \}   \\[1mm]
   \textnormal{Case 4}:& \quad  H_*(X,\partial X)/H_*(X)=\{{0},0,0,\mathbb{Z},\mathbb{Z}_{\textnormal{gcd}(K',L')},  \mathbb{Z}_{K'}\oplus \mathbb{Z}_{L'},0,\mathbb{Z}  \}\,.
\end{aligned}
\end{equation}

\newpage

\bibliographystyle{utphys}
\bibliography{G2ConesArXiv}

\end{document}